\documentclass{aa}  
\usepackage{graphicx}
\usepackage{txfonts}
\usepackage{hyperref}   
\hypersetup{colorlinks=true,linkcolor=blue,citecolor=blue,filecolor=blue,urlcolor=blue}
\newcommand{\alpacc}{$\alpha_\mathrm{acc}$}
\newcommand{\mdot}{$\dot{m}$}
\newcommand{\mseed}{$M_\mathrm{seed}$}
\newcommand{\rseed}{$R_\mathrm{seed}$}
\newcommand{\xd}{$X_\mathrm{d}$}
\newcommand{\mdotyr}{M$_{\sun}$/yr}
\newcommand{\msun}{M$_{\sun}$}
\newcommand{\rsun}{R$_{\sun}$}
\newcommand{\lsun}{L$_{\sun}$}
\newcommand{\mj}{M$_\mathrm{J}$}
\newcommand{\tkh}{$\tau_\mathrm{KH}$}
\newcommand{\tacc}{$\tau_\mathrm{acc}$}
\newcommand{\tce}{$T_\mathrm{bce}$}
\newcommand{\teff}{$T_\mathrm{eff}$}
\begin{document} 

\title{Protostellar accretion in low mass metal poor stars and the cosmological lithium problem}

\author{Emanuele Tognelli\inst{1}
          \and
          Pier Giorgio Prada Moroni\inst{1,2}
          \and
          Scilla Degl'Innocenti\inst{1,2}
          \and
          Maurizio Salaris\inst{3}
          \and
          Santi Cassisi\inst{2,4}
          }
\institute{Dipartimento di Fisica, Universit\`a di Pisa, Largo Bruno Pontecorvo 3, I-56127, Pisa, Italy\\
         \email{ema.tog@gmail.com}
         \and
         INFN - Sezione di Pisa, Largo Bruno Pontecorvo 3, I-56127, Pisa, Italy\\
         \email{pier.giorgio.prada.moroni@unipi.it}\\
         \email{scilla.deglinnocenti@unipi.it}
                 \and
         Astrophysics Research Institute, Liverpool John Moores University, 146 Brownlow Hill, Liverpool L3 5RF, UK
         \and
         INAF - Osservatorio Astronomico d'Abruzzo, Via M. Maggini, I-64100, Teramo, Italy
        }
\date{Received 18 September, 2019; accepted 24 April, 2020}

\abstract
{The cosmological lithium problem, that is, the discrepancy between the lithium abundance predicted by the Big Bang nucleosynthesis and the one observed for the stars of the 'Spite plateau', is one of the long standing problems of modern astrophysics. Recent hints for a possible solution involve lithium burning induced by protostellar mass accretion on Spite plateau stars. However, to date, most of the protostellar and pre-main sequence  stellar models that take mass accretion into account have been computed at solar metallicity, and a detailed analysis on the impact of protostellar accretion on the lithium evolution in the metal-poor regime, which is relevant for stars in the Spite plateau, is completely missing.}
{The purpose of this paper is to fill this gap, analysing, in detail, for the first time the effect of protostellar accretion on low metallicity low-mass stars with a focus on pre-main sequence lithium evolution.}
{We computed the evolution from the protostar to the main-sequence phase of accreting models with final masses equal to 0.7 and 0.8~\msun, and three metallicities $Z=0.0001$, $Z=0.0010$, and $Z=0.0050$, corresponding to [Fe/H]$\sim -2.1$, $-1.1$ (typical of Spite plateau stars), and [Fe/H]$\sim -0.42$, respectively. We followed the temporal evolution of the chemical composition by considering nuclear burning, convective mixing, and diffusion. The effects of changing some of the main parameters affecting accreting models, that is the accretion energy (i.e. cold versus hot accretion), the initial seed mass \mseed{} and radius \rseed{}, and the mass accretion rate \mdot{} (also considering episodic accretion), have been investigated in detail. }
{As for the main stellar properties and in particular the surface $^7 Li$ abundance, hot accretion models converge to standard non-accreting ones within 1 Myr, regardless of the actual value of \mseed{}, \rseed{}, and \mdot. Also, cold accretion models with a relatively large \mseed{} ($\gtrsim 10~$\mj) or \rseed{} ($\gtrsim 1~$\rsun) converge to standard non-accreting ones in less than about 10-20~Myr. However, a drastically different evolution occurs whenever a cold protostellar accretion process starts from small values of \mseed{} and \rseed{} (\mseed$\sim 1~$\mj, \rseed$\lesssim 1~$\rsun). These models almost entirely skip  the standard Hayashi track evolution and deplete lithium before the end of the accretion phase. The exact amount of depletion depends on the actual combination of the accretion parameters (\mdot, \mseed{},  and \rseed), achieving in some cases the complete exhaustion of lithium in the whole star. Finally, the lithium evolution in models accounting for burst accretion episodes or for an initial hot accretion followed by a cold accretion phase closely resemble that of standard non-accreting ones.}
{To significantly deplete lithium in low-mass metal poor stars by means of protostellar accretion, a cold accretion scenario starting from small initial \mseed{} and \rseed{} is required. Even in this extreme configuration leading to a non-standard evolution that misses almost entirely the standard Hayashi track, an  unsatisfactory fine tuning of the parameters governing the accretion phase is required to deplete lithium in stars of  different mass and metallicity -- starting from the Big Bang nucleosynthesis abundance -- in such a way as to produce the observed Spite plateau.}

\keywords{Stars: abundances - Stars: evolution - Stars: formation - Stars: pre-main sequence - Stars: protostars}

\maketitle

\section{Introduction}
The Big Bang nucleosynthesis (BBN) is one of the underpinnings of the standard Big Bang cosmological model \citep[see, e.g.][]{cyburt16, coc17}. A good agreement is found between BBN calculations and primordial abundances of deuterium and helium inferred from observations of cosmological clouds at high redshift on the line of sight of distant quasars \citep[for deuterium, see e.g.][]{pettini} and observations of ${\rm H_{II}}$ regions of compact blue galaxies \citep[for $^4 He$, see e.g.][]{aver}.

The situation is different for the primordial $^7 Li$ abundance\footnote{Throughout the paper,  Li denotes the abundance of $^7 Li$. BBN predicts an abundance of $^6 Li$ orders of magnitude lower than the $^7 Li$ one. Also observationally the abundance of $^6 Li$ in metal poor main sequence stars appears to be negligible \citep{lind13}.}. Estimates of the cosmological Li abundance are based on measurements on the so-called Spite plateau \citep[see, e.g.][and references therein]{spite82, cp05, asplund06, melendez10b, sbordone10}. In brief, field halo main sequence stars with [Fe/H]$ < -$1.5 and \teff{} above $\sim$5900~K do show remarkably uniform surface Li abundances\footnote{Here, we use the spectroscopic notation $A(Li)={\rm log}(N(Li)/N(H))$+12 where $N(X_i)$ is the numerical abundance of the element $X_i$.} $A(Li)=2.1$-$2.4$, the exact value varies from author to author, depending on the adopted \teff{}  scale\footnote{Observations of Spite plateau abundances typically determine the sum of $^7 Li$ and $^6 Li$ abundances but, as mentioned previously, the $^6 Li$ abundance is negligible.}.

In this [Fe/H] regime, Galactic chemical evolution does not predict a change of Li in the intergalactic medium compared to the  cosmological value \citep[e.g.][]{romano01}, and also metal poor low-mass main sequence (MS) stars in the \teff{} range of the Spite plateau are expected (from standard stellar evolution models) to have negligibly depleted Li during the pre-main  sequence (PMS) phase. If convection is the only element transport mechanism during the MS, a plateau is indeed expected and the observed Spite plateau $A(Li)$ value should be consistent with the BBN one. BBN calculations in the framework of the current standard $\Lambda$CDM cosmological model however give values between $A(Li)=2.67$ \citep[see, e.g.][]{cyburt16} and $A(Li)=2.75$ \citep{pitrou18}, about 0.3-0.6~dex larger than the Spite plateau value. This is the so-called cosmological lithium problem.

A number of potential avenues to solve this discrepancy have been proposed over the years. They range from new physics that modifies BBN results \citep[see, e.g.][and references therein]{fields11, gpp16, coc17}, the reassessment of BBN reaction rates \citep[e.g.][]{damone18}, and the astration of 30\%-50\%  of halo matter in Population III stars during galaxy formation \citep{piau06}; however, the problem is still open.

At the same time, an additional element transport mechanism that, from first principles, is expected to be effective in stars --atomic diffusion-- adds a further layer of complexity to the interpretation of Spite plateau abundances. Atomic diffusion makes Li slowly sink below the convective envelope of MS stars \citep[see, e.g.][and references therein]{richard05}, and this surface depletion is expected to increase with increasing \teff{}. The fact that the observed abundances are then expected to be lower than the initial ones can explain the discrepancy with BBN, but the dependence of Li depletion on \teff{} means that a Spite plateau should not exist starting from a constant BBN Li abundance. 

Potential solutions to this problem envisage the inclusion of MS mass loss with appropriately chosen  (higher than solar mass loss) rates \citep{swenson95, vauclair95}, or an  ad-hoc (in the sense that it is not directly derived from the calculated efficiency of any physical processes active inside stars) turbulent transport below the convective envelope \citep{richard05}. When both processes are combined with atomic diffusion, they induce a uniform Li depletion with \teff{} in the appropriate temperature range, producing a plateau as observed; however, at $A(Li),$ the values are lower than the initial one \citep[see][for details]{vauclair95, richard05}. In this way, when starting from the measured plateau abundances, the cosmological Li discrepancy is mitigated, and in case of the 'ad hoc' turbulence models which were introduced by \citet{richard05}, the cosmological Li discrepancy is almost erased \citep{spite12}.

Depending on the precise value of $A(Li)$ assumed for the Spite plateau, the turbulence needed to match the BBN Li abundance --and produce a theoretical Li-abundance plateau above 5900~K and [Fe/H]=$-$1.5 -- may or may not bring additional Li to the Li-burning regions during the MS, compared to the case of pure atomic diffusion. To this purpose, \citet{msb12} measured the surface Li abundance in a sample of red giant branch (RGB) stars at the completion of the first dredge up, with metallicities typical of Spite plateau stars. After correcting for the effect of the first dredge up by using theoretical stellar models with atomic diffusion, these authors derived an initial $A(Li)$ for these objects that is $\sim$0.3-0.4~dex lower than the BBN abundance. Whether this discrepancy can be ascribed to the extra Li burned during the MS by the turbulence necessary to produce a Spite plateau that is consistent with BBN has yet to be established.

Fairly recently, a number of authors \citep[see, e.g.][and references therein]{asplund06, bonifacio07, aoki09, sbordone10, bonifacio12} have found that Spite plateau (in the sense of \teff{} and [Fe/H] ranges) stars below [Fe/H]$\sim -$3.0 display a steady decrease of $A(Li)$ (and an associated large spread) with decreasing [Fe/H]. At the moment, there is no clear explanation for this. We can only notice that atomic diffusion is expected to increase the depletion of surface Li with decreasing [Fe/H] during the MS evolution of low mass metal poor stars because of their increasingly thinner convective envelopes.

Finally, a different idea, which like all proposed mechanisms needs to make more or less ad hoc assumptions, has recently been put forward by \citet{fu15}, involving the PMS evolution of Spite plateau stars. According to these authors, Li evolution during the PMS phase could be regulated by the competition between the destroying effect of a very efficient convective overshooting at early times during the Hayashi track evolution and a late residual accretion 
of pristine matter, which contains Li, that is modulated by UV photoevaporation caused by the central accreting star. The photoevaporation controls the final abundance depending on the mass, and the combination of these effects in stars with different initial masses tends to produce a uniform Li abundance just after the PMS, at a value below the initial BBN one. During the MS evolution, atomic diffusion is then forced to be inhibited in the outermost region of the model envelopes to preserve a plateau of Li abundances. The decrease of $A(Li)$ for [Fe/H] below $\sim-$3.0 is potentially explained by a failed or weaker late accretion.

It should be clear from this brief overview that all hypotheses based on stellar physics and evolution that are used to explain the cosmological Li discrepancy need to resort, to different degrees, to fine tuning and/or ad hoc assumptions. Here we want to add an additional important element to the discussion, namely the effect on the PMS and MS Li abundances of the accretion phase  during the protostellar evolution, which is not usually included in PMS calculations such as \citet{fu15} models. As shown by \citet{baraffe10}, for low-mass models at solar metallicity, the accretion phase could alter the starting value of $A(Li)$ at the beginning of the PMS and MS, reducing it below the initial value at the beginning of accretion. If also confirmed for metal poor low-mass models, this would open an additional avenue of research to explain the discrepancy between Spite plateau and BBN abundances. 

To date, the initial accretion from a circumstellar disc onto a central seed in low mass stars has mainly been modelled at solar metallicity \citep[see e.g.][]{hartmann97,siess97,baraffe09,hosokawa09,baraffe10,hosokawa10,hosokawa11,baraffe12,kunitomo17,kunitomo18,haemmer19} and only in a few cases, especially for massive stars, in metal-poor models \citep[e.g.][]{hosokawa09b}. On the other hand, only a few attempts have been made to study the effect of the protostellar evolution on surface lithium abundance in low-mass metal-poor stars  \citep{dantona14, tognelli15c}. The purpose of this paper is to fill such a gap by analysing in detail the effect of the protostellar accretion phase on low metallicity stars, with a particular focus on PMS Li burning. 

The paper is organised as follows. Section~\ref{models} describes our calculations and is followed by Sect~\ref{standard}, which summarises PMS and surface Li evolution in standard constant mass evolutionary calculations. Sections~\ref{cold}, ~\ref{hot}, and ~\ref{metals} present our exploration of the parameter space regarding models including protostellar accretion and the effect of changing the initial metallicity in the calculations. A summary and conclusions follow in Sect.~\ref{summary}.

\section{The models}
\label{models}
In standard stellar evolution calculations, PMS evolution begins once a completely formed, fully convective, bright and expanded star has already attained its final MS mass. However, stars form from a parental cloud accreting matter on a central seed until eventually the accretion ceases. Such an accretion process (i.e. protostellar evolution) should be treated, at least during the very early stages, by hydrodynamical codes because of the presence of shock fronts. However, hydrodynamical simulations show that at some point, the protostar evolution becomes hydrostatic \citep{larson69,larson72}; from that point on, it is possible to model the subsequent evolution using hydrostatic models. 

A crucial issue in protostellar evolution is the way in which the accretion proceeds. Roughly speaking, the star can be deeply embedded into a cloud, surrounded by a thick  envelope of mass that accretes on the whole stellar surface (i.e. spherical accretion) or it can accrete from a disc on a limited part of the stellar surface (disc accretion). The first scenario has been analysed in the pioneering works by \citet{stahler86} and \citet{palla91}, and recently revised by \citet{hosokawa09}. A more common approach is the disc accretion \citep[see e.g.][]{hartmann97,siess97,baraffe09}. The advantage of the latter is that it is simpler to model in a stellar evolution code. In addition, it can be used to simulate the main effects of a spherical accretion scenario, assuming what is called hot accretion \citep{baraffe09,hosokawa11}. In this situation, the matter that falls on the star carries an amount of internal energy per unit mass larger than that in the surface stellar matter, thus providing an external and additional energy source to the star. 

The stellar models discussed here were computed with the Pisa version of the FRANEC evolutionary code \citep[][]{deglinnocenti08,dellomodarme12}, adopting the same input physics and parameters as well as using the same treatment for protostellar mass accretion described in \citet[][]{tognelli15c, tognelli18}. A key input physics that is relevant for the  protostellar and PMS evolution is the set of reaction rates, in particular those for the deuterium \citep[$^2$H$(p,\gamma)^3$He from][]{descouvemont04} and lithium burning channels. We took the following lithium burning reactions into account: $^7$Li$(p,\alpha)\alpha$ \citep[rate from][]{lamia12} and the channel $^7$Li$(\alpha,\gamma)^{11}$B \citep[inefficient during the protostellar phase, rate from ][]{nacre}. Information on the other reaction rates can be found in the previously mentioned papers.

We followed the evolution from the protostar to the MS phase of models with final masses equal to M=0.7 and 0.8~\msun, and three different chemical compositions are considered to be representative of the metal poor Galactic populations.  The selected masses are typical values of population II halo stars with ages between $\sim12$ and $\sim13.5$~Gyr on the Spite plateau \citep{sc:01}. More precisely, as initial metallicity and helium abundances we considered the pairs of values ($Z=0.0001$, $Y=0.2487$), ($Z=0.0010$, $Y=0.2505$), and ($Z=0.0050$, $Y=0.2585$), which correspond by adopting the \citet[][]{asplund09} solar metal distribution to [Fe/H]$\sim -2.1$, $-1.1$ (typical of Spite plateau stars) and [Fe/H]$\sim -0.42$, respectively. We set the initial deuterium mass fraction to \xd=$4\times 10^{-5}$  \citep[consistent with recent cosmological deuterium mass fraction measurements, see e.g.][]{pettini08, pettini, cooke18}, and an initial lithium abundance $A(Li) = 2.67$, which is consistent with cosmological nucleosynthesis calculations, corresponding to a lithium mass fraction of $X_\mathrm{Li} \approx 2.46 \times 10^{-9}$ for $Z=0.0001$, $X_\mathrm{Li} \approx 2.44 \times 10^{-9}$ for $Z=0.0010$, and $X_\mathrm{Li} \approx 2.41 \times 10^{-9}$ for $Z=0.0050$. It is important to notice, however, that for the present differential analysis, the actual initial lithium content is inconsequential, and the results described in the following are independent of the adopted initial $A(Li)$.

We accounted for disc mass accretion during the protostellar phase following the same formalism described in detail by \citet{siess97}. Accretion affects the value of stellar mass, gravitational energy, chemical composition, and, when the matter is accreted with non-negligible internal energy content, the thermal energy of the star as well. In our calculations the accreted material, as well as accretion energy, was added uniformly and instantaneously to the convective envelope. It is important to notice that all of the models considered in this work have a convective envelope, which is thin or thick, during the protostellar accretion phase. 

The accreted material is uniformly redistributed in each mesh within the convective envelope of the star. More specifically, if we denote with $M_\mathrm{add}\equiv$\mdot$\Delta t$ the mass accreted after a time step $\Delta t$, at a given mass accretion rate \mdot{}, and the mass inside the convective envelope
of the star before the accretion with $M_\mathrm{ce}$, then the mass in the generic layer j after a time $\Delta t$ ($\Delta m_\mathrm{j}^\mathrm{post-acc}$) is simply given by 
\begin{equation}
\Delta m_\mathrm{j}^\mathrm{post-acc} = \bigg(1 + \frac{M_\mathrm{add}}{M_\mathrm{ce}}\bigg) \Delta m_\mathrm{j}^\mathrm{pre-acc}\equiv(1+f_\mathrm{j}) \Delta m_\mathrm{j}^\mathrm{pre-acc}
\label{eq:ma}
\end{equation}
where $\Delta m_\mathrm{j}^\mathrm{pre-acc}$ is the mass contained in layer j before the time step $\Delta t$.

We assumed that the accreted matter, with the same chemical composition of the stellar original matter, is instantaneously mixed. Additionally, we followed the temporal evolution of the chemical composition by considering the effects of nuclear burning, convective mixing, and diffusion. 

The material accreted from the inner boundary layer of the disc could, in principle, bring some energy into the star. We define the accretion luminosity $L_\mathrm{acc}$ (i.e. the energy per unit time that the accreted material carries into the star) by adopting the same formalism as \citet{siess97} as
\begin{equation}
L_\mathrm{acc} = \frac{1}{2} \alpha_\mathrm{acc}\frac{G M_\star \dot{m}}{R_\star}\equiv \epsilon_\mathrm{acc} \dot{m}
\label{eq:acc_ene}
\end{equation}
where, $M_\star$ and $R_\star$ denote the stellar mass and radius and \alpacc{} is a free parameter that sets the fraction of energy that is actually deposed into the star.  In the case of disc accretion, the value of \alpacc{} is varied within the range  [0,~1] \citep{siess97,baraffe09}. From the previous equation, we can define $\epsilon_\mathrm{acc}$ as the accretion energy per unit mass deposed into the star.

In the following, we refer to cold accretion when \alpacc=0 and hot accretion when \alpacc=1. For \alpacc=0 (cold accretion), the star accretes material with a negligible energy content, compared to the energy per unit mass of the stellar matter. In this case, the protostar changes its structure due to the mass increase and possibly due to a chemical composition variation, but no additional thermal energy is provided by the accretion. The adoption of \alpacc$\sim 0$ can be justified as long as the infalling matter is free to radiate its energy before hitting the star. This is expected to occur when the matter surrounding the star is optically thin \citep[see e.g. Appendix B in][]{baraffe12}.

On the other hand, for \alpacc=1 (hot accretion), the matter is accreted with an energy content larger than that of the stellar material \citep[see e.g.][]{hartmann97}, and the protostar is supplied with additional energy coming from an external source, that is, the disc. In this case it is assumed that the infalling matter is not capable of efficiently losing its internal energy before reaching the stellar surface. Such a configuration can also be used to describe a protostar that is deeply embedded in an optically thick cloud \citep[e.g.][]{hartmann97,baraffe12}. 

The accretion energy $\epsilon_\mathrm{acc}$ (if $\ne 0$) is uniformly redistributed in the region where the mass is accreted (i.e. in the convective envelope) in exactly the same way as the accreted material, according to the following expression,
\begin{equation}
\epsilon_\mathrm{acc,j} = \frac{\alpha_\mathrm{acc}}{2}\frac{GM_\star}{R_\star} f_\mathrm{j}=\epsilon_\mathrm{acc} f_\mathrm{j}
\label{eq:la}
.\end{equation}

Besides \alpacc, accretion models depend on several other parameters; this reflects the present uncertainty in modelling this early protostellar phase. The most important parameters are the mass of the initial accreting stellar model \mseed, the radius of the initial model \rseed{}, and the mass accretion rate \mdot. Reasonable values for these quantities can be constrained from observations and theoretical arguments, but their potential range is quite large. We discuss the range of values adopted for each of these parameters in the next sections. We set the zero-point of time (age of the star $t_0=0$) at the first model when the protostellar accretion begins, hence $M(t=0) = $\mseed{} and $R(t=0)=$\rseed. We are aware that setting $t_0=0$ at the beginning of the accretion neglects the time necessary for the formation of the second Larson core \citep[$\sim10^3$-$10^5$~yr, see e.g.][]{larson69}, but such an offset is orders of magnitude smaller than the PMS timescale ($\ga 10^7$~yr), and thus it has no impact on the PMS and MS evolution.

\subsection{Seed mass and radius}
We started the accretion on an initial model with characteristics similar to those expected for the second Larson core hydrostatic structure \citep{larson69,larson72}, which we refer to in the following as $M_\mathrm{seed}$ and $R_\mathrm{seed}$. The currently adopted \mseed{}, which is mainly derived for solar-like chemical composition, ranges between about 0.001-0.020~\msun{} and \rseed{} ranges from 0.5-10~\rsun{} \citep[see e.g.][]{larson69,masunaga00,machida2008,baraffe12,tomida13,vaytet2013}. Such values can be assumed to also be representative of metal poor seed models \citep[e.g.][]{omukai10}. Here, we only chose two values for \mseed, namely \mseed=0.001 and 0.010~\msun, which correspond to about 1 and 10 Jupiter masses ($M_J$), respectively. In the following, we refer to these cases as large Larson core (LLC) for \mseed=10~\mj, and small Larson core (SLC) for \mseed=1~\mj. Our reference choice for \rseed{} is 3.0~\rsun{}, which is compatible with the initial radius values obtained by \citet[][]{larson69} for a mass
of the order of 0.010~\msun. However, we tested the dependence on the initial radius by also adopting \rseed=0.5 and 1.5~\rsun{} \citep[e.g.][]{larson69}.

\subsection{Mass accretion rate}
Observations show that the mass accretion rate \mdot{} changes by orders of magnitude during the protostellar and the PMS phases. It ranges from values of about $10^{-3}$~\mdotyr{} for efficiently accreting young stars, to values below  $10^{-10}$~\mdotyr{} for stars older than a few million years, for which the accretion disc is dissipating \citep{hartmann96,muzerolle00,calvet05,muzerolle05,ingleby14,hartmann16}. Moreover, \mdot{} is also expected to vary on shorter timescales. Observations and modelling of accretion discs show that \mdot{} is characterised by periods of intense accretion (bursts) where \mdot{} can reach values as large as $10^{-3}~$\mdotyr, separated by periods of quiescence where \mdot{} drops \citep{hartmann96,muzerolle05b,vorobyov05,vorobyov06,bae13,ingleby14}. Clearly, due to the computational and observational difficulties and to the expected star to star variability, it is still not possible to derive a precise time dependence for matter accretion. Here we are not interested in adopting a complex time-dependent value of \mdot, which, as we show in the following, is not the parameter with a major impact on accreting protostellar models. Thus, as reference choice, we made the assumption of a constant accretion rate with \mdot=~$10^{-5}$~\mdotyr; a value similar to the one largely used in the literature \citep{baraffe09,hosokawa11,kunitomo17}. Then, we analysed the effect of using different accretion rates, namely \mdot=~$10^{-6}$, $5\times 10^{-6}$ and $5\times 10^{-5}$~\mdotyr. In addition to a constant \mdot{}, we also adopted a simple bursts accretion history, as discussed in Section~\ref{bursts}.
\begin{figure*}[!h]
\centering
\includegraphics[width=0.49\linewidth]{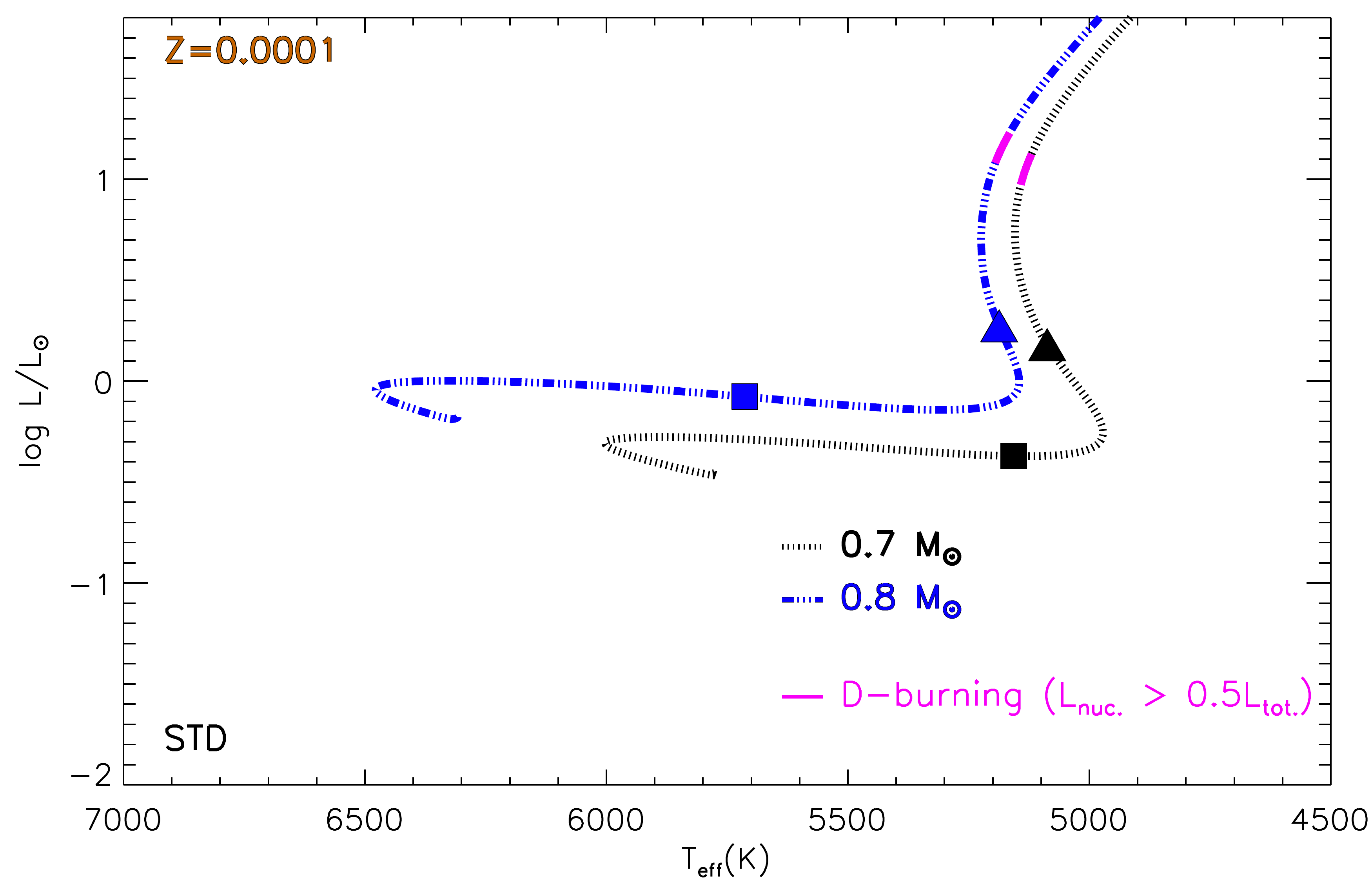}
\includegraphics[width=0.49\linewidth]{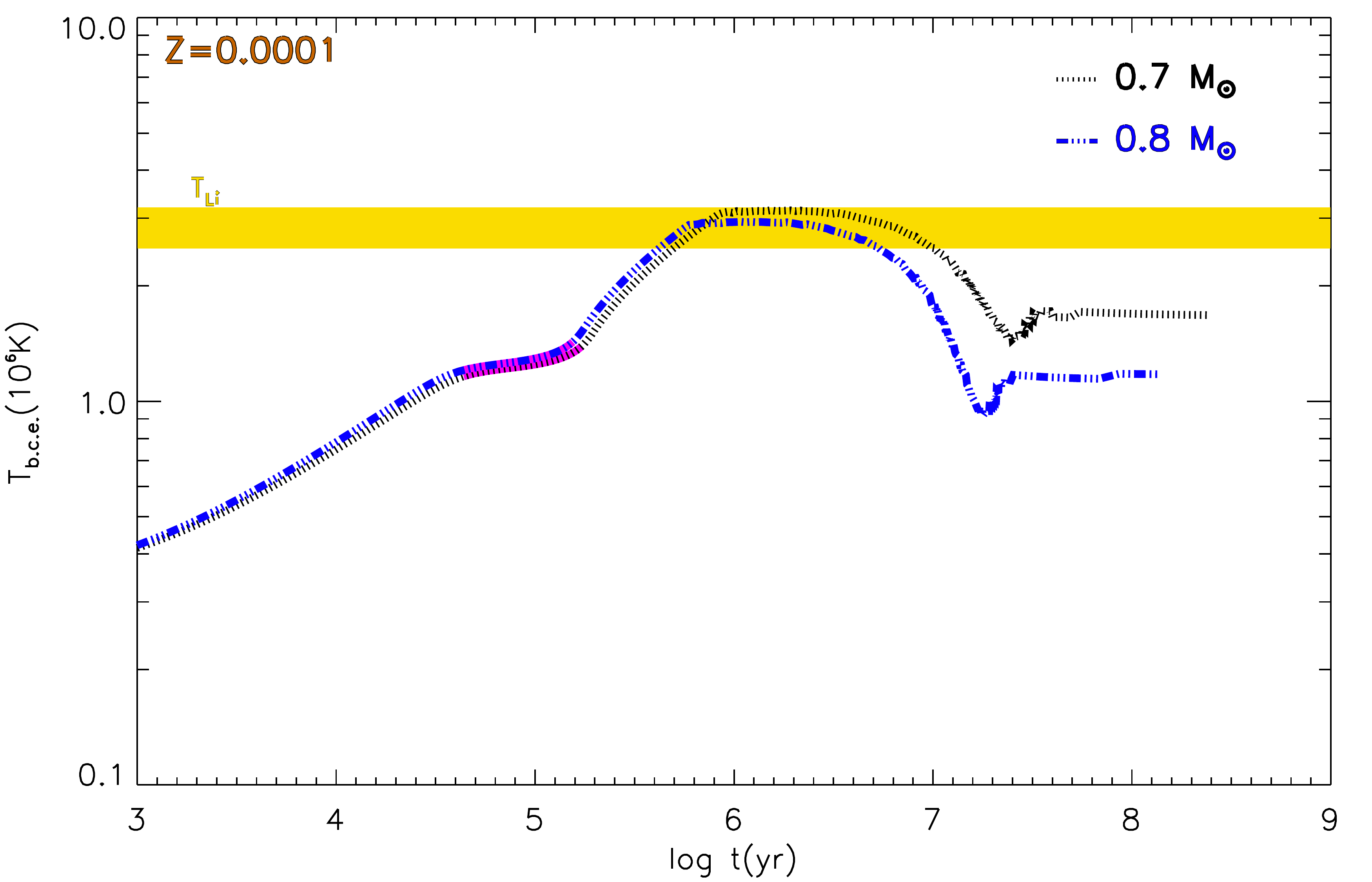}
\caption{PMS evolution of standard non-accreting models with masses $M=0.7$ and 0.8~\msun, and  $Z=0.0001$. Left panel: Evolution in the HR diagram.\ We marked the position of the models at 1~Myr (filled triangles) and 10~Myr (filled squares). Right panel: Evolution of the temperature at the base of the convective envelope as a function of age. The yellow shaded zone shows the Li-burning temperature range ($T_\mathrm{Li}=2.5$-$3.2\times 10^6$~K, see text). We also marked the region where the deuterium burning supplies more than 50~percent of the total luminosity (thick magenta line).}
\label{fig:std}
\end{figure*}

\section{Standard non-accreting models}
\label{standard}

Before investigating the effect of mass accretion on the evolution of the surface lithium content of low-mass metal poor PMS models, we briefly recall the main aspects of PMS evolution of standard non-accreting models, which begin their evolution as fully convective structures on the Hayashi track.

Figure~\ref{fig:std} shows the Hertzsprung-Russell (HR) diagram (left panel) and the temporal evolution of the temperature at the bottom of the convective envelope (\tce, right panel) for two standard models of 0.7 and 0.8~\msun{} metal poor stars ($Z=0.0001$) during the PMS phase until their zero age main sequence (ZAMS). During the early PMS, the evolution follows the Hayashi track, the locus of fully convective, chemically homogeneous models. In this phase the star contracts on thermal timescales (Kelvin-Helmholtz timescale\footnote{We made an order of magnitude estimate of the Kelvin-Helmholtz timescale by using the definition \tkh$=GM^2/2RL$, where $M$ is the mass, $R$ is the radius, $L$ is the luminosity of the star, and $G$ is the gravitational constant.}, \tkh), until a central temperature ($T_\mathrm{c}$) of about $10^6$~K is reached and deuterium is ignited ($\log (L/$\lsun$)\approx 0.9$-1.3~dex). D-burning generates enough energy to almost halt the gravitational contraction.\ Additionally, given the strong dependence of the energy generation on the temperature ($\epsilon_D \approx T^{12}$), it also prevents a further increase of $T_\mathrm{c}$ until deuterium is almost completely exhausted in the whole (fully convective) structure. This plateau of the central temperature can be seen in the right panel of Fig.~\ref{fig:std} at $\log t(yr) \sim 4.6$-5.2.

\begin{figure}
\centering
\includegraphics[width=0.98\columnwidth]{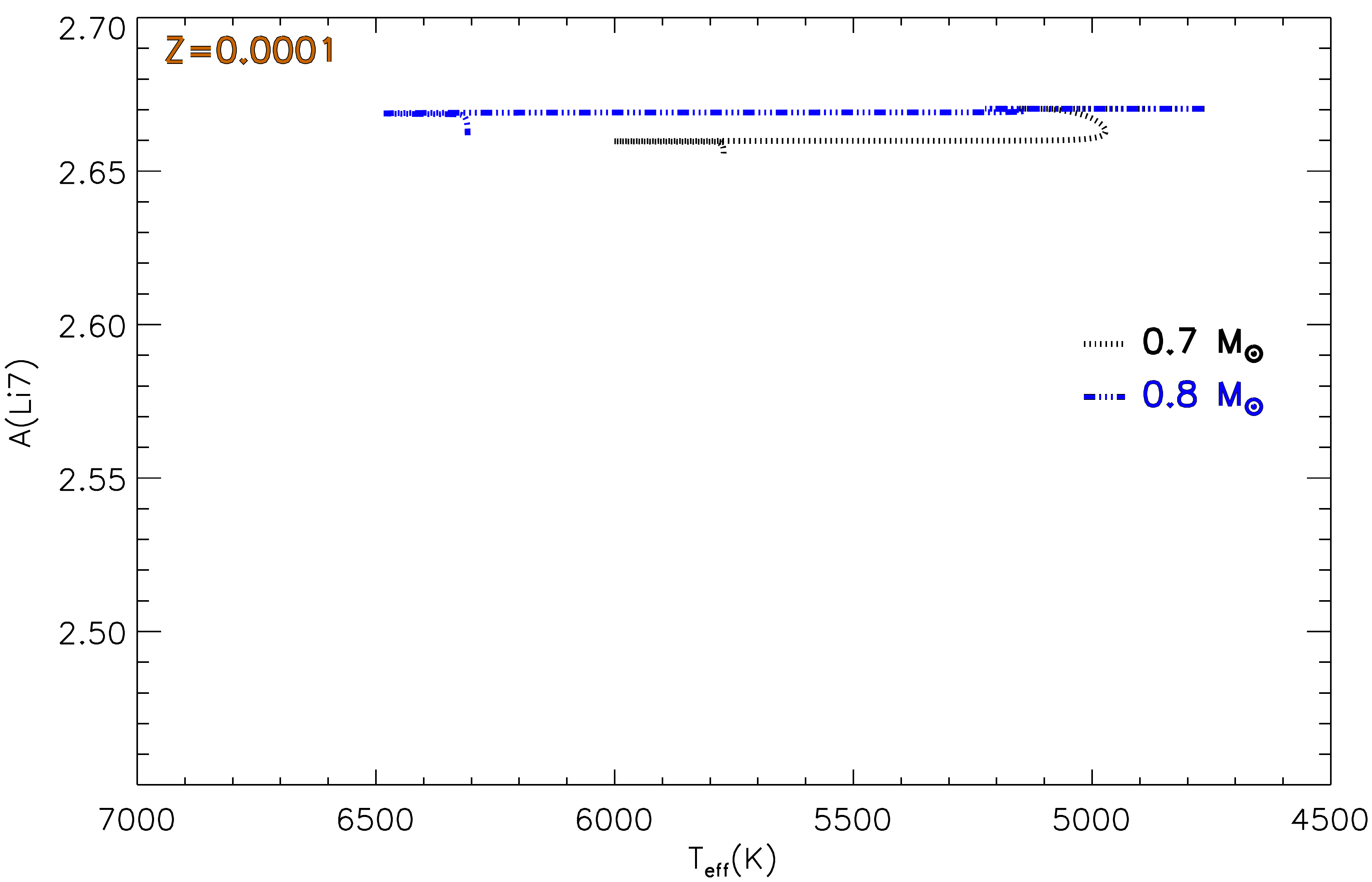}
\caption{Surface lithium abundance $A(Li)$ evolution as a function of the effective temperature until the ZAMS, for the non-accreting 0.7 and 0.8~\msun{}, $Z=0.0001$ models of Fig.~\ref{fig:std}.}
\label{fig:std_li}
\end{figure}

Once the deuterium has been exhausted, the gravitational contraction starts again: $T_\mathrm{c}$ rises and in turn, the radiative opacity decreases. Eventually, a radiative core forms and   the surface is chemically decoupled from the inner regions. As the central temperature rises, the base of the convective envelope withdraws outwards and \tce{} decreases, as is shown in the figure. Since lithium burning is only efficient\footnote{We determined an approximate value of the lithium burning temperature by taking the temperature at which the lithium abundance in the convective envelope has been reduced by 2\% with respect to the initial value. We found a temperature range between 2.5 and 3.2$\times 10^{6}$~K. Such a range does not appreciably change if a lithium variation of 1\% or 3\% is considered.} at $T\ga 2.5$-$3.2\times 10^6~\mathrm{K}\equiv T_\mathrm{Li}$ (depending on the stellar mass and chemical composition), the surface lithium abundance can only be depleted if the base of the convective envelope is hotter than $T_\mathrm{Li}$. The level of the surface Li depletion hence depends on the value of the difference $\Delta T_\mathrm{Li,burn.}\equiv T_\mathrm{bce} - T_\mathrm{Li}$ and on how long this difference is greater than 0. For the models displayed, $\Delta T_\mathrm{Li, burn.}$ is not very large (\tce{} barely reaches $3\times 10^6~$K) and it quickly decreases. As an example, at $\log t (yr) \sim 6.7$ for the $M=0.7$~\msun{} model and at $\log t(yr) \sim$ 7 for the $M=0.8$~\msun{} one, \tce{} is below the minimum temperature required to burn lithium in the PMS. As a consequence, the models experience a very small surface lithium depletion. Figure~\ref{fig:std_li} shows the evolution of the surface $A(Li)$ as a function of \teff{} (a plane useful to directly compare models and observations) for the same calculations of Fig.~\ref{fig:std}. The PMS evolution corresponds to the top horizontal line with \teff{} in the range of [4800, 6000]~K for the $M=0.7$~\msun{} calculations and [4800, 6500]~K for the $M=0.8$~\msun{} ones. The 0.8~\msun{} model shows a depletion of 0.01~dex, while the 0.7~\msun{} one does not deplete surface lithium; this clearly shows that during the PMS phase such models essentially do not destroy lithium.

\begin{figure*}
\centering
\includegraphics[width=0.49\linewidth]{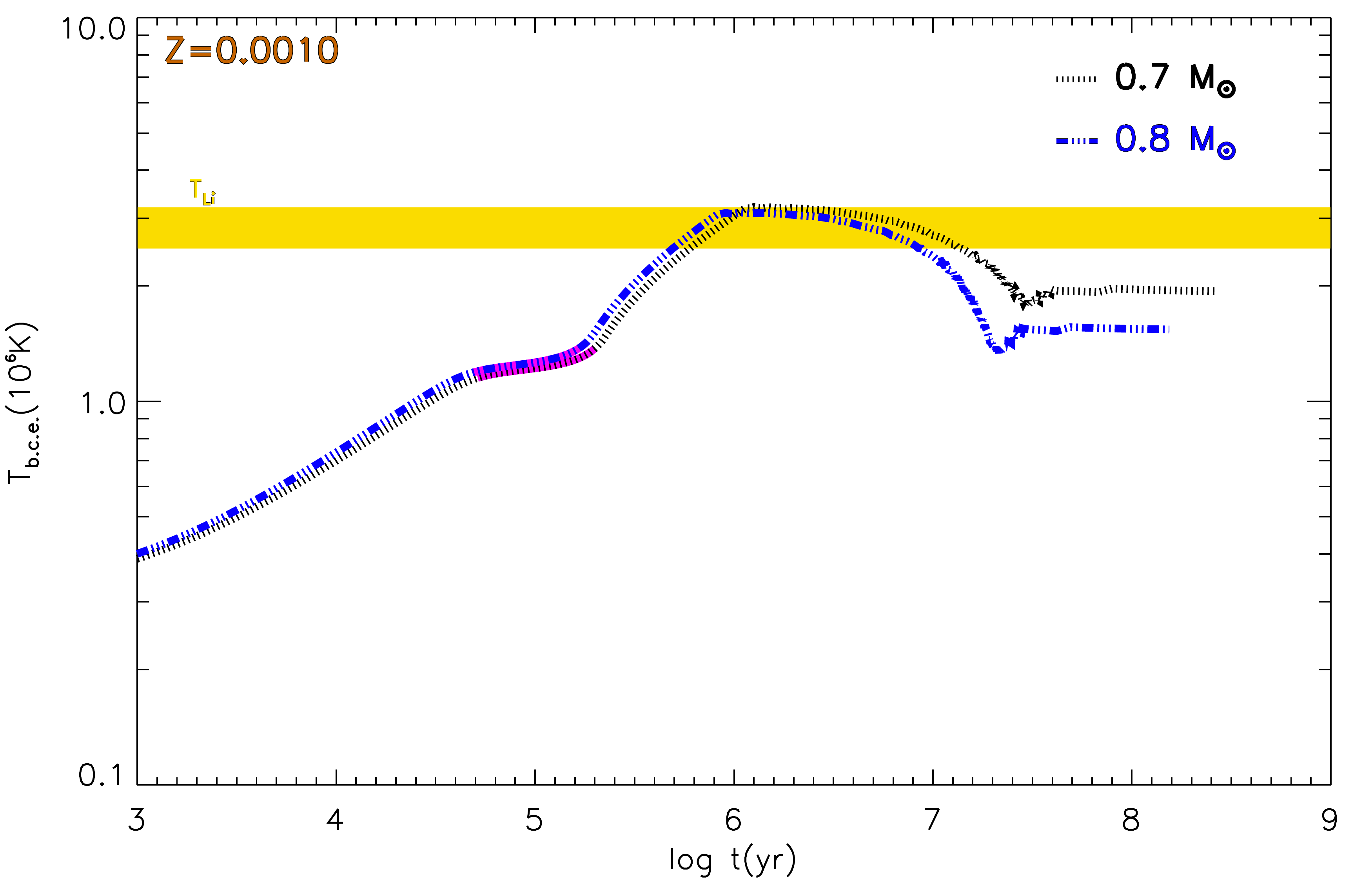}
\includegraphics[width=0.49\linewidth]{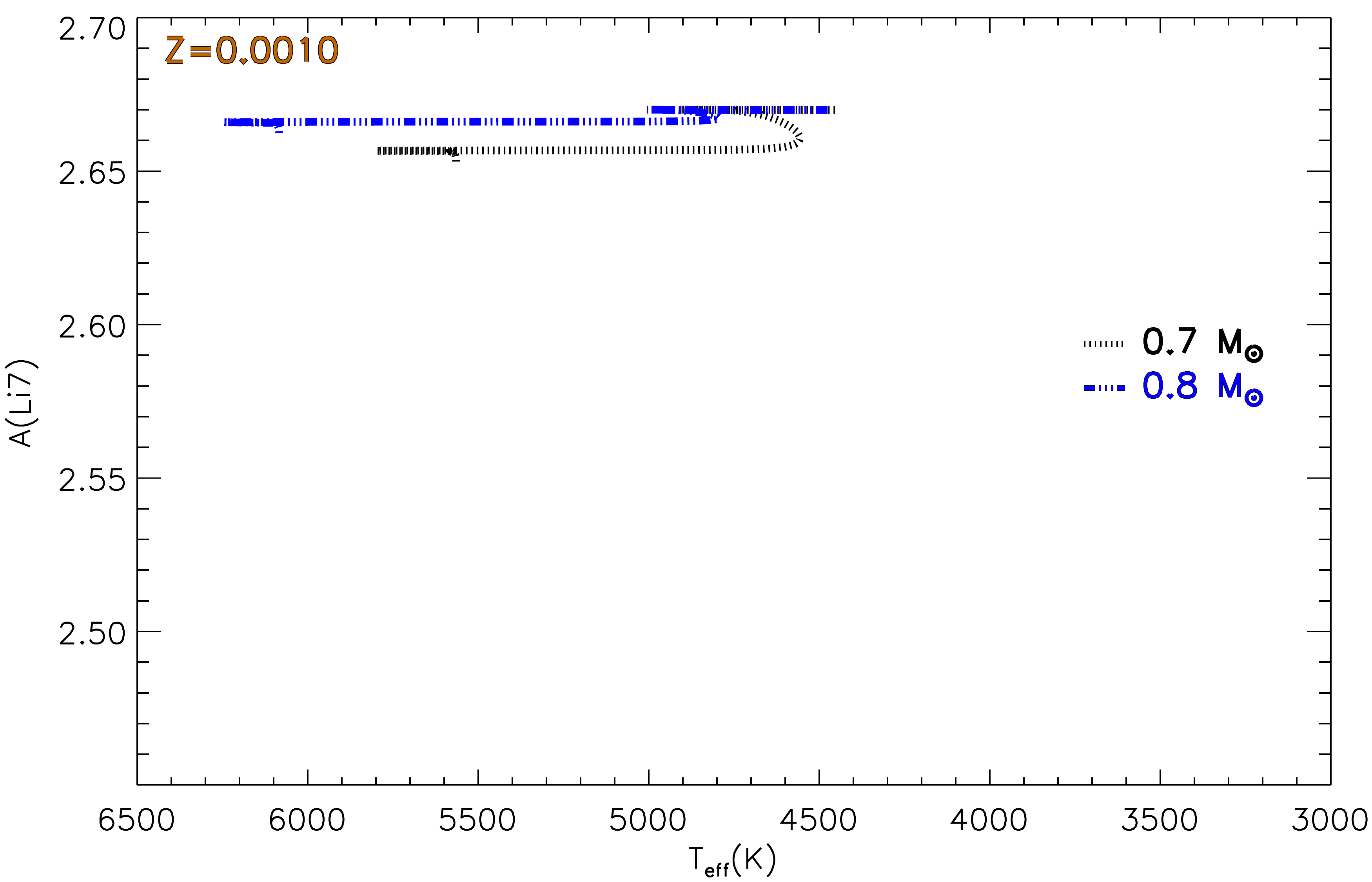}
\includegraphics[width=0.49\linewidth]{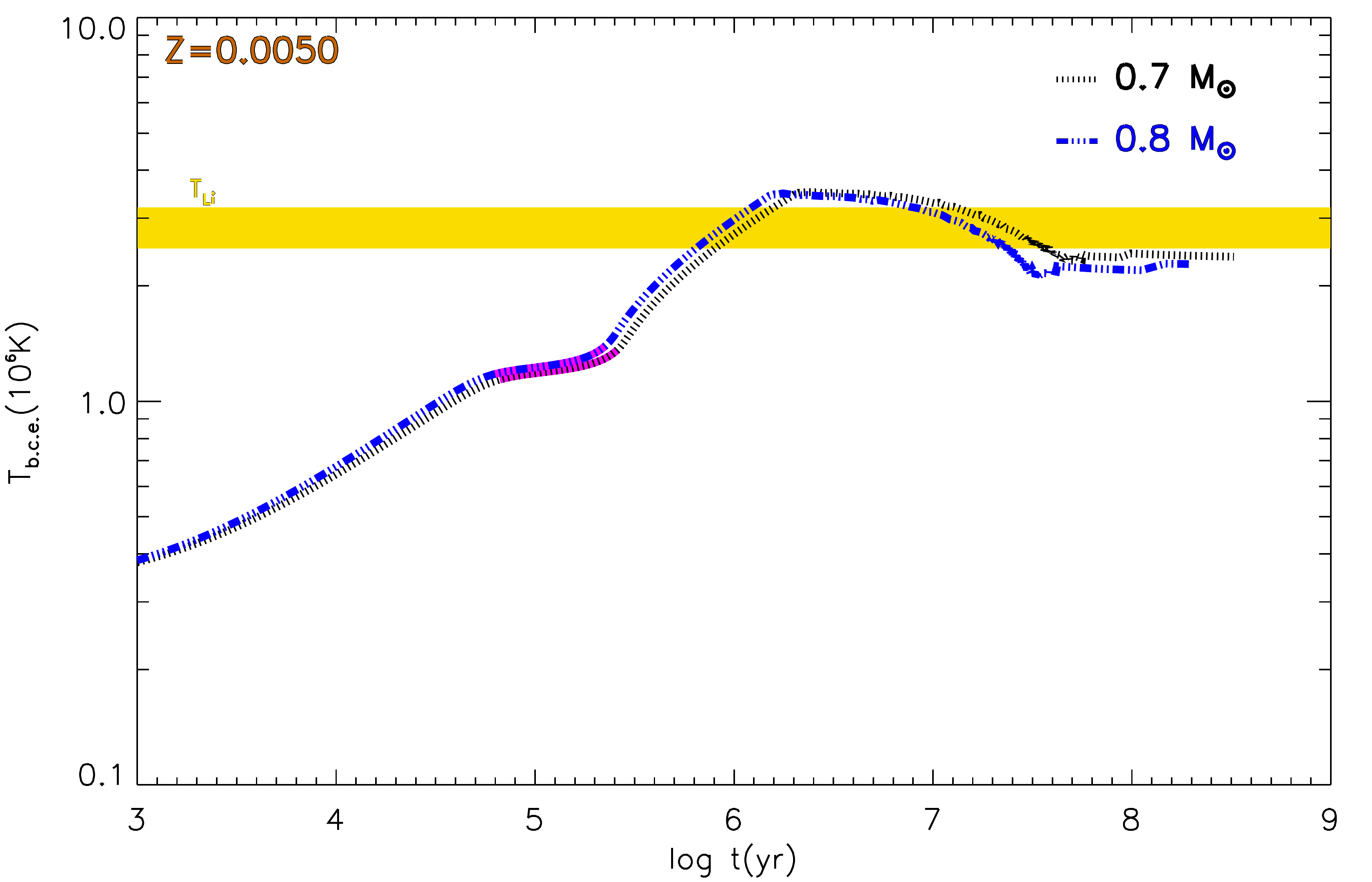}
\includegraphics[width=0.49\linewidth]{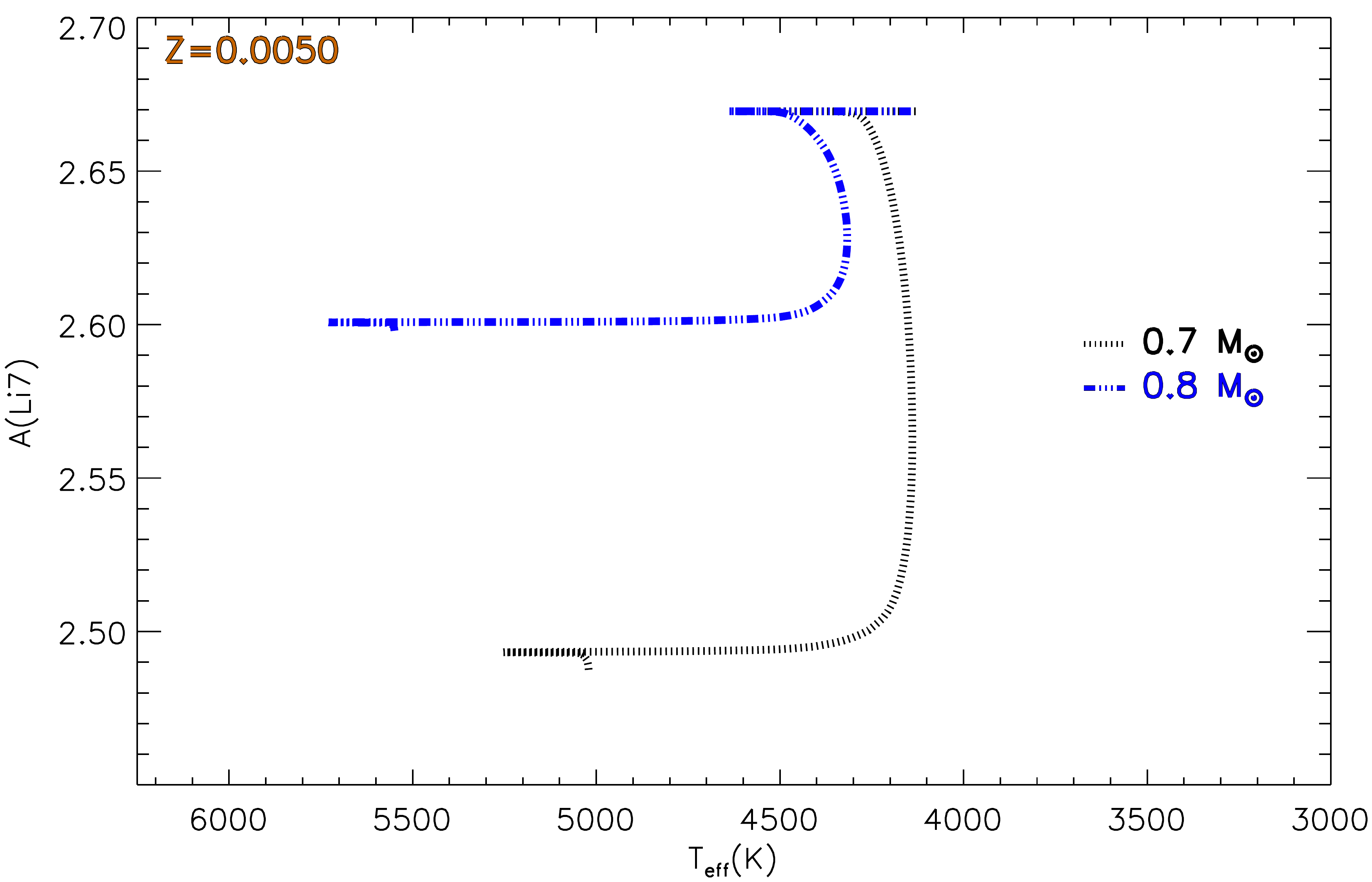}
\caption{Temporal evolution of the temperature at the bottom of the convective envelope,  \tce{} (left panel),  and evolution of the surface lithium abundance as a function of effective temperature (right panel) for standard non-accreting models with $Z=0.0010$ (top panels) and $Z=0.0050$ (bottom panels).}
\label{fig:std_li_0p001_005}
\end{figure*}

Figure~\ref{fig:std_li_0p001_005} shows the temporal evolution of \tce{} and the surface lithium abundance as a function of the effective temperature for metal richer models with $Z=0.001$ and $Z=0.005,$ respectively. Qualitatively, the larger metallicity leads to an increase in the radiative opacity, a thicker convective envelope that reaches hotter regions, and in turn a more efficient surface lithium depletion during the PMS. Such a metallicity effect is small in the case with $Z=0.0010$, where, for both masses, the depletion is just a bit larger than what was obtained for the $Z=0.0001$ case. The metallicity effect is more evident for the $Z=0.0050$ set. In this case, the surface lithium at the end of the PMS in the 0.7 and 0.8~\msun{} models is depleted by 0.15~dex (0.7~\msun) and 0.08~dex (0.8~\msun), respectively.

\section{Results for cold accretion models}
\label{cold}
We now start by presenting our cold accretion models, which were calculated by adopting different values for \rseed, \mseed{}, and \mdot{} at $Z=0.0001$.

\subsection{Dependence on \rseed}
\begin{figure*}
\centering
\includegraphics[width=0.49\linewidth]{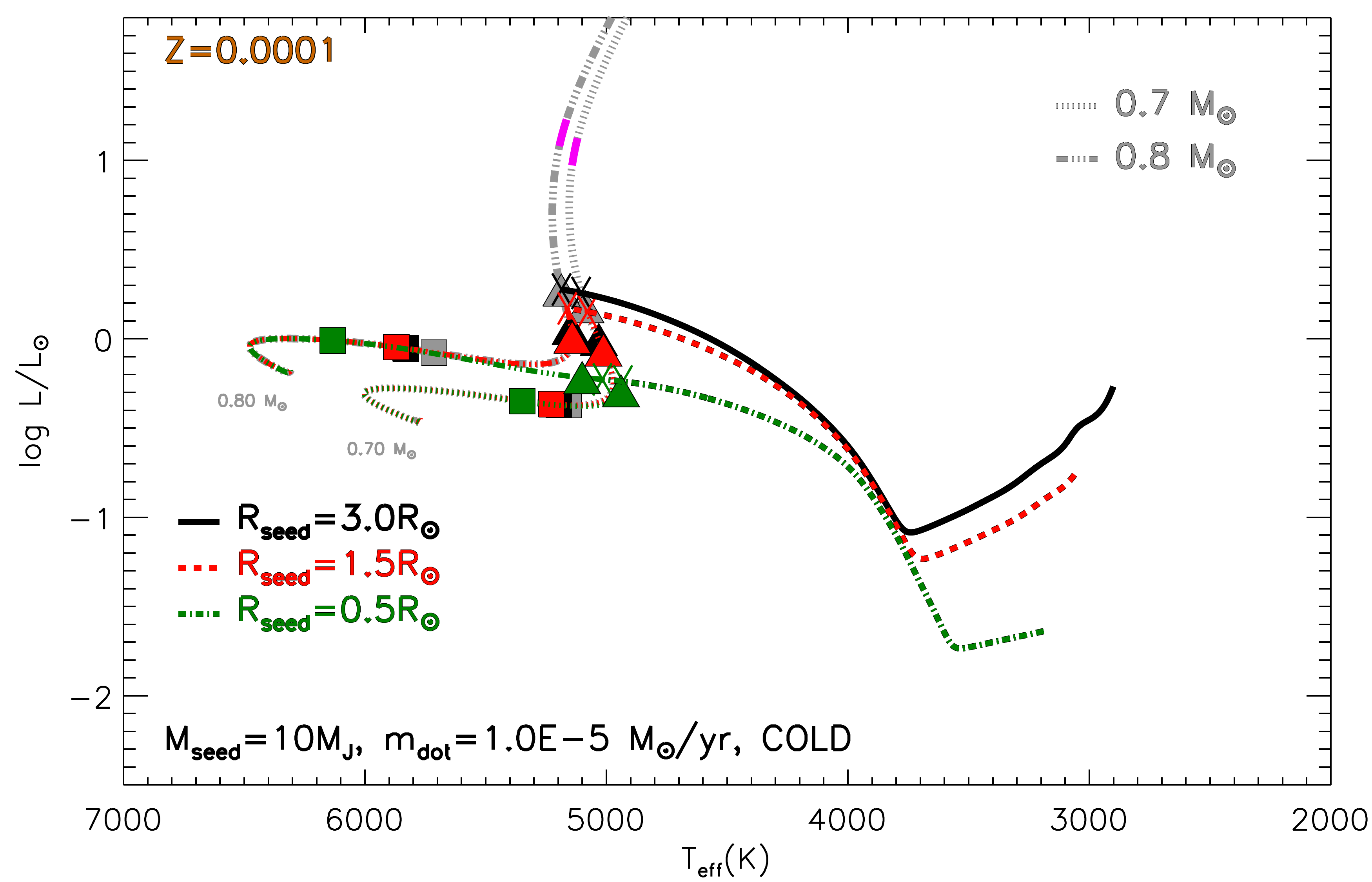}
\includegraphics[width=0.49\linewidth]{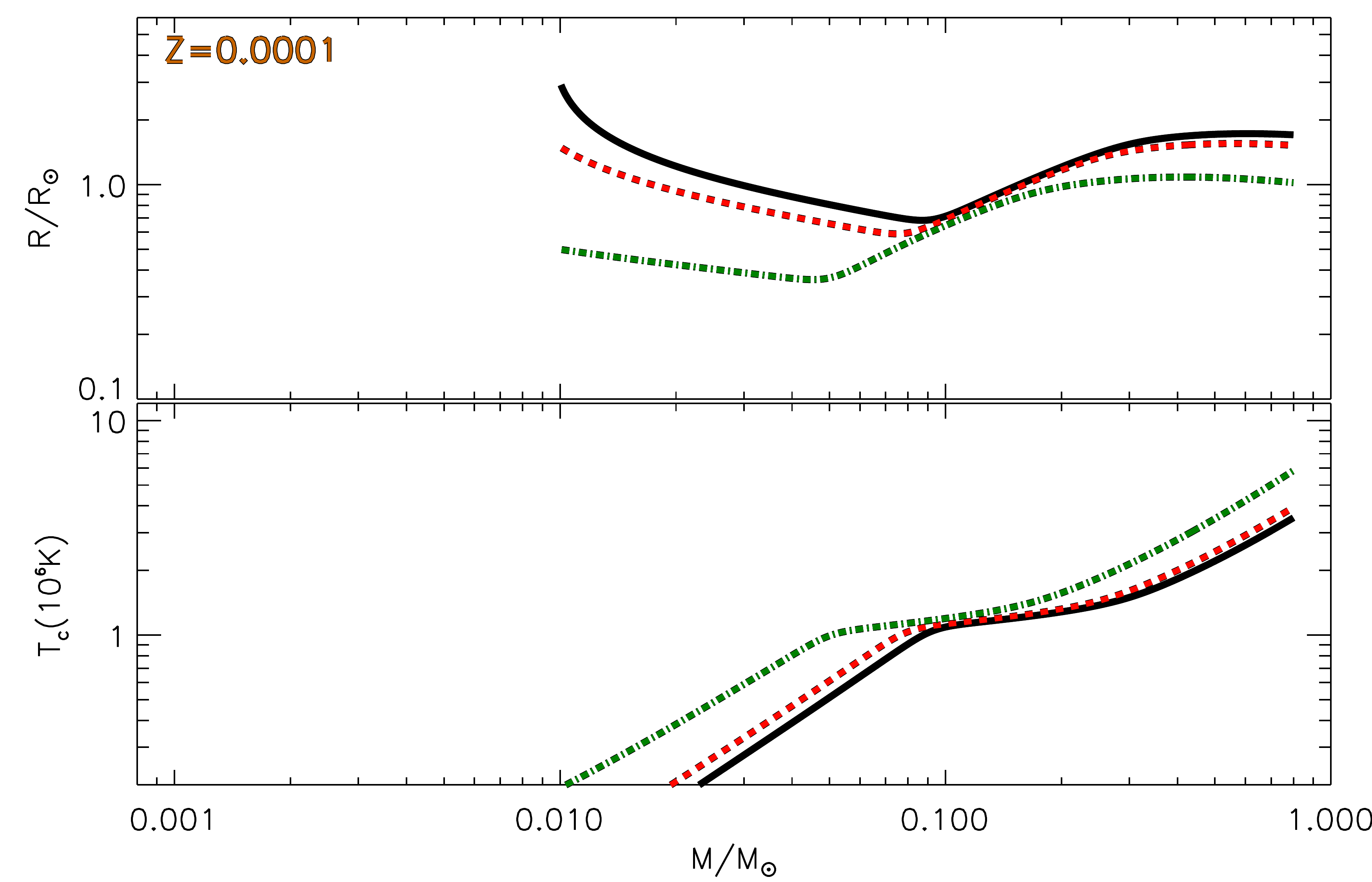}
\includegraphics[width=0.49\linewidth]{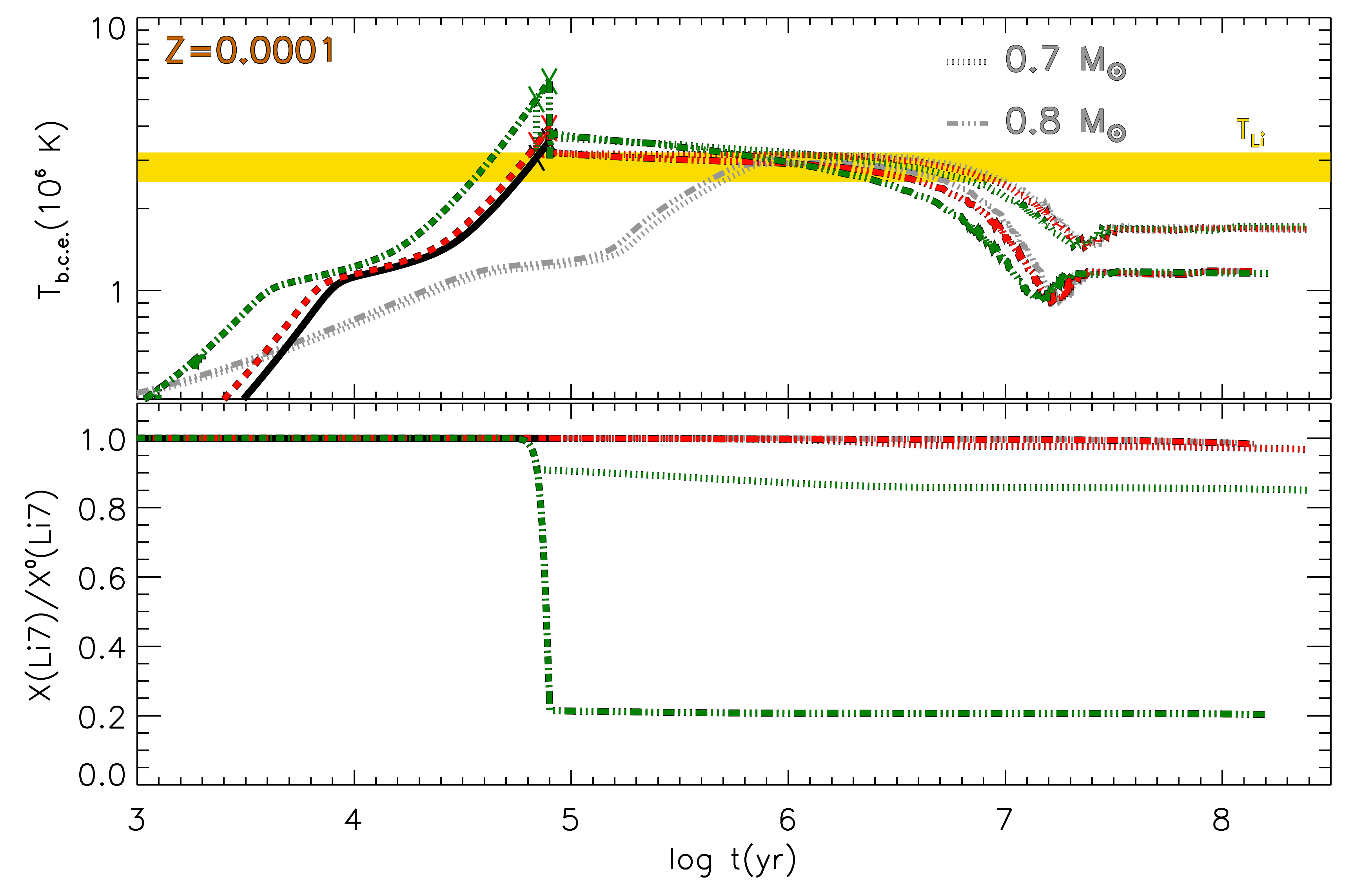}
\includegraphics[width=0.49\linewidth]{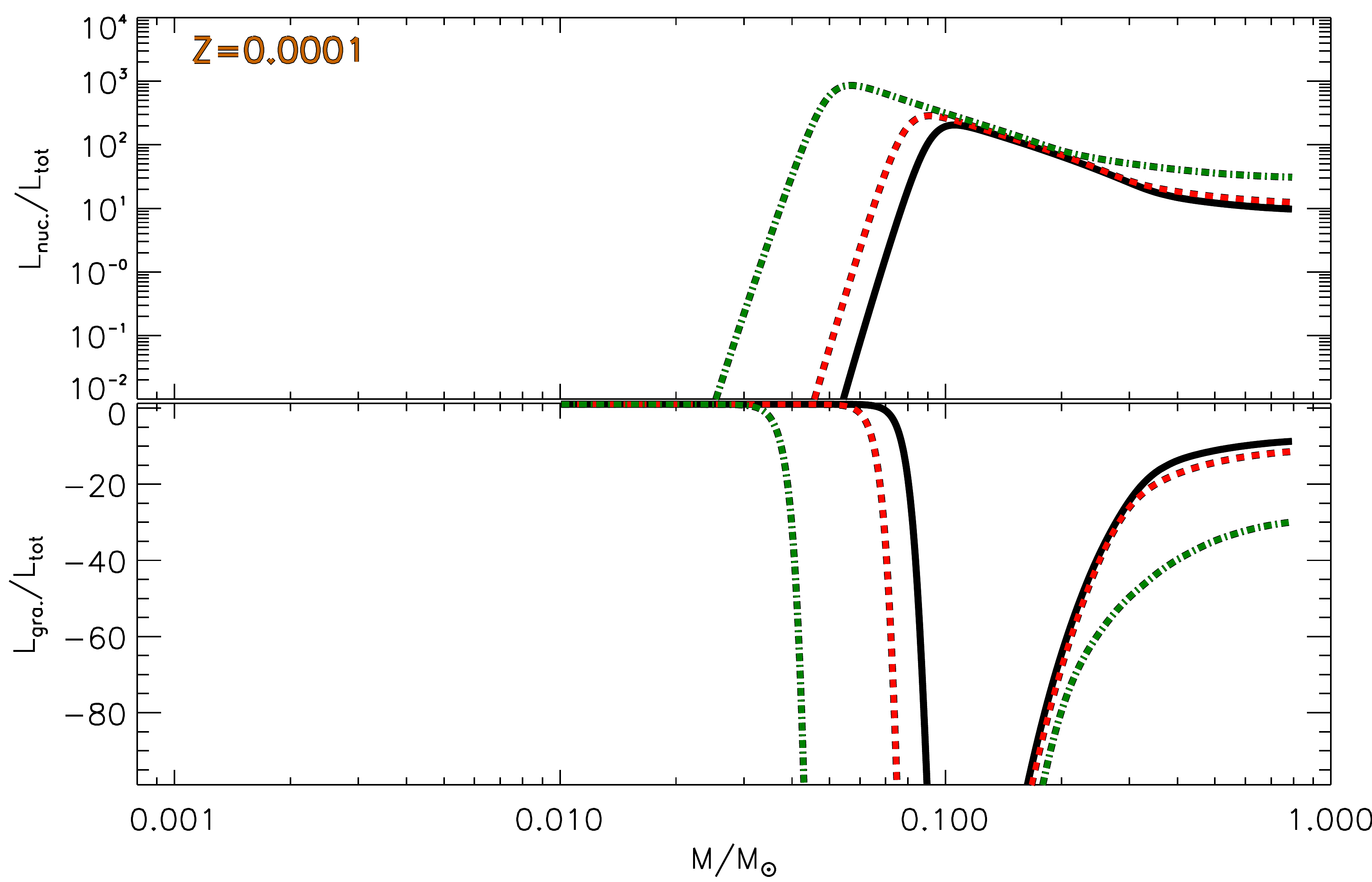}
\caption{Evolution of LLC accreting models with $Z=0.0001$, final masses of $M_\mathrm{fin}=0.7$ and 0.8~\msun, \mdot=$10^{-5}$~\mdotyr, \mseed=10~\mj, and \rseed=0.5, 1.5, and 3.0~\rsun. Standard non-accreting models are also shown for comparison purposes (grey lines). Top left panel: HR diagram. The position of models at 1~Myr (filled triangles) and 10~Myr (filled squares), the region corresponding to the deuterium burning phase (thick magenta line), and the end of the accretion phase (crosses) are marked. Top right panel: Evolution of the total radius (in solar units) and central temperature (in units of $10^6$~K) as a function of the actual value of the total mass during the protostellar phase. Bottom left panel: Time evolution of the temperature at the bottom of the convective envelope and surface Li (divided by the initial one) for the accreting and non-accreting models (the latter are shown with grey lines). The horizontal yellow shaded region denotes the Li burning temperature range. Bottom right panel: Contribution to the total luminosity of the nuclear ($L_\mathrm{nuc.}$) and gravitational luminosity ($L_\mathrm{gra.}$) as a function of the actual value of the total mass.}
\label{fig:rini}
\end{figure*}
\begin{figure}
\centering
\includegraphics[width=0.98\linewidth]{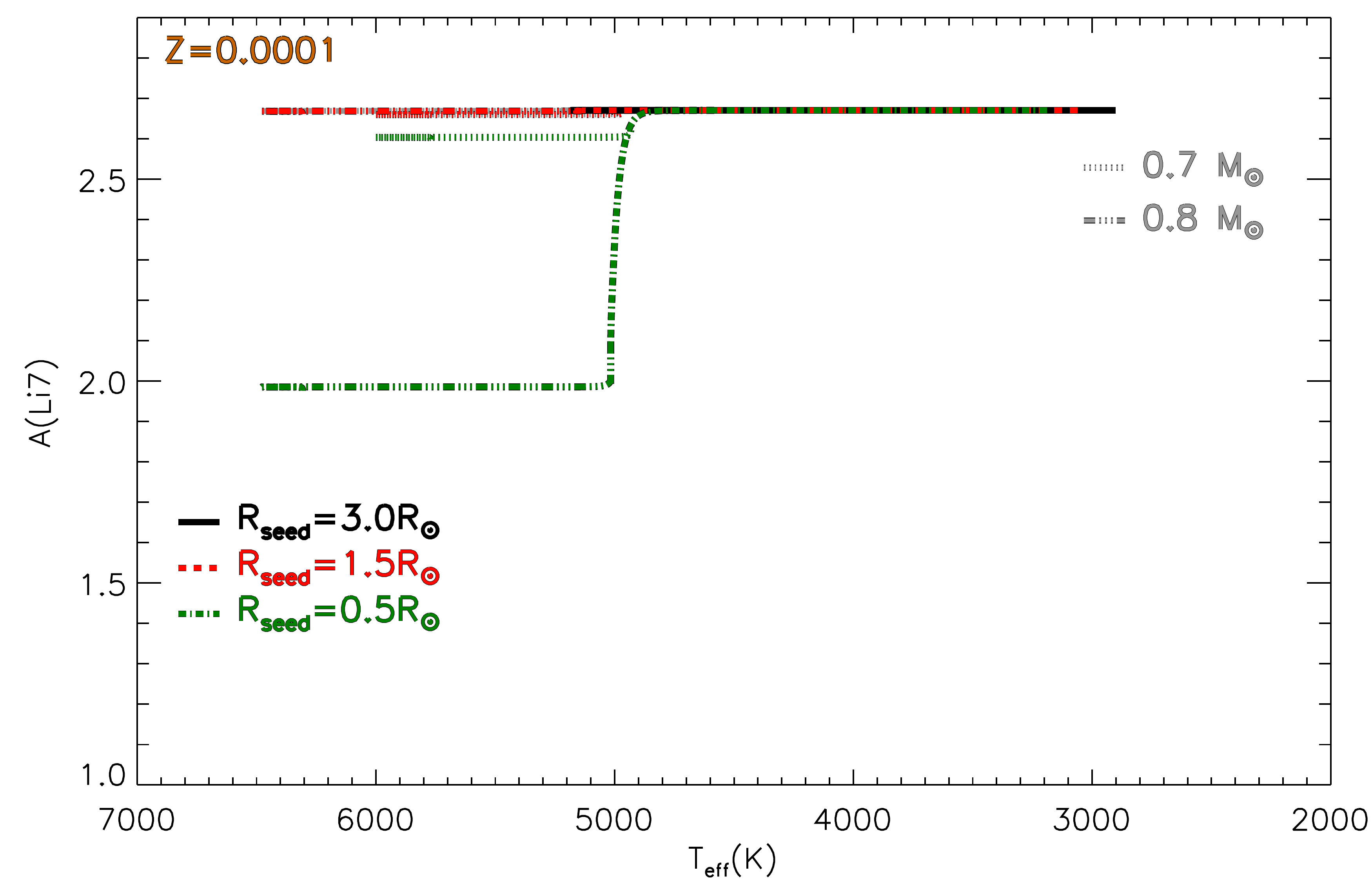}
\caption{Surface $A(Li)$ as a function of \teff{} for models of Fig.~\ref{fig:rini}.}
\label{fig:rini_li}
\end{figure}

Figure~\ref{fig:rini} shows the results of our calculations for a metallicity of $Z=0.0001$ (corresponding to [Fe/H]$\sim-$2.1, which is typical of Spite plateau stars) for the LLC case with \mseed=10~\mj, \mdot=$10^{-5}$~\mdotyr, final masses of $M=0.7$ and 0.8~\msun, and three values of \rseed, namely \rseed=0.5, 1.5, and 3.0~\rsun. For comparison purposes, we also display some results for the standard non-accreting models.

The evolution of these accreting objects can be summarised as follows. Accretion starts on a cold object (\teff$\sim 3000$~K) which, during the first part of the evolution, contracts. In this phase, the main effect of accretion is to add matter, thus making the model more compact. Models are relatively cold and fully convective, nuclear burning is inefficient, and the models continue to be chemically homogeneous. The consequence of the contraction and mass accretion is an increase in the temperature in the whole structure. As the central temperature reaches about $10^6$~K, deuterium burning ignites and releases enough nuclear energy to first halt the gravitational contraction and then lead to an expansion.  In the HR diagram, the D-burning phase corresponds to the minimum in luminosity at \teff{} of about 3600-3800~K. The effect of the deuterium burning energy on the model can be seen in the top right panel of Fig.~\ref{fig:rini}, which shows the radius and the central temperature as a function of the actual value of the total mass. The minimum of the radius during accretion approximately corresponds to the point where D-burning becomes efficient (see also the nuclear luminosity in the bottom right panel). From this point on, the radius increases. It is also interesting to notice that, before D-burning, the central temperature increases steeply until it reaches about $10^6$~K, when it abruptly changes its slope (for $M \in $[0.05, 0.1]~\msun, depending on \mdot) and becomes almost flat during D-burning. 

As long as the accretion is active, fresh deuterium is continuously fed to the fully convective model, thus the deuterium abundance after each time step is determined by two opposite mechanisms: nuclear burning, which destroys deuterium in the star, and accretion that supplies fresh deuterium from the disc. As mass accretion continues, the rate of expansion reduces. During the whole protostellar accretion phase, these models remain fully convective and they are thus chemically homogeneous.

Accretion is stopped when the models reach the requested final mass, either 0.7 or 0.8~\msun. From this moment on, they \lq{relax}\rq\, and reach the standard counterpart in the HR diagram on timescales of the order of \tkh. As we discuss later, the value of \tkh{} at the end of accretion affects the age of the model at a given position in the HR \citep[see e.g.][]{baraffe09}.

The quantitative details of this general picture are very sensitive to the value of \rseed. The smaller \rseed{}, the larger the impact of accretion on the model evolution compared to the non-accreting case. By adopting a large \rseed{}, the initial \tkh{} timescale of the star is small. This time scale has to be compared with the characteristic accretion time scale, which is \tacc$\propto 1/$\mdot. \ If \tkh{} is smaller than \tacc, the evolution is mainly driven by  the surface radiative losses, and only marginally by the accretion of matter. Thus, the star is expected to contract and get fainter. At some point, \tkh{} raises over \tacc{} and the evolution is mainly determined by the accretion process. Thus, models computed with very large initial radii tend to converge at the same structure regardless of the initial \rseed{} value \citep[see also][]{baraffe09,baraffe10,baraffe12}. 

On the contrary, a small initial radius implies a large initial \tkh{}, and the protostellar evolution is mainly determined by the mass accretion. The effect of the adopted \rseed{} on the evolution in the HR diagram and the total radius is shown in the top left and top right panels of Fig.~\ref{fig:rini}. Models computed with \rseed= 3.0~\rsun{} and 1.5~\rsun{} both converge in the HR diagram and in the radius versus mass diagram soon after deuterium ignition, despite the difference of a factor of two in the initial \rseed{}. The situation is completely different when \rseed=0.5~\rsun. The model is much more compact and hotter than the others during the whole accretion phase. As a result, at the end of the accretion phase, models are far from the corresponding standard tracks. This result is similar to what is discussed by \citet{hartmann16} in the case of solar metallicity stars. Cold accretion models with a relatively small initial \rseed{} systematically produce PMS stars that are fainter than those observed in young clusters and stellar associations. This seems to indicate that small \rseed{} in cold accretion models could be ruled out by the comparison with observations available for solar metallicity PMS stars. Unfortunately, the lack of PMS stars in metal-poor environments prevents us from making a similar comparison in the HR or colour-magnitude diagram, and only the indirect effects of \rseed{} on MS stars (as those caused on the surface Li) can be used to test the consistency of such models.

In order to have an estimate of the impact of the accretion on the age of the models once the accretion ended, we plotted the models at 1~Myr and 10~Myr in Fig.~\ref{fig:rini} for the standard and accretion tracks. The difference is relatively large if one considers a model at 1~Myr (about 0.2~dex difference in luminosity) because of the large \tkh$\sim 10$-15~Myr. However, as models relax, they reach the standard tracks over time scales of a few \tkh{}, and they eventually join the standard models along the Henyey (radiative core) PMS track. It is important to remark that such cold accretion models skip most of the evolution along the Hayashi track.

The bottom left panel of Fig.~\ref{fig:rini} shows the temporal evolution of \tce{} and the temporal evolution of the surface lithium abundance divided by the initial lithium abundance in the protostar. The maximum value of \tce{} is attained by the model with \rseed=0.5~\rsun, while in the other cases (\rseed=1.5 and 3~\rsun) \tce{} is only slightly larger than $T_\mathrm{Li}$. The largest PMS Li depletion occurs for \rseed=0.5~\rsun, while the protostellar lithium depletion is negligible for the other
values of \rseed. The total variation of surface lithium abundance during the protostar phase in the model with \rseed=0.5~\rsun{} strongly depends on the value of the final mass: It is about 15\% for $M_\mathrm{fin}=$0.7~\msun{} and 80\% for $M_\mathrm{fin}=$0.8~\msun. Figure~\ref{fig:rini_li} shows the surface lithium abundance in the ($A(Li)$ versus \teff) diagram for models with protostellar accretion. As already discussed, the largest effect is achieved for \rseed=0.5~\rsun{} with a depletion of about 0.08~dex for $M_\mathrm{fin}=$0.7~\msun{} and 0.7~dex for $M_\mathrm{fin}=$0.8~\msun, while the other cases do not show an appreciable lithium depletion.
\begin{figure*}
\centering
\includegraphics[width=0.49\linewidth]{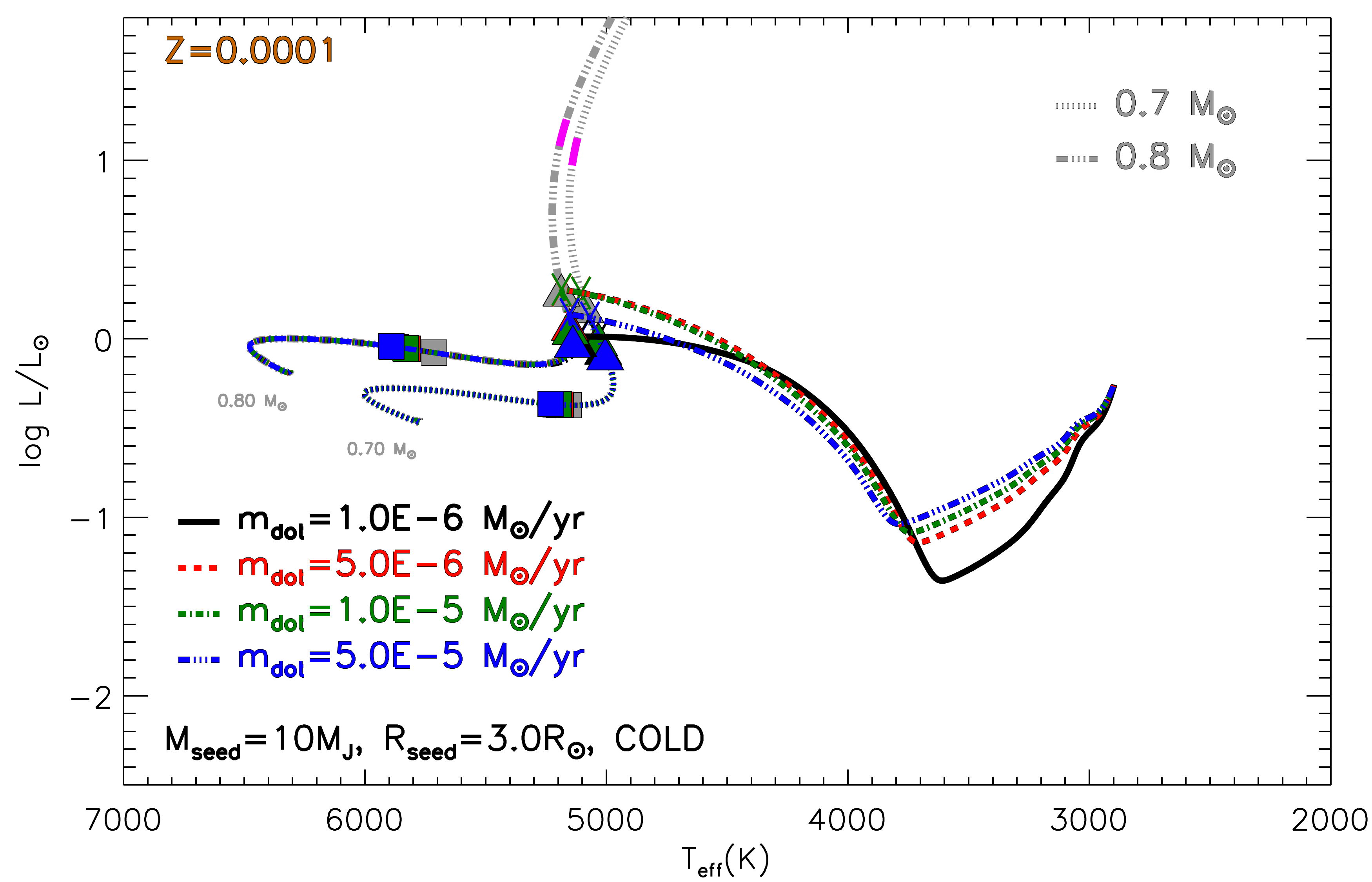}
\includegraphics[width=0.49\linewidth]{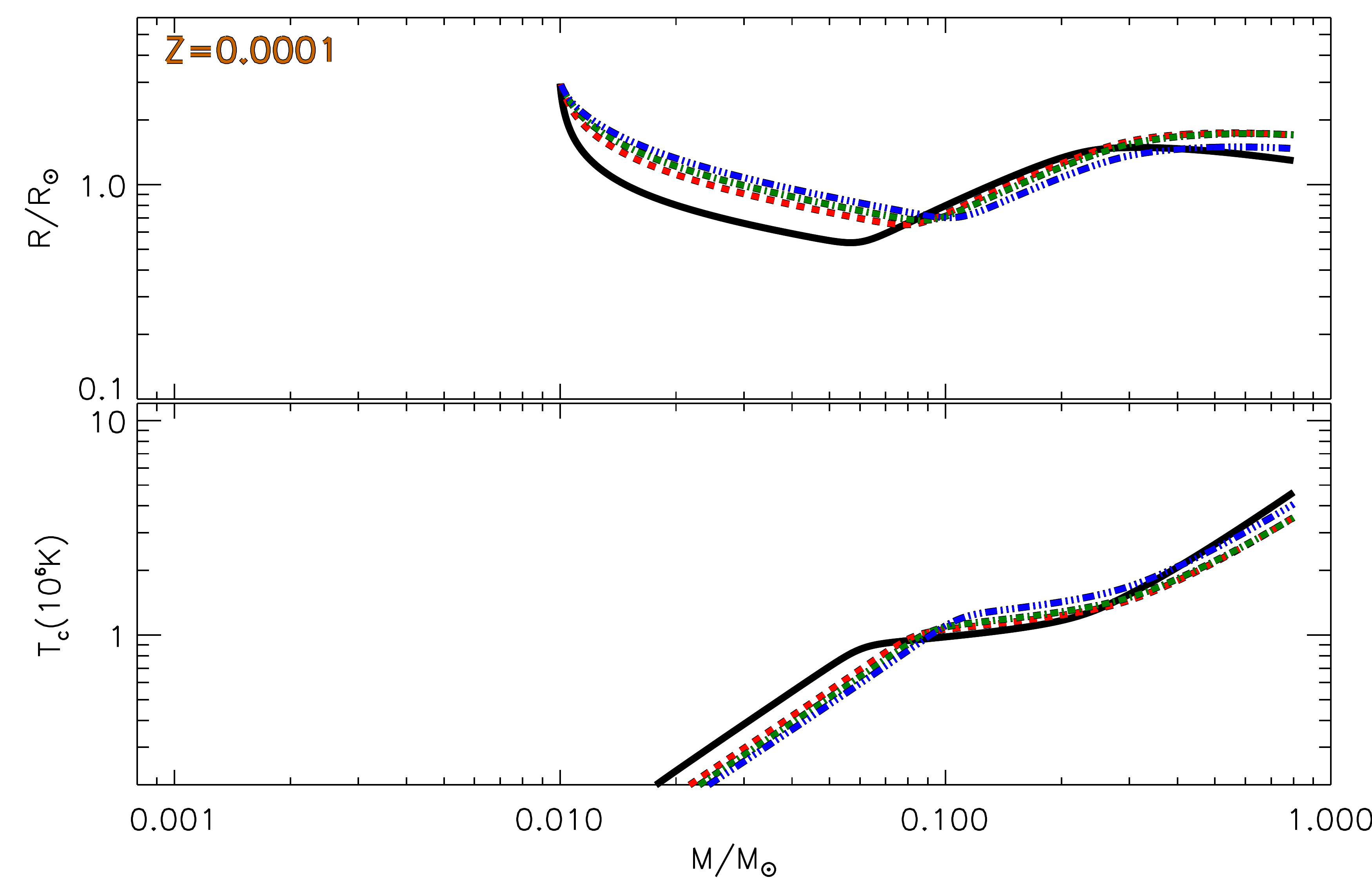}
\includegraphics[width=0.49\linewidth]{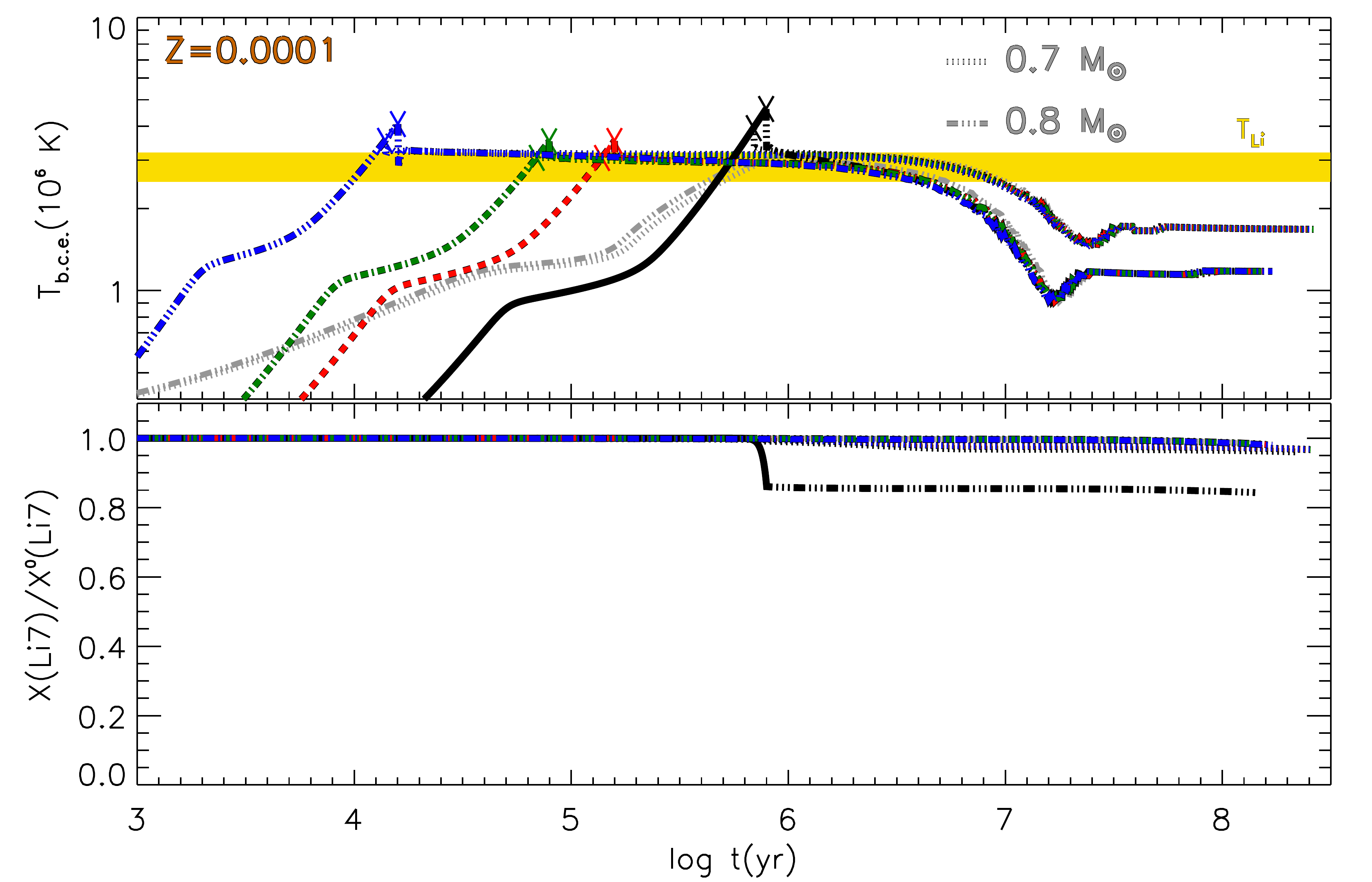}
\includegraphics[width=0.49\linewidth]{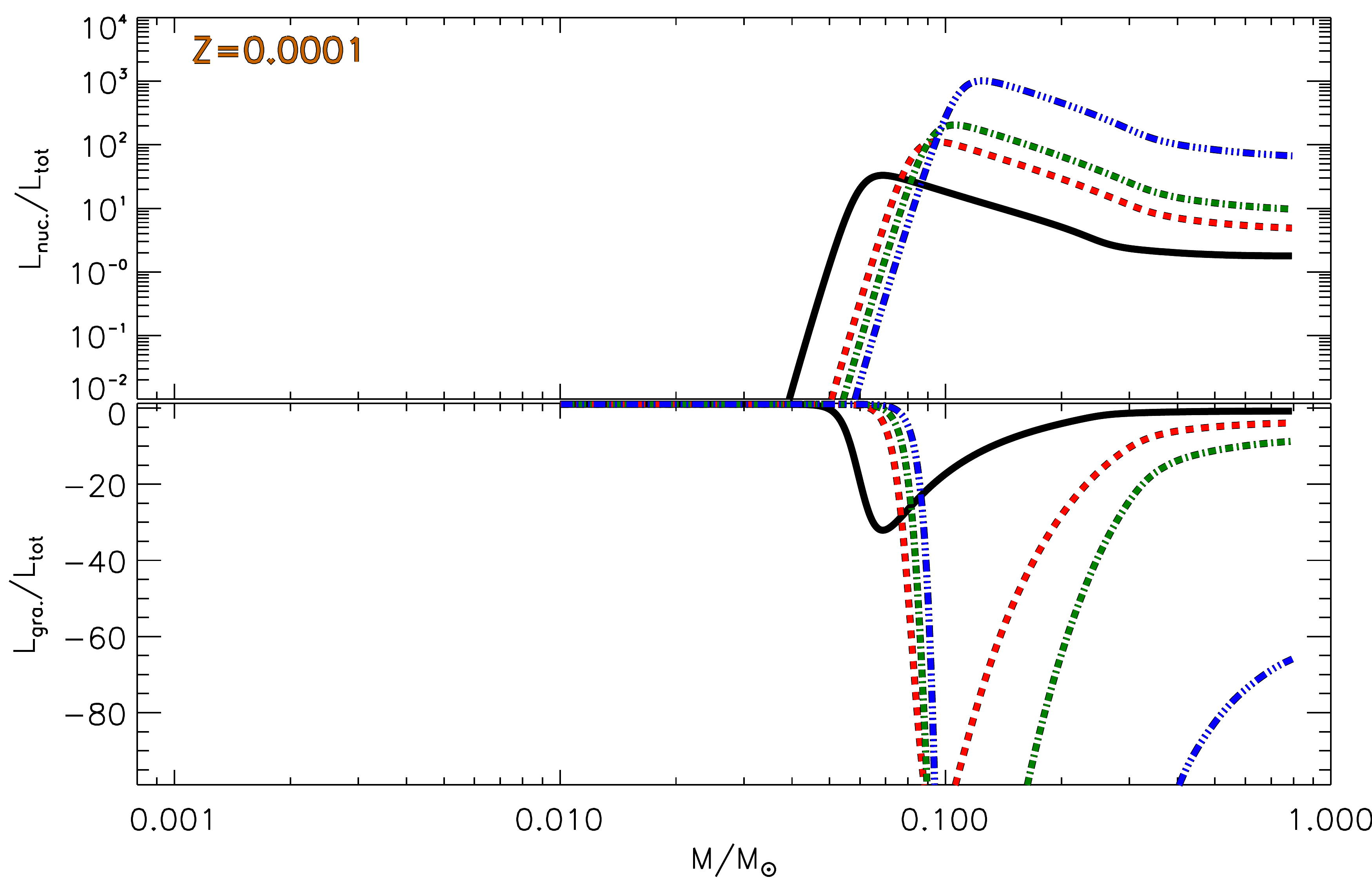}
\caption{Effect of \mdot{} on LLC models (\mseed=10~\mj, \rseed=3~\rsun): \mdot=$1 \times 10^{-6}, 5 \times 10^{-6}, 10^{-5}, 5 \times 10^{-5}$ \mdotyr. The symbols are the same as in Fig.~\ref{fig:rini}.}
\label{fig:mdot}
\end{figure*}
\subsection{Dependence on \mdot}
The case of varying \mdot{} is displayed in Fig.~\ref{fig:mdot}, which shows the results for $Z=0.0001$ for the LLC case with \mseed=10~\mj, \rseed=3~\rsun{}, final masses $M=0.7$ and 0.8~\msun{}, and accretion rates of \mdot=$1 \times 10^{-6}, 5 \times 10^{-6}, 10^{-5}, 5 \times 10^{-5}$~\mdotyr. The general evolution during the accreting phase is not significantly affected  by the exact value of \mdot, at least in the explored interval. Regardless of the adopted \mdot, at the end of the accretion phase, models with protostellar accretion are very close to  each other. From this moment on, they {relax}  and reach the standard counterpart in the HR diagram. The largest difference in the HR position occurs for the 1~Myr model. In all of the accretion calculations, the 1~Myr model is fainter by about 0.2~dex than the  standard track. The difference reduces at 10~Myr in the 0.7~\msun{} case, while for the 0.8~\msun{} it is still appreciable, but eventually the differences vanish on timescales of the order of about 20~Myr.

Once relaxed, the model evolution in the HR diagram is the same as for standard calculations with the same mass. It is important to notice that the accretion tracks join the standard ones almost at the end of the Hayashi track, which is close to the region where a radiative core develops and well below the region corresponding to deuterium burning in standard PMS models. These accreting models largely  miss the Hayashi track evolution. 

The effect of using different values for \mdot{} has a small impact on the evolution of the model. The largest differences, which are however relatively small, occur when a low accretion rate is adopted, for example, \mdot=$10^{-6}$~\mdotyr. The effect of such a low \mdot{} value is that the deuterium ignition is postponed, which occurs in a less massive model at a smaller radius if compared to what was obtained for a larger \mdot{} (top right and bottom right panels in Fig.~\ref{fig:mdot}). The expansion during deuterium burning is similar for the different \mdot{} values, while the duration of the expansion in mass is shorter if \mdot=$10^{-6}$~\mdotyr{} is used.\ This is the case since accreting matter supplies the star with a lower amount of deuterium. When accretion ends, models with \mdot=$10^{-6}$~\mdotyr{} join the standard track at a lower luminosity (and smaller radius). Interestingly, if the accretion rate increases in the range of $10^{-6}$-$10^{-5}$~\mdotyr, then the luminosity (and the radius) of the model left at the end of the accretion phase increases too. But at larger accretion rates, for example, $5\times 10^{-5}$~\mdotyr, the model luminosity (and radius) begins to decrease. However, the figure shows that the tracks are only marginally affected by \mdot{} even by changing the accretion rate by an order of magnitude.

The bottom left panel of Fig.~\ref{fig:mdot} shows the evolution of the temperature at the bottom of the convective envelope and the surface lithium abundance. It is important to notice that the protostar is fully convective in the whole protostellar phase, so \tce{} is the same as $T_\mathrm{c}$, and the structure is chemically homogeneous. 

During the accretion phase, \tce{} progressively increases and it reaches values that are slightly larger than $T_\mathrm{Li}$. However, because of the limited amount of time during which \tce$>T_\mathrm{Li}$ and the continuous replenishment of lithium from the accreted matter, Li is not efficiently depleted. The maximum depletion occurs in the 0.8~\msun{} model for \mdot=$10^{-6}$~\mdotyr{} as a result of the reduced amount of lithium supplied to the star by accretion, the slightly larger values of \tce, and the longer evolutionary timescales. In this case, the surface Li abundance after the PMS is $\sim 15$\% lower than the initial one, which corresponds to a depletion of about 0.08~dex in $A(Li$); whereas, for the other values of \mdot{} and the final mass, the PMS Li-depletion is negligible, as in the standard case. Given the small impact on the surface lithium we do not show the figure $A(Li)$ versus \teff.
\begin{figure*}
\centering
\includegraphics[width=0.49\linewidth]{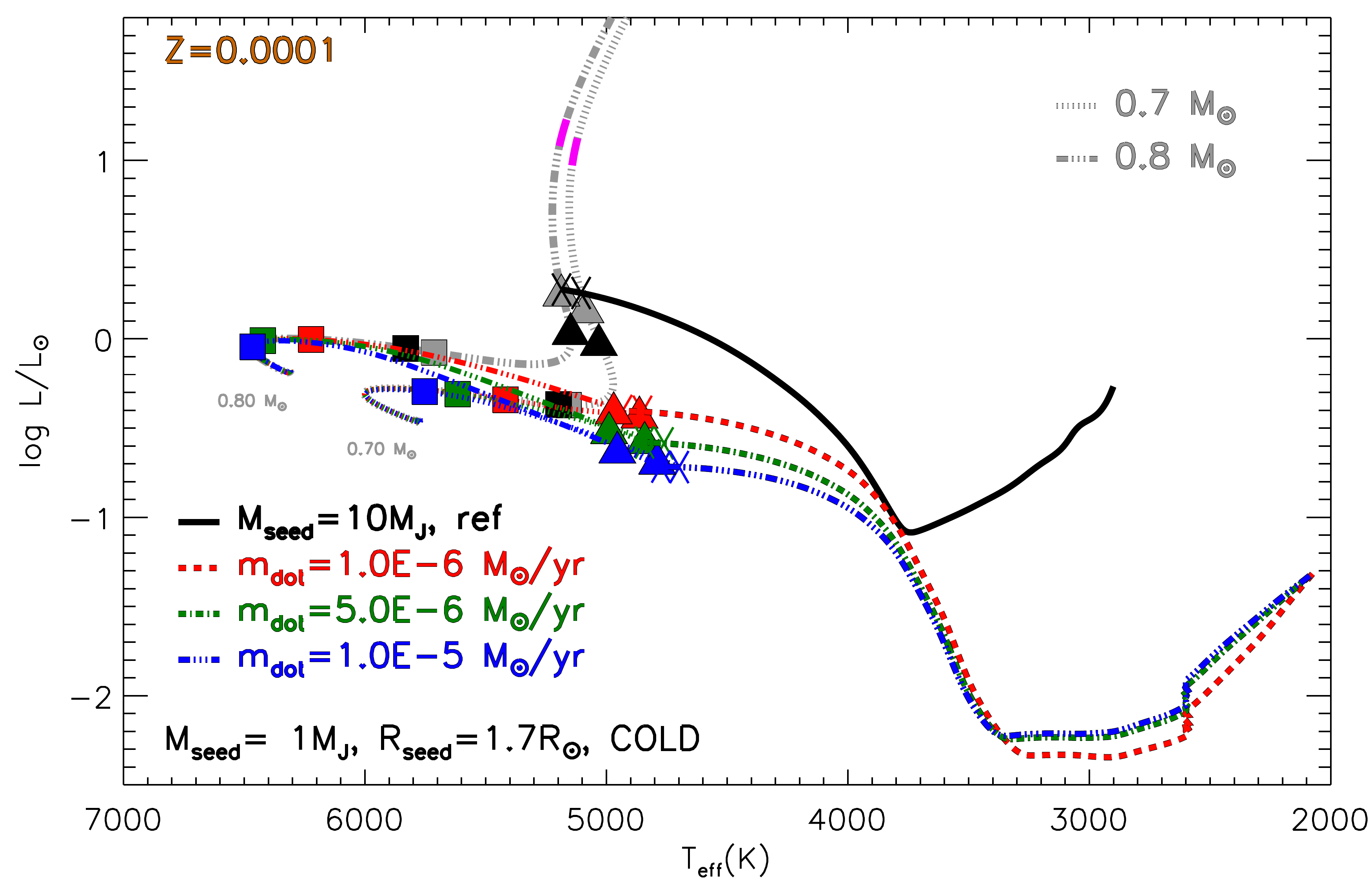}
\includegraphics[width=0.49\linewidth]{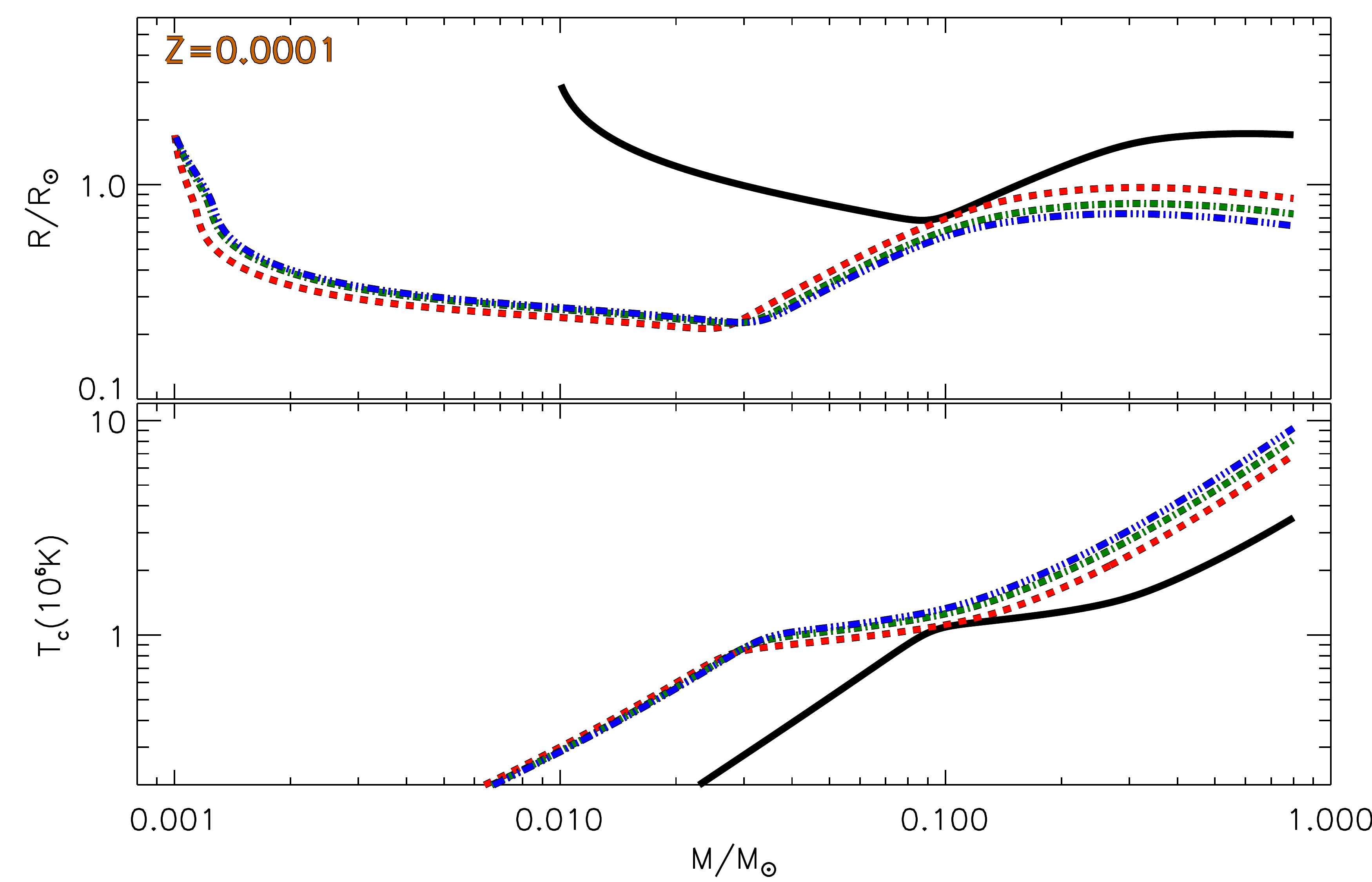}
\includegraphics[width=0.49\linewidth]{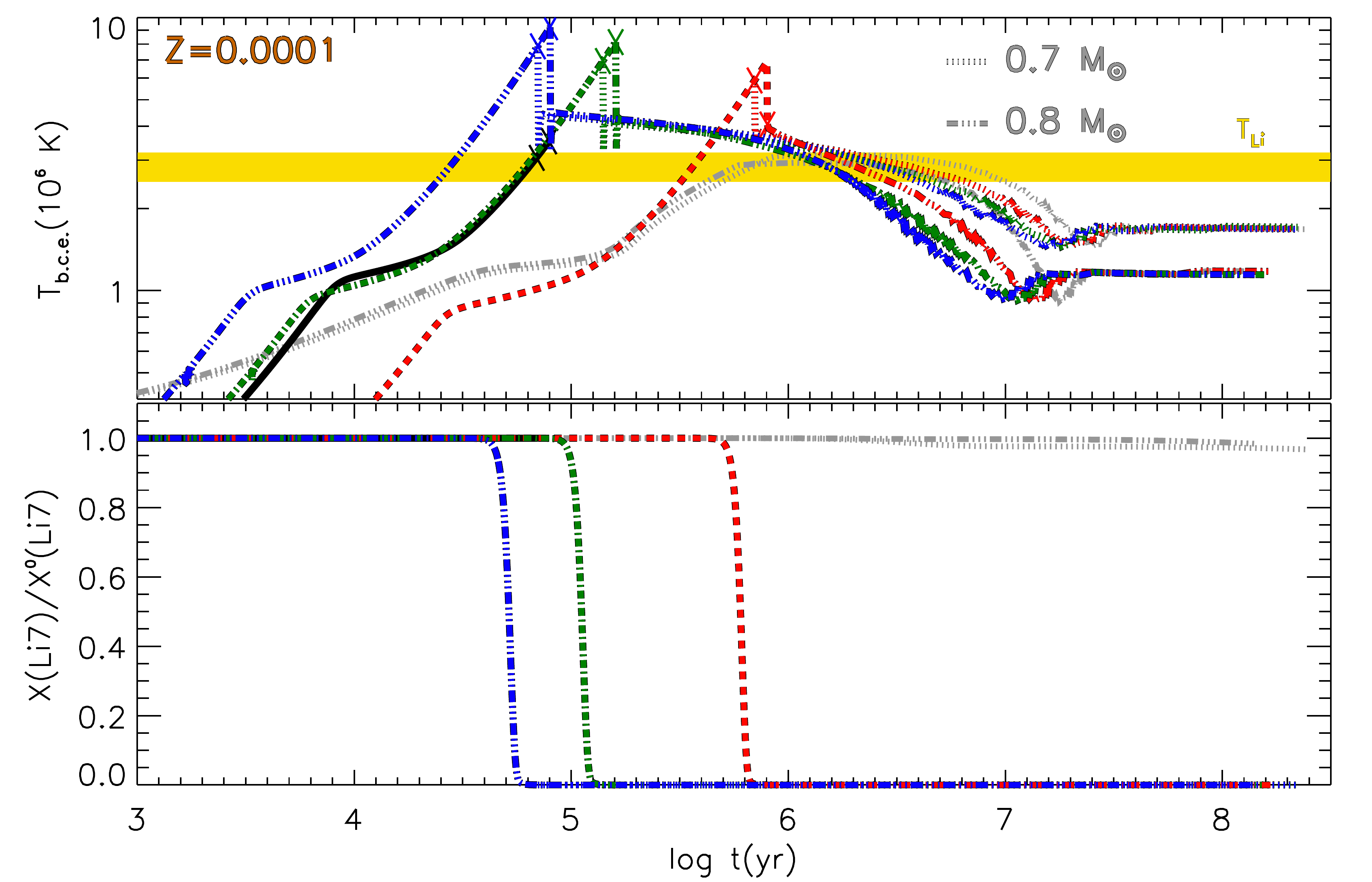}
\includegraphics[width=0.49\linewidth]{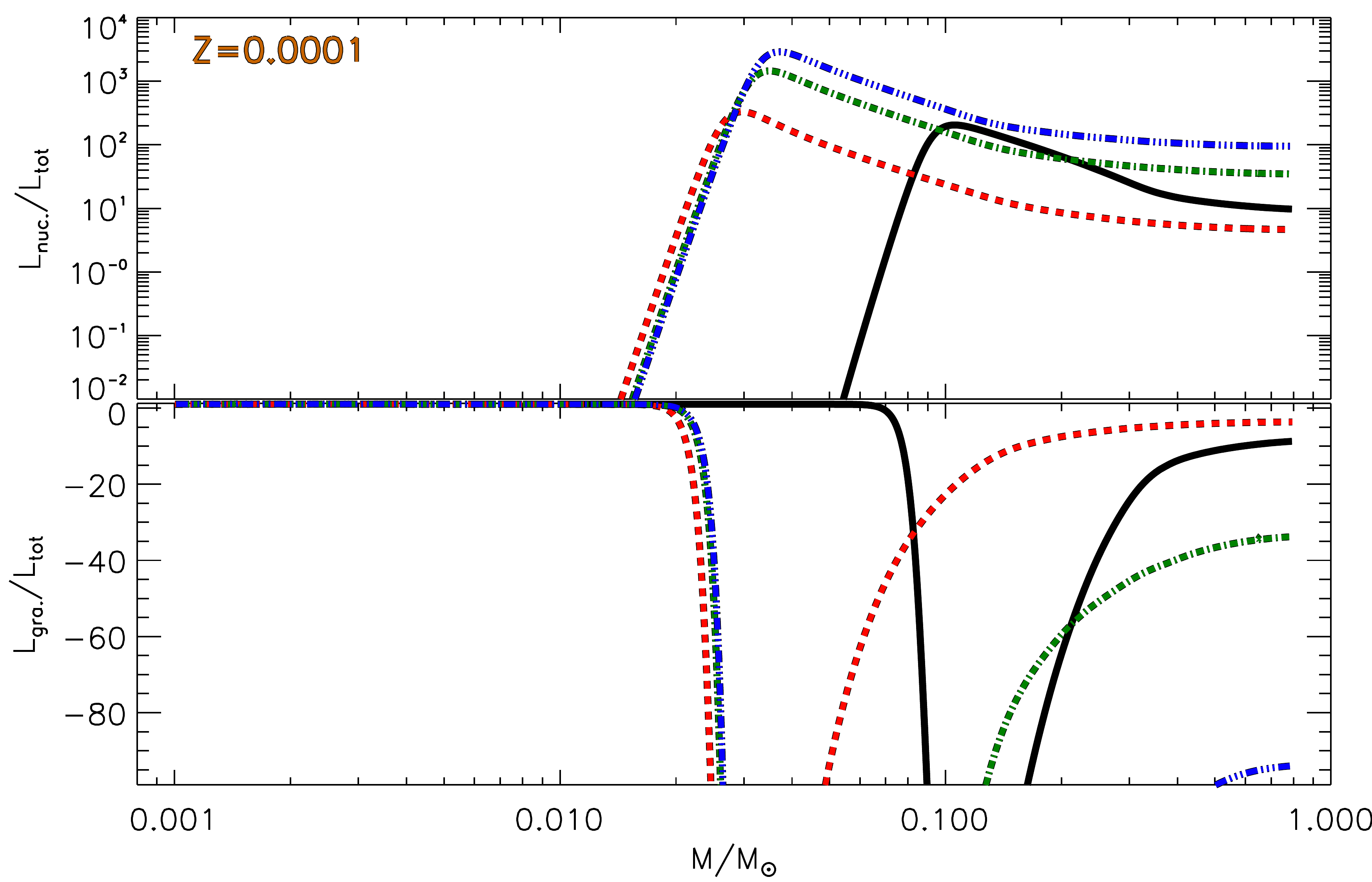}
\caption{Effect of \mdot{} on SLC models (\mseed=1~\mj, \rseed=1.7~\rsun). For comparison purposes, the reference LLC model of Fig.~\ref{fig:rini} (\mseed=10~\mj, \rseed=3~\rsun, \mdot=$10^{-5}~$\mdotyr) is also shown. Symbols are the same as in Fig.~\ref{fig:mdot}.}
\label{fig:mdot_mbassa}
\end{figure*}
\begin{figure}
\centering
\includegraphics[width=0.98\linewidth]{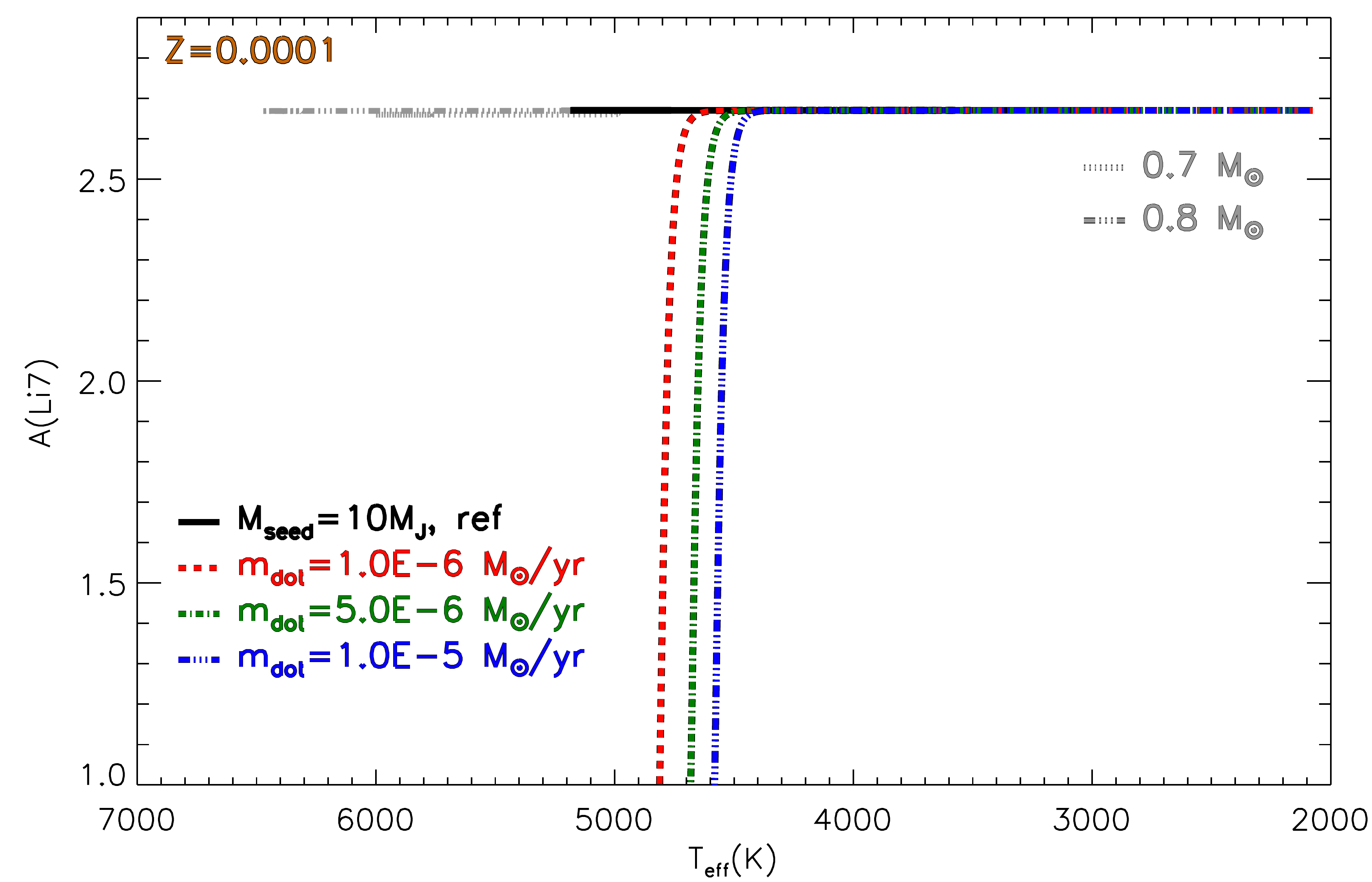}
\caption{Surface $A(Li)$ as a function of \teff{} for SLC models of Fig.~\ref{fig:mdot_mbassa}.}
\label{fig:mdot_mbassa_li}
\end{figure}

\subsection{Dependence on \mseed}

In the previous sections, we have fixed the seed mass to 10~\mj{}. Here we analyse how the value of \mseed{} affects the protostellar evolution, testing a lower value of 1~\mj{}, which is consistent with the lower limit expected for the second Larson core (which we refer to as SLC models).

Figure~\ref{fig:mdot_mbassa} shows SLC calculations with the same values of \mdot{} as in Fig.~\ref{fig:mdot}, for models with $Z=0.0001$, $M=$0.7 and 0.8~\msun{} and \mseed=1~\mj, \rseed=1.7~\rsun . It is important to notice that for convergence issues, \rseed{} is half the value employed in the previously described calculations, but it is fully compatible with the value found by \citet{larson69} for masses of the order of 1~\mj. Moreover, as we have shown, a variation of \rseed{} between 3 and 1.5~\rsun{} only has a small effect on the model evolution (see Fig.~\ref{fig:rini}). Therefore,  the differences between the SLC and the LLC calculations, with the same \mdot , are mainly due to the effect of varying \mseed. 

The SLC models start their evolution at a much lower effective temperature compared to LLC models (\teff$\approx 2000$~K) and similarly to LLC calculations, the first part of the evolution is characterised by a contraction that increases the temperature until D-burning ignites (minimum in luminosity is at about 3400~K, actual mass is about 0.03~\msun, see bottom right panel). The energy released by D-burning leads to a steep radius expansion, which is clearly visible in the radius evolution shown in the top right panel of Fig.~\ref{fig:mdot_mbassa}. However, the model is hotter in the interior and more compact than the LLC case at the actual same mass. This leads to a fainter location in the HR diagram. When accretion ends, the model is relatively far from the standard track with the same mass,  radius, effective temperature, and luminosity. From that point on, the model progressively relaxes and approaches the standard track with the corresponding mass, but due to the small luminosity and radius, \tkh{} is relatively large; \tkh$\sim 20$-60~Myr is the largest value corresponding to the largest \mdot. As a consequence, the accretion tracks join the standard ones close to the zero age main sequence (ZAMS). 

Thus, SLC models essentially miss the Hayashi track evolution. Also, the age is more substantially affected by protostellar accretion because of the large \tkh{} involved. The 1~Myr models are very far from the standard ones, about 0.5-0.7~dex in luminosity and about 200~K in \teff, while the 10~Myr ones show no difference in luminosity (being along the Henyey track which is horizontal in the HR diagram).\ However, they display a large difference in \teff{}, from about 200~K to 700~K depending on \mdot{} and $M_\mathrm{fin}$.

Also, the evolution of \tce{} is strongly affected by the protostellar accretion in SLC models, as is shown in the bottom left panel of Fig.~\ref{fig:mdot_mbassa}. For actual masses of about 0.2-0.3~\msun{} during the accretion phase (the largest mass value corresponds to the lowest \mdot), \tce{} reaches values that are much larger than $T_\mathrm{Li}$, which are maintained for about 1-2~Myr. This means that \tce$> T_\mathrm{Li}$ even after the end of protostellar accretion, which ceases at about 7-8$\times 10^4$~yr (\mdot=$10^{-5}$~\mdotyr), 1.4-1.6$\times 10^5$~yr (\mdot=$5\times 10^{-6}$~\mdotyr), and  7-8$\times 10^5$~yr (\mdot=$10^{-6}$~\mdotyr). The effect on the surface lithium abundance is shown in the bottom left panel of Fig.~\ref{fig:mdot_mbassa}. The protostars are fully convective during the accretion phase and consequently lithium is completely destroyed in the whole structure before the end of the protostellar phase in all SLC cases. The supply of fresh lithium due to accretion is not enough to counterbalance the efficient nuclear burning. In the observational plane (Fig.~\ref{fig:mdot_mbassa_li}), SLC protostellar accretion models predict stars to have no residual surface lithium for effective temperatures larger than about 4600-4900~K.
\begin{figure*}
\centering
\includegraphics[width=0.49\linewidth]{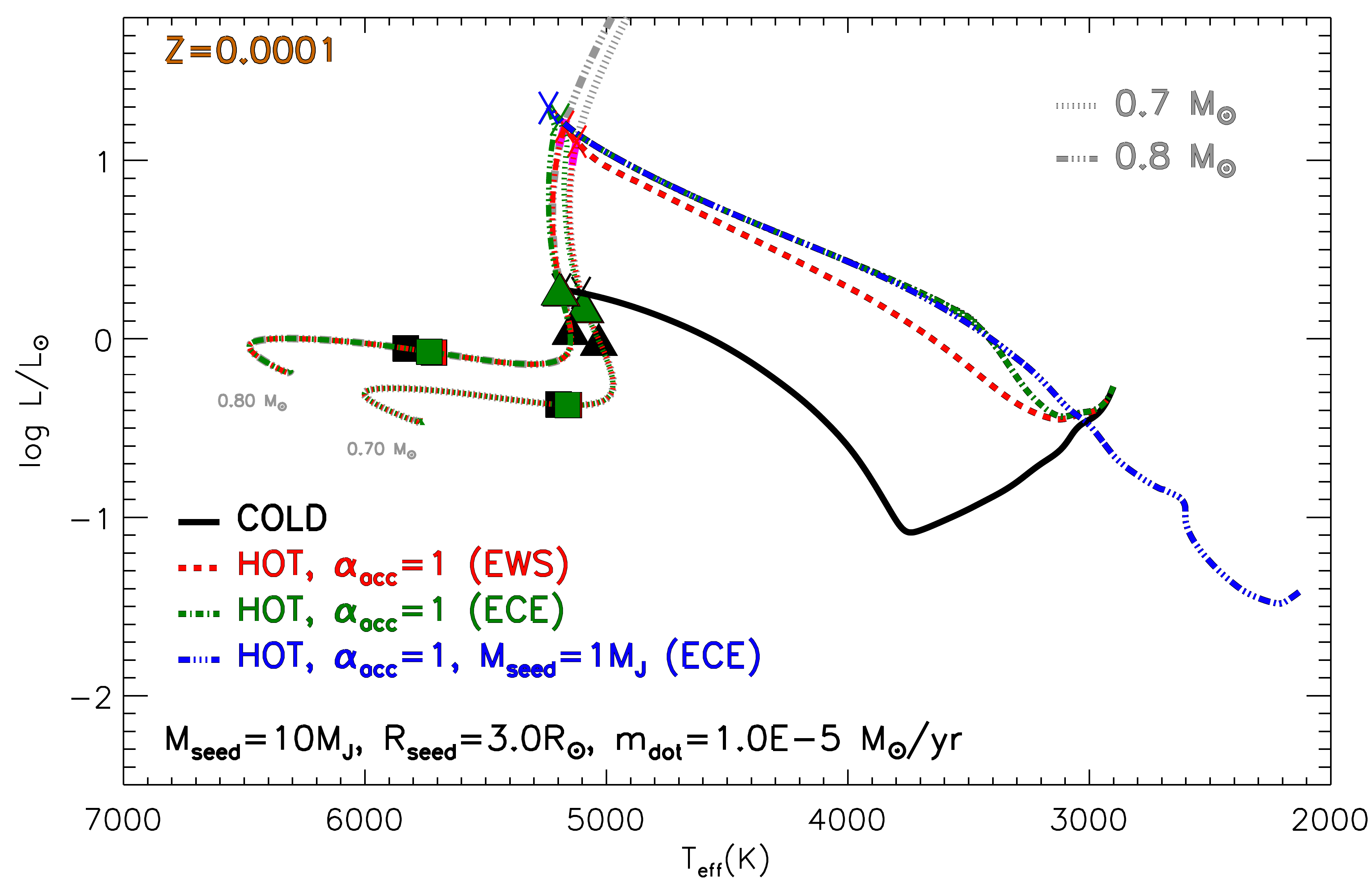}
\includegraphics[width=0.49\linewidth]{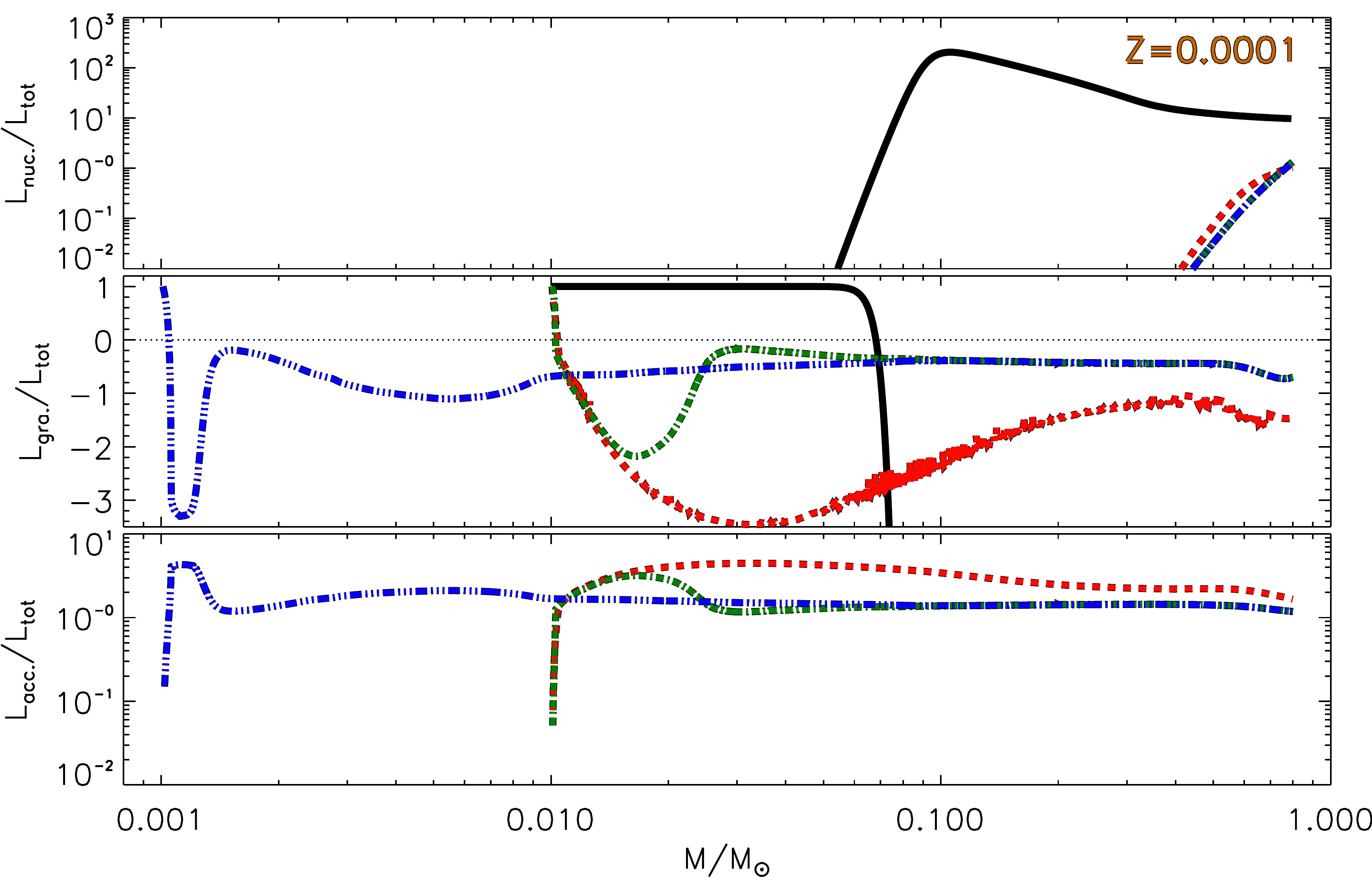}
\includegraphics[width=0.49\linewidth]{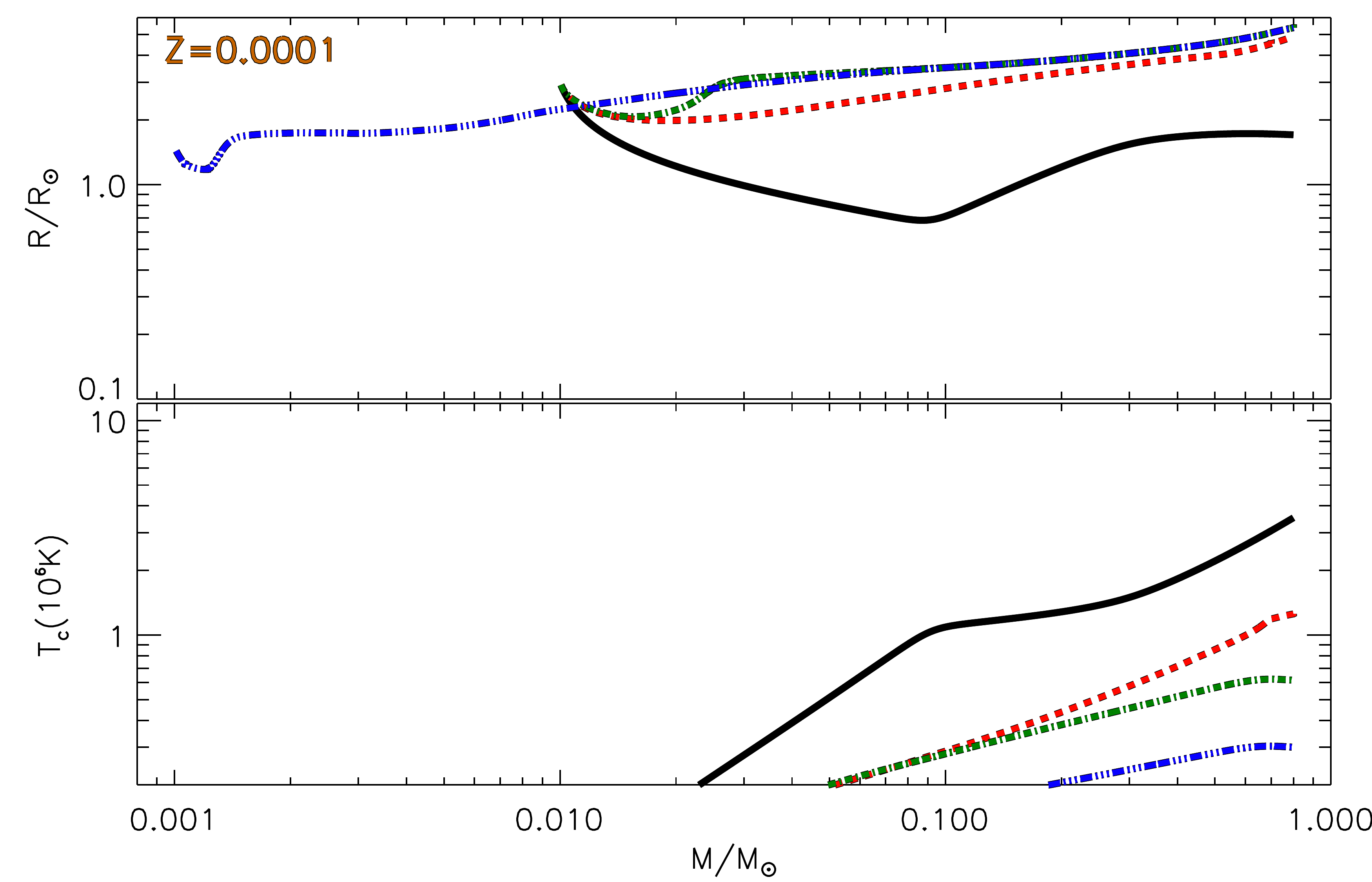}
\includegraphics[width=0.49\linewidth]{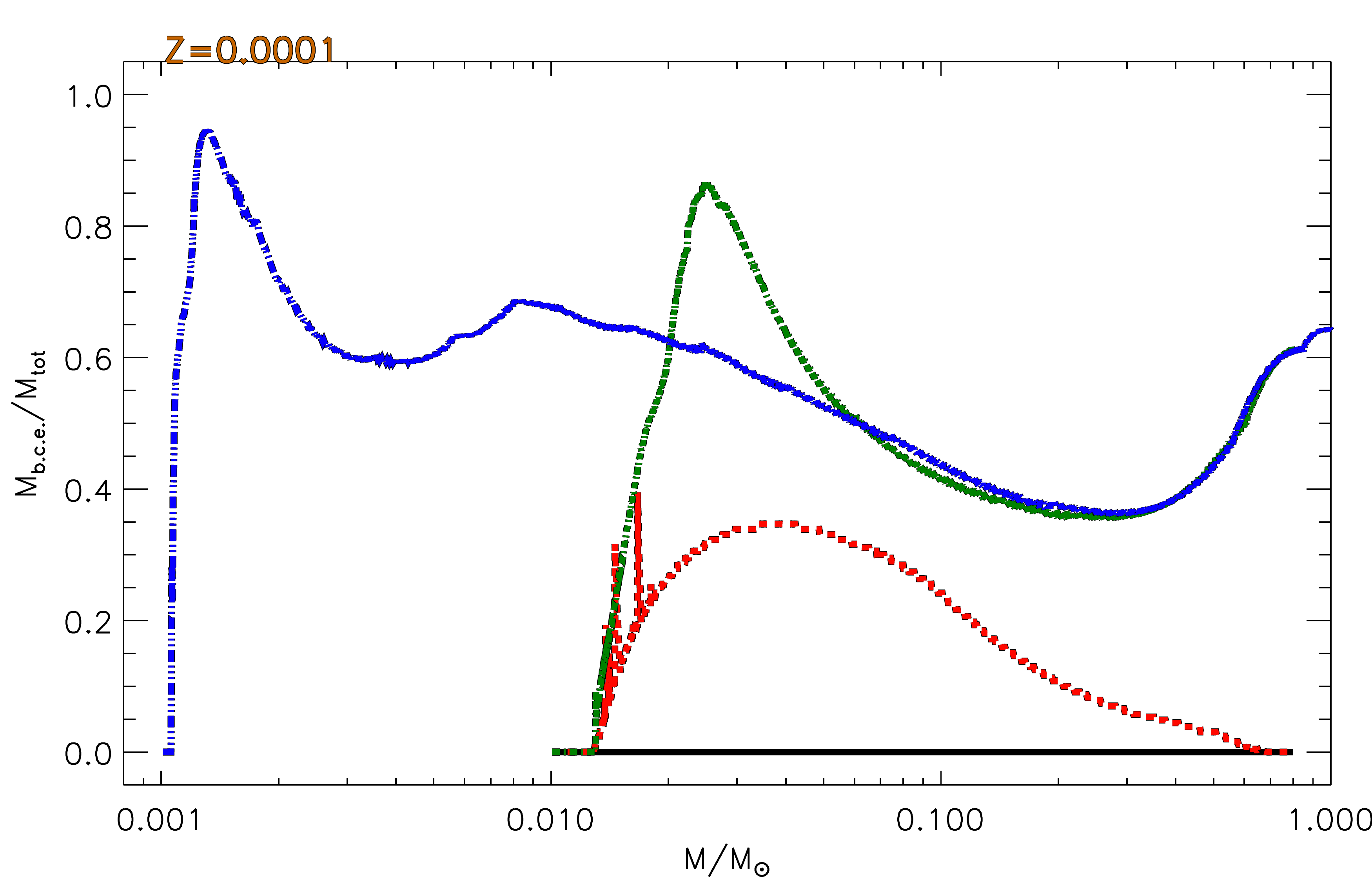}
\caption{Evolution of hot accreting models with $Z=0.0001$, final masses of $M_\mathrm{fin}=0.7$ and 0.8~\msun, and \mdot=$10^{-5}$~\mdotyr. Standard non-accreting models (grey lines) and reference cold LLC accreting models (\mseed=10~\mj, \rseed=3~\rsun, \mdot=$10^{-5}~$\mdotyr, black line) are also shown for comparison purposes. Top left panel: HR diagram. The position of models at 1~Myr (filled triangles) and 10~Myr (filled squares), deuterium burning region (thick magenta line), and the end of the accretion phase (cross) are marked. Top right panel: Relative contribution to the total luminosity of the nuclear burning ($L_\mathrm{nuc.}$), gravitational contraction, and expansion ($L_\mathrm{gra.}$) as well as accretion ($L_\mathrm{acc.}$) as a function of the stellar mass during the protostellar accretion phase. Bottom left panel: Evolution of the surface radius (in solar units) and central temperature (in units of $10^6$~K) as a function of the stellar mass. Bottom right panel: Relative mass coordinate at the bottom of the convective envelope as a function of the stellar mass.}
\label{fig:hot}
\end{figure*}

\section{Results for hot accretion models}
\label{hot}

So far, we have analysed the impact of protostellar accretion on PMS models in the cold accretion scenario when the accreted material contains negligible amounts of internal energy. On the other hand, the accreted matter could retain part of its thermal energy before reaching the surface of the protostar. In this case, that is, the hot accretion scenario, the matter accreted has a non-negligible internal energy content, which is redistributed within the star, thus  providing an external energy source. We assumed that the accretion luminosity $L_\mathrm{acc}$ defined in eq.~\ref{eq:acc_ene} is different from 0, exploring the extreme case \alpacc=1. This corresponds to the case of accretion from a disc, with the whole accretion energy being transferred to the protostar. 

Given the accretion rates considered here, accreting hot models develop a radiative core during the protostellar phase, which is different from the cold cases where the protostars are fully convective. In this situation, the infalling matter is uniformly redistributed within the external convective layers and not throughout the whole star. Similarly to the matter, the accreted energy is also supposed to be distributed in the convective envelope (see eq.\ref{eq:ma} and \ref{eq:la} in Sect.~\ref{models}). However, in the literature, some authors prefer to redistribute both the accreted matter and the accretion energy in the whole star \citep[e.g.][]{baraffe12}. To check the effect of the energy redistribution inside the stars, we tested two cases. In one case, the accretion energy was uniformly redistributed throughout the whole structure (hereafter EWS, Energy distributed in the Whole Star); in the other case, the energy was only redistributed (uniformly) within the convective envelope (hereafter ECE, Energy distributed in the Convective Envelope). 

Figure~\ref{fig:hot} shows the evolution of models computed starting from \mseed=10~\mj, \rseed=3.0~\rsun with \mdot=$10^{-5}$~\mdotyr{} for EWS and ECE hot cases. The evolution in the HRD is shown in the top left panel of the figure. Hot accretion models follow a path that corresponds to more extended and brighter objects during the whole protostellar evolution, compared to the cold accretion models. After a quick gravitational contraction (where the radius changes from 3 to 2~\rsun, bottom left panel in Fig.~\ref{fig:hot}) the models expands again going back to approximately the initial radius. The rate of  expansion depends on the extension of the region where the accretion energy is deposed. In the EWS case, the expansion begins at a larger value of the actual total mass compared to the ECE case. The reason is that while in EWS models the accretion energy continues to be redistributed throughout the whole structure, in the ECE case it is deposed in a thinner region (just the convective envelope). If the same amount of energy is spread in a thinner region, that region reacts with a stronger expansion, leading to larger radii for a given actual total mass. 

To clearly visualise the three energy sources in hot models, in the top right panel of Fig.~\ref{fig:hot} we show the contribution to the total luminosity of the nuclear burning, gravitational, and accretion energy. The difference between hot and cold models is clearly visible by comparing the black (cold) and the green or red sequences (hot models). While in the cold case the energy comes from the gravitational contraction ( $L_\mathrm{gra.}/L_\mathrm{tot} = 1$, before D-burning ignites), in hot models the gravitational energy becomes rapidly negative, marking the expansion of the star. Such expansion corresponds to the injection of accretion energy, as witnessed by the rise in $L_\mathrm{acc.}/L_\mathrm{tot}$ at $M\ga 0.001$~\msun. As the accretion proceeds, the contraction produces a temperature increase that leads to the D-burning onset in cold models. For this model and masses larger than about 0.06-0.07~\msun{}, the main energy source is the deuterium burning ($L_\mathrm{nuc.}/L_\mathrm{tot} \gg 1$). The fact that $L_\mathrm{nuc.}/L_\mathrm{tot} \gg 1$ means that most of the deuterium energy is absorbed inside the star and transformed into gravitational energy. On the contrary, hot models ignite deuterium at larger masses, about 0.4-0.6~\msun{}, which depends on if they are EWS or ECE.\ However, deuterium energy never exceeds twice the surface luminosity. It is interesting to notice that in ECE models (green line), both gravitational and accretion luminosities stabilise earlier than in EWS (red line) ones. This happens because, as stated above, the accretion energy is redistributed in a thinner region in the ECE case, so a more rapid expansion occurs which leads to a more rapid stabilisation of the accretion process. It is important to notice that when D-burning onsets, hot models are sustained by two energy sources: D-burning and accretion energy.

Figure ~\ref{fig:hot} also shows an accreting model starting from a lower \mseed{} value (blue line), namely \mseed=1~\mj. Qualitatively, ECE models with low seed mass show the same evolution as ECE models with \mseed=10~\mj, which is contrary to what happened in the cold cases, where \mseed{} played a crucial role in the protostellar evolution.

ECE and EWS models differ in the extension of the convective envelope, as is shown in the bottom right panel of Fig.~\ref{fig:hot}, where we plotted the relative mass coordinate at the bottom of the convective envelope (i.e. $M_\mathrm{b.c.e.}/M_\mathrm{tot}$)\footnote{As we plotted the mass coordinate at the bottom of the convective envelope $M_\mathrm{b.c.e.}/M_\mathrm{tot}$, a model with a thinner convective envelope (at fixed mass) shows a larger value for this quantity.}. Considering the models with \mseed=10~\mj, for an actual mass of about 0.02-0.03~\msun, which corresponds to the point where radius expansion starts, the mass in the convective envelope is very different in the ECE and EWS cases.\ In the former, the star has a convective region that reaches a relative mass coordinate of about 0.8-0.85 at most (it contains less than about 15-20~percent of the stellar mass); whereas, in the EWS case, it reaches 0.30-0.35 (70-75~percent of the stellar mass is in the convective envelope). In addition, the convective envelope in ECE models is significantly thinner in mass than that in EWS ones during the whole protostellar evolution. In ECE models, it barely reaches a maximum relative mass depth of $M_\mathrm{b.c.e.}/M_\mathrm{tot}\approx$~0.37-0.4 (when $M_\mathrm{tot} \sim $0.2-0.3~\msun) before the convection withdraws outwards; whereas, in the EWS ones, the convective envelope gets deeper and deeper as the mass increases ($M_\mathrm{b.c.e.}/M_\mathrm{tot}\approx 0.2$ for $M_\mathrm{tot} \sim0.1 $\msun~and $M_\mathrm{b.c.e.}/M_\mathrm{tot}\la 0.05$ for $M_\mathrm{tot}\sim0.4~$\msun). In particular, in the EWS case, the protostar is essentially  fully convective for masses  larger than 0.6-0.65~\msun, which is different for the ECE case where the protostar is never fully convective. Such a large difference in the extension of the convective envelope between ECE and EWS models reflects the difference at the same mass in the radius, with ECE models having a radius that is systematically larger than that of the EWS ones. In Fig.~\ref{fig:hot} we overplotted, just for comparison purposes, the evolution of cold models that were computed by adopting the same \mseed, \rseed, and \mdot {} to clearly shown that the difference between hot and cold models is very large. 

One of the effects of such cold and expanded structures in the hot accretion cases (both ECE and EWS) is that, at the end of the accretion, models are left very close to the Hayashi tracks of standard models with the same final mass, and  more importantly, they are quite luminous. The models right after the protostellar phase (when accretion is completed) are at the beginning of the D-burning phase of standard tracks (magenta segment is in the top left panel of Fig.~\ref{fig:hot}). In this evolutionary stage, the star has a relatively small \tkh{} (about $10^5$~yr), so it has plenty of time to relax to approach the non-accreting tracks, and the location of the 1~Myr models is the same for hot accretion and standard non-accreting tracks. Moreover, because deuterium burning essentially happens on the Hayashi track after the accretion phase, any temporal offset with standard calculations is erased. As such, the position of the 1 and 10~Myr models computed including hot protostellar accretion shows no appreciable differences with respect to that of the standard models.
\begin{figure}
\centering
\includegraphics[width=0.98\columnwidth]{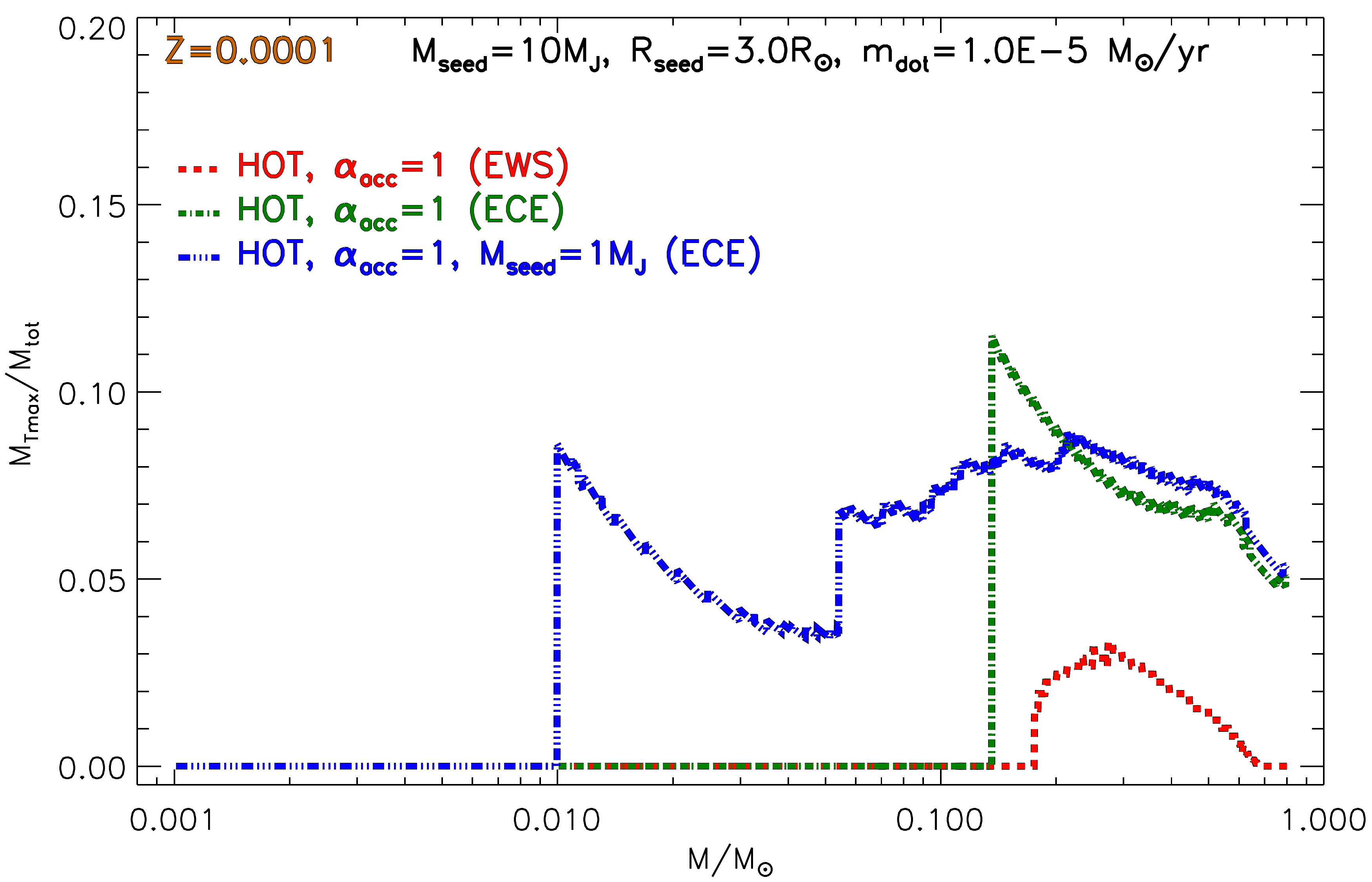}
\caption{Location of the maximum temperature inside the star (i.e. relative mass coordinate $M_{T\mathrm{max}}/M_\mathrm{tot}$) as a function of the actual stellar mass during the protostellar accretion phase for the EWS and ECE hot models of Fig.~\ref{fig:hot} computed using \mseed=1 and 10~\mj.} 
\label{fig:tmax}
\end{figure}

There is an additional point that deserves to be discussed in reference to hot accretion models, which is the temperature profile inside the star. Figure~\ref{fig:tmax} shows the location (relative mass coordinate) of the maximum temperature inside the accreting star  as a function of the actual stellar mass for the three models with hot accretion shown in Fig.~\ref{fig:hot}: EWS and ECE with \mseed=10~\mj{} and ECE with \mseed=1~\mj, all were computed with \alpacc=1 and \mdot=$10^{-5}$~\mdotyr. Because of the presence of an external energy source during the protostellar accretion, the maximum temperature ($T_\mathrm{max}$) region is off centre in both ECE and EWS cases. The position of $T_\mathrm{max}$ depends on the actual value of the stellar mass (i.e. on time) and on the way in which the accretion energy is distributed into the stellar structure. In particular, in the ECE case, $T_\mathrm{max}$ is more external than in the EWS case. This is caused by the fact that in ECE models, all accretion energy is distributed within the convective envelope and consequently it only directly warms up these layers. Then, the heat flows from the convective envelope towards the centre. Qualitatively, the resulting thermal profile is thus the result of the heat flow due to the accretion energy, which would produce the hottest regions close to the convective envelope, in addition to the intrinsic temperature gradient determined by the hydrostatic condition, which requires the maximum temperature at the stellar centre. The two combined effects produce a maximum temperature that is off centre. On the contrary, in the EWS case, the accretion energy is distributed throughout the whole star.\ Thus the entire structure is heated up, albeit not uniformly,  depending on the mass enclosed in the individual layers. In these models, $T_\mathrm{max}$is  much closer to the centre than in the ECE ones. 
\begin{figure*}
\centering
\includegraphics[width=0.49\linewidth]{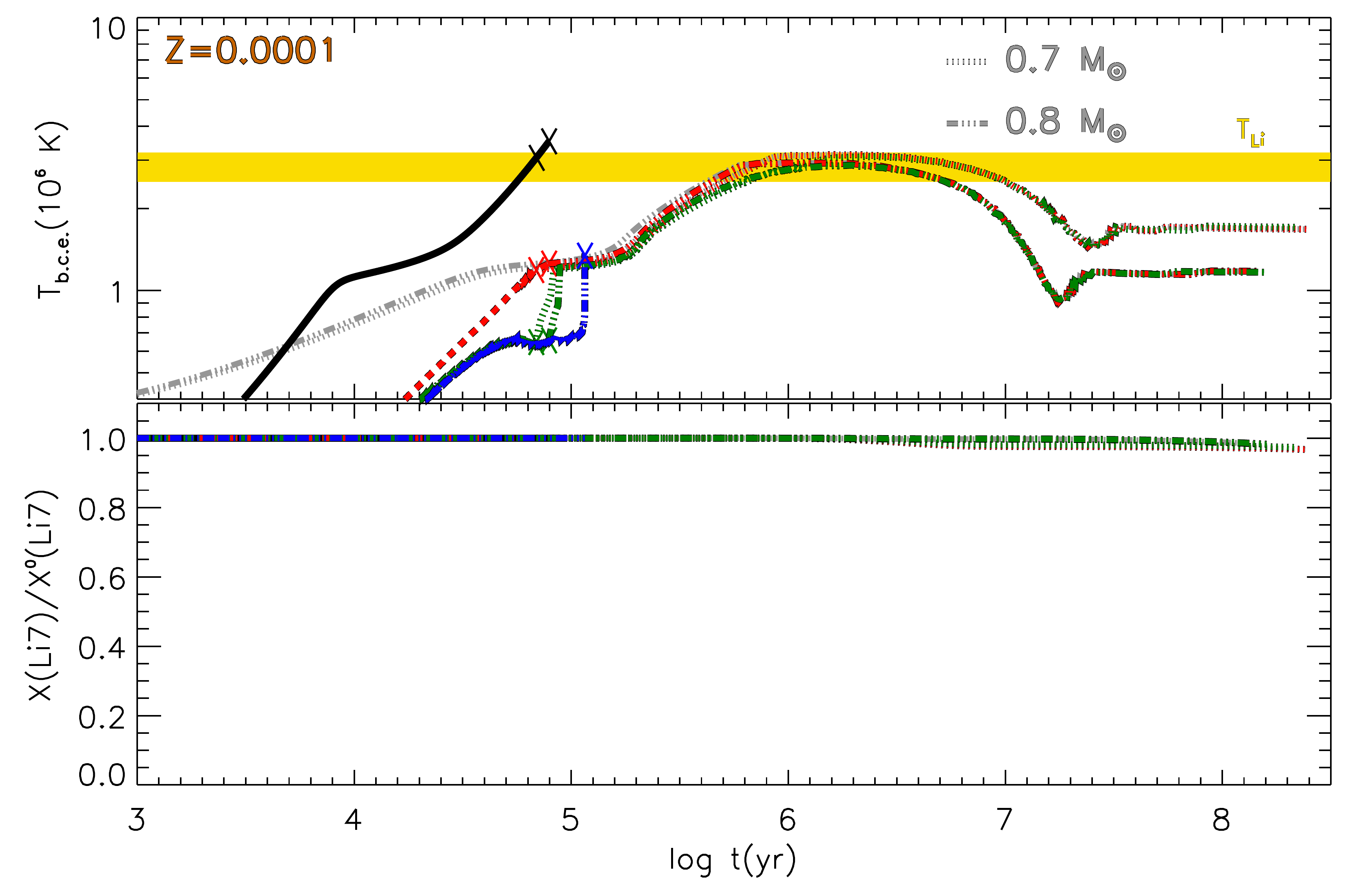}
\includegraphics[width=0.49\linewidth]{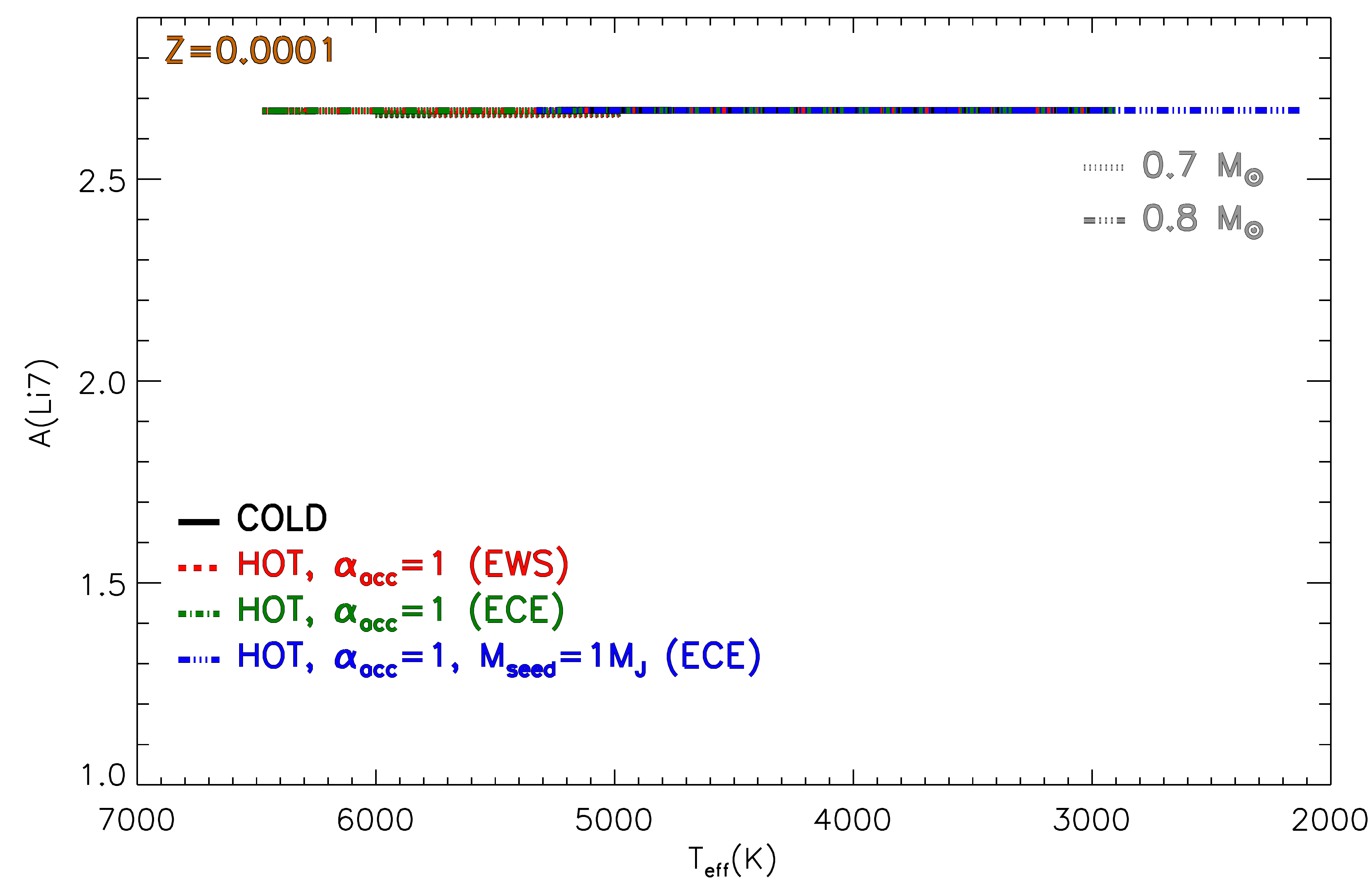}
\caption{Temporal evolution of the temperature at the bottom of the convective envelope and surface lithium abundance (divided by the initial one, left panel) and surface $A(Li)$ as a function of \teff{} (right panel) for the same hot ECE and EWS models shown in Fig.~\ref{fig:hot}. The evolution of the reference cold accretion model (LLC, black line) is only shown during the accretion phase. The crosses mark the end of the accretion phase.}
\label{fig:hot_li}
\end{figure*}

PMS lithium evolution is not affected by the inclusion of the protostellar hot accretion scenario. The reason is that during the protostellar phase, the temperature at the bottom of convective envelope in such models is lower than that in cold accretion ones, as shown in Fig.~\ref{fig:hot_li}. In this case, \tce{} barely reaches 1-1.5$\times 10^6$~K; whereas in ECE, \tce{} is even smaller, thus too low to destroy Li. Moreover, at the end of the accretion, models are left on the Hayashi line of the corresponding standard track, and the PMS lithium depletion is the same as in standard models.

We tested the impact of changing \mseed{} on hot accretion models. Figure~\ref{fig:hot} also shows the evolution of ECE models that start from \mseed=1~\mj. It is clear that the models with \mseed=1~\mj{} converge with those with \mseed=10~\mj{} in all diagrams, hence the value of the adopted seed mass value is inconsequential for the subsequent PMS evolution, providing that \mseed$< M_\mathrm{fin}$ (the final mass of the star).
\begin{figure}
\centering
\includegraphics[width=0.98\columnwidth]{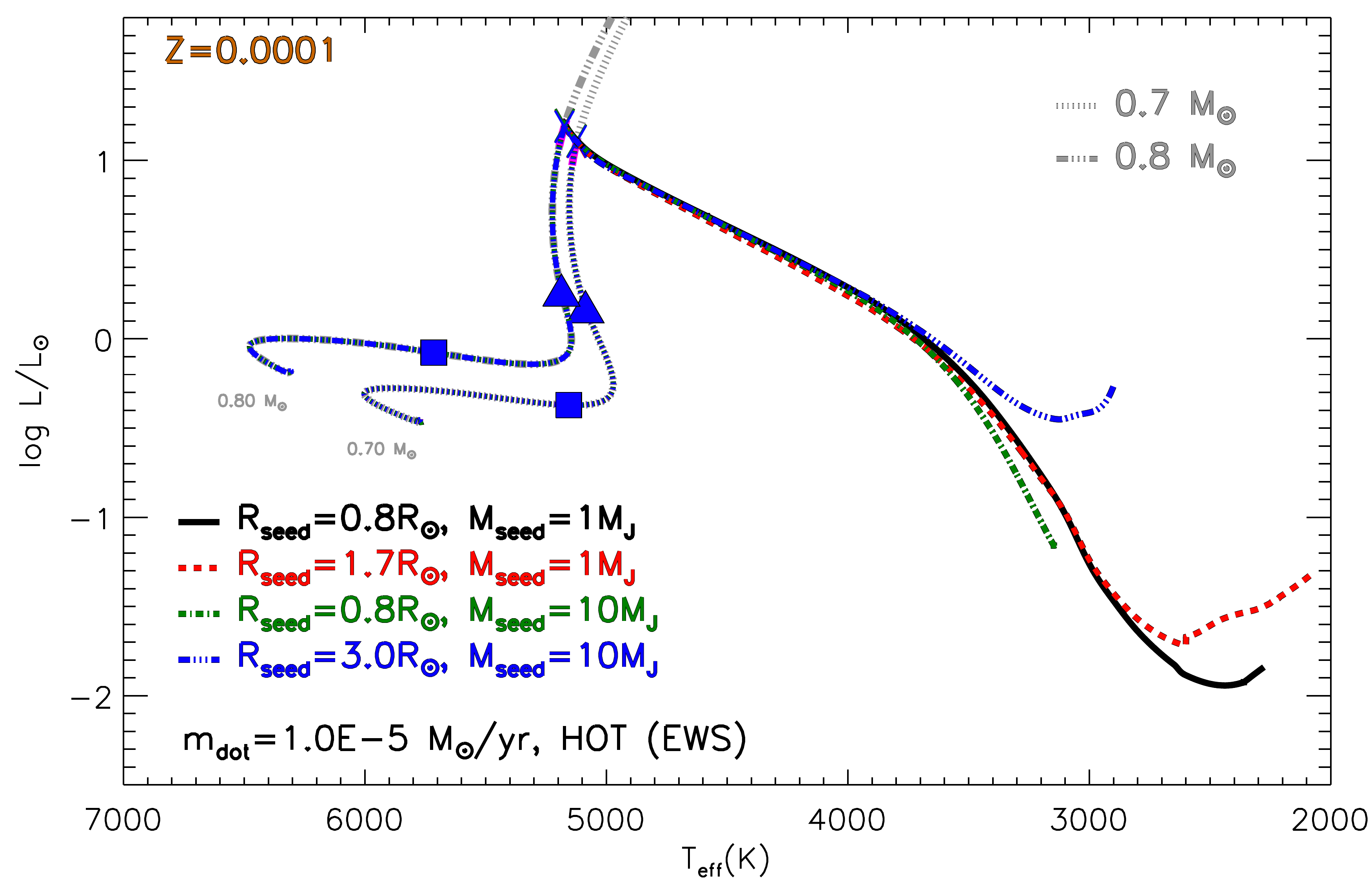}
\caption{Evolution in the HR diagram of hot accreting models with different values of \rseed{} and \mseed. Symbols are the same as in Fig.~\ref{fig:hot}.}
\label{fig:rseed_hot}
\end{figure}

As a final test, we analysed the effect of changing \rseed{} on hot EWS  models\footnote{Because of convergence difficulties, we computed the models using the EWS configuration.}. Figure~\ref{fig:rseed_hot} shows the effect on the HR diagram of a variation of \rseed{} on models with \mseed=1 and 10~\mj. In this case, the initial radius only affects the beginning of the protostellar accretion, but the \rseed{} difference is progressively absorbed by the models during the protostellar evolution. In particular, the effect of the adopted \rseed{} vanishes well before the stars end the accretion. We recall that in hot models, the structure stays expanded mainly because of the accretion energy. As a consequence, if the accretion occurs on a seed mass with a  small \rseed, then the accretion energy is large ($L_\mathrm{acc} \propto 1/R_{\star}$). This leads to a large amount of accretion energy that makes the star expand. On the contrary, if \rseed{} increases, $L_\mathrm{acc}$ decreases and the star mainly contracts. The presence of the external energy source $L_\mathrm{acc}\propto 1/R_{\star}$ acts as feedback, which keeps the star in a region of bright and expanse objects. Similar to the other hot accretion cases, a PMS lithium abundance does not depend on the adopted \rseed{} value.

\subsection{Burst accretion scenario}
\label{bursts}
\begin{figure*}
\centering
\includegraphics[width=0.49\linewidth]{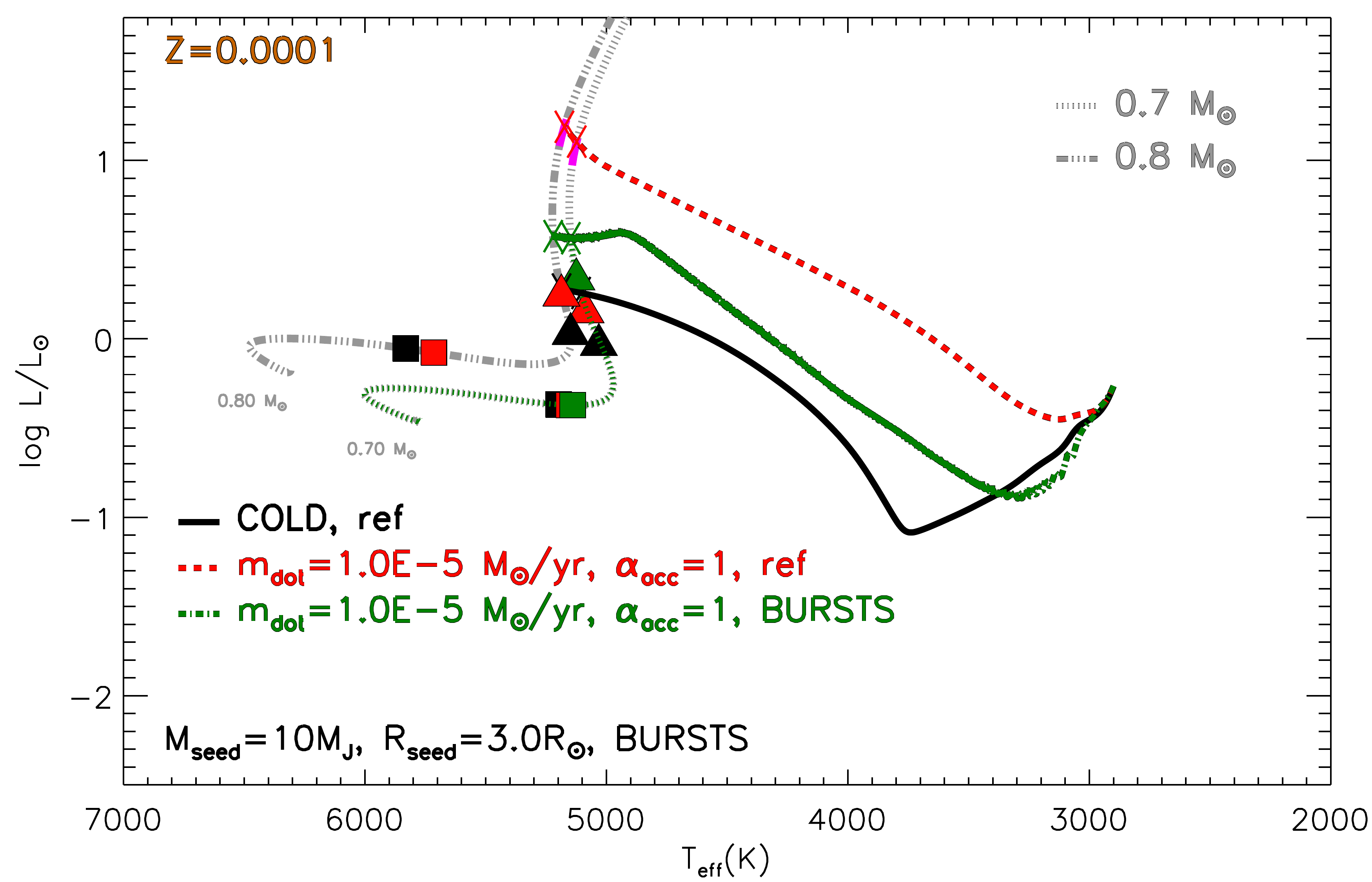}
\includegraphics[width=0.49\linewidth]{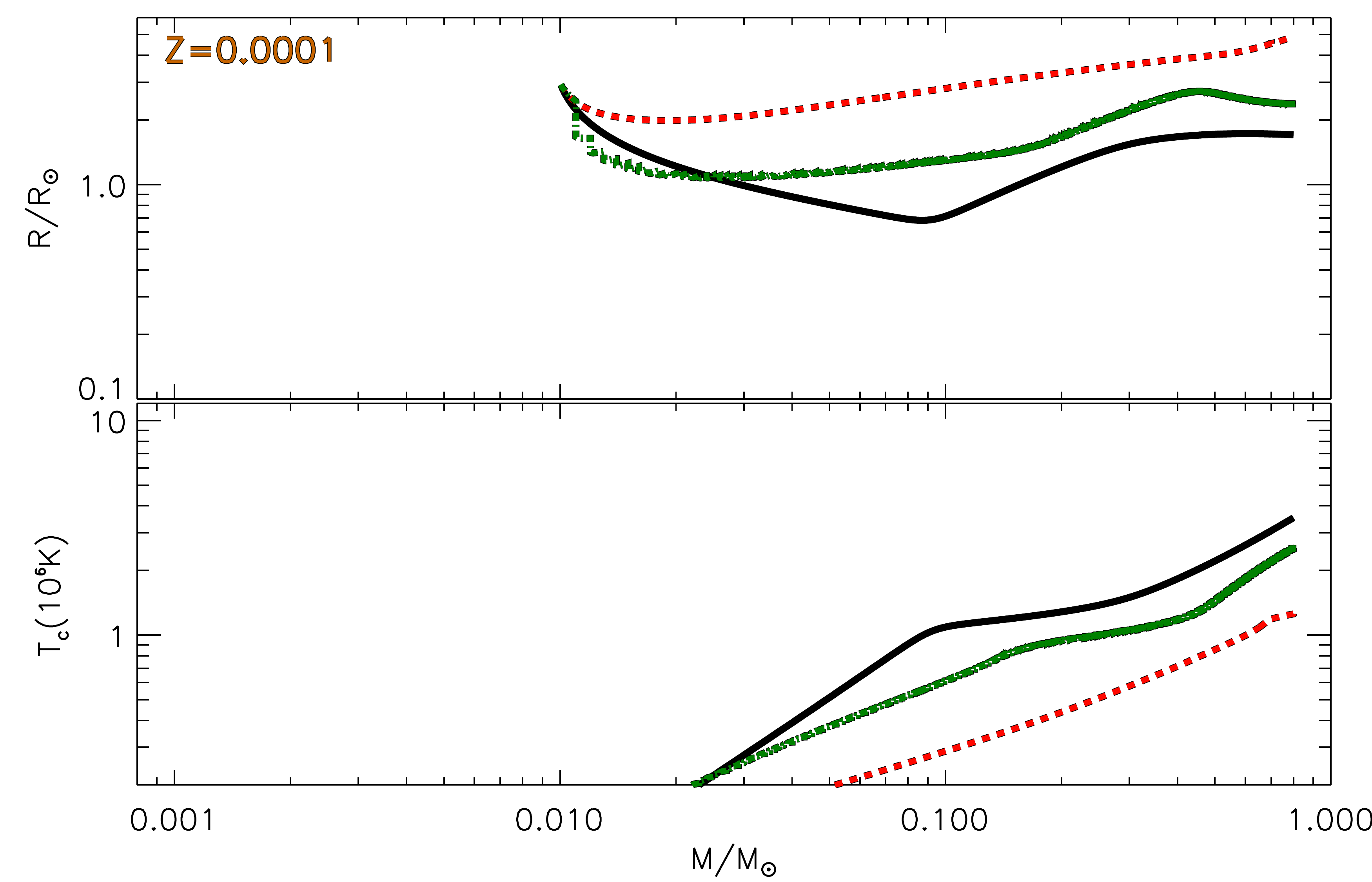}
\caption{Evolution of hot bursts accreting models with $Z=0.0001$, final masses of $M_\mathrm{fin}=0.7$ and 0.8~\msun, \mseed=10~\mj, \rseed=3~\rsun{}, and \mdot=$10^{-5}$\mdotyr{} (green line). Standard non-accreting models  (grey lines), reference cold LLC (black line), and the hot (red line) accreting models with \mseed=10~\mj, \rseed=3\rsun{}, and \mdot=$10^{-5}$~\mdotyr{} are shown for comparison purposes. Left panel: HR diagram with the position of models at 1~Myr (filled triangles) and 10~Myr (filled squares), deuterium burning region (thick magenta line), and the end of the accretion phase (crosses) are marked. Right panel: Evolution of the surface radius (in solar units) and central temperature (in units of $10^6$~K) as a function of the actual total mass. Hot and cold reference accretion cases (red and black lines) are only shown during the accretion phase for comparison purposes.}
\label{fig:hot_burst}
\end{figure*}
\begin{figure}
\centering
\includegraphics[width=0.98\linewidth]{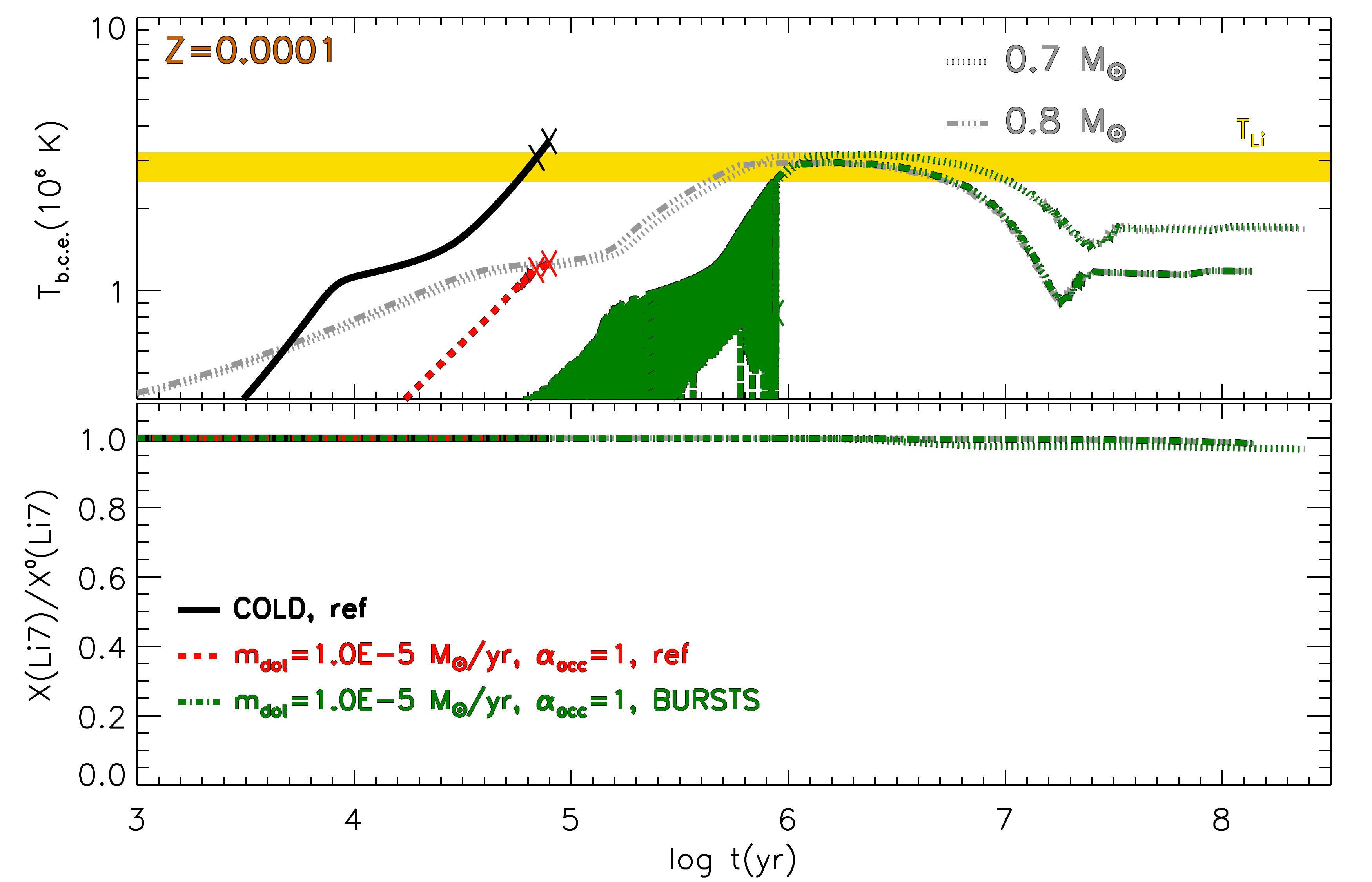}
\caption{Temporal evolution of the temperature at the bottom of the convective envelope and surface lithium abundance (divided by the initial value) for the same bursts, hot and cold LLC accretion models are shown in Fig.~\ref{fig:hot_burst}. The crosses mark the end of the accretion phase.}
\label{fig:hot_burst_li}
\end{figure}

In the previous sections we assumed a constant mass accretion rate. However, there are both observational and theoretical indications that accretion occurs in episodes of high mass 
accretion rates (bursts), which are separated by intervals of quiescence when accretion rates drop to low values \citep[see e.g.][and references therein]{hartmann96,vorobyov05,enoch09,hartmann18,hsieh18}. Thus, we also investigated the possibility of having episodic accretion in the hot case. In this simple test, we have assumed that the star does not accrete material for an interval of time $\Delta t_q$ (quiescent phase), after which it is subjected to a burst of accretion with \mdot$\ne 0$ for an interval of time of $\Delta t_b$. We assumed that the bursts occur with \alpacc=1 (hot accretion). If not, the results are very similar to the cold accretion cases already discussed. In this simple test, the accretion history consists of a succession of bursts and quiescent phases during the entire protostellar evolution. Our simplified treatment of a time dependent \mdot{} value is  intended to only check whether bursty accretion modifies the evolution compared to the constant accretion case. 

For the computation of this class of models, we assumed the following parameters: \mdot=$10^{-5}$~\mdotyr, $\Delta t_q =1000$~yr, $\Delta t_b=100$~yr, and there is EWS hot accretion during the bursts \citep[similar to one of the cases analysed by][]{baraffe09}. Figure~\ref{fig:hot_burst} shows the HR diagram as well as the radius and central temperature evolution along with the hot one with constant accretion. In the HR diagram, the resulting evolutionary track lies in between the hot and the cold case. The accretion leaves a model on the Hayashi track, but it is well below the D-burning. The age is only mildly affected by the presence of the bursts; indeed the 1~Myr bursts model is slightly more luminous than the standard one, while at 10~Myr the differences have been erased. The difference at 1~Myr is mainly caused by the different accretion history. In the hot case, we assumed a constant accretion rate; whereas, in the bursts model, the accretion is zero for most of the time because we adopted a duration of the bursts about 1/10 the quiescent phase. So, at the same age, the mass of the star in the hot and bursts model are very different. Indeed, to reach the final mass of 0.7 or 0.8~\msun, the bursts model takes about 1~Myr, which in comparison to the case of a constant accretion rate is about 7-8$\times 10^4$~yr. 

For what concerns the radius evolution (right panel of Fig.~\ref{fig:hot_burst}), it first decreases because of the gravitational contraction, then it expands when the D-burning gets efficient (actual mass $M\sim 0.15$~\msun). Models are almost fully convective during the quiescent phase; whereas, during the bursts, the models become partially radiative because of the large amount of accretion energy that has been deposed in the star. This produces a rapid variation of the position in the bottom of the convective envelope and consequently of the value for the temperature at this point. Figure~\ref{fig:hot_burst_li} shows the temporal evolution of \tce{} and the surface lithium in bursts models. The mean value of \tce{} is similar to the cold accretion case, reaching a maximum of about 2-2.5$\times10^6~$K, but for a time that is too short to produce an appreciable effect on lithium. The bursts model therefore does not produce any effect on the protostellar surface lithium abundance, and consequently the PMS Li evolution is the same as in standard non-accreting models.

\subsection{Hot plus cold protostellar accretion}
\begin{figure*}
\centering
\includegraphics[width=0.49\linewidth]{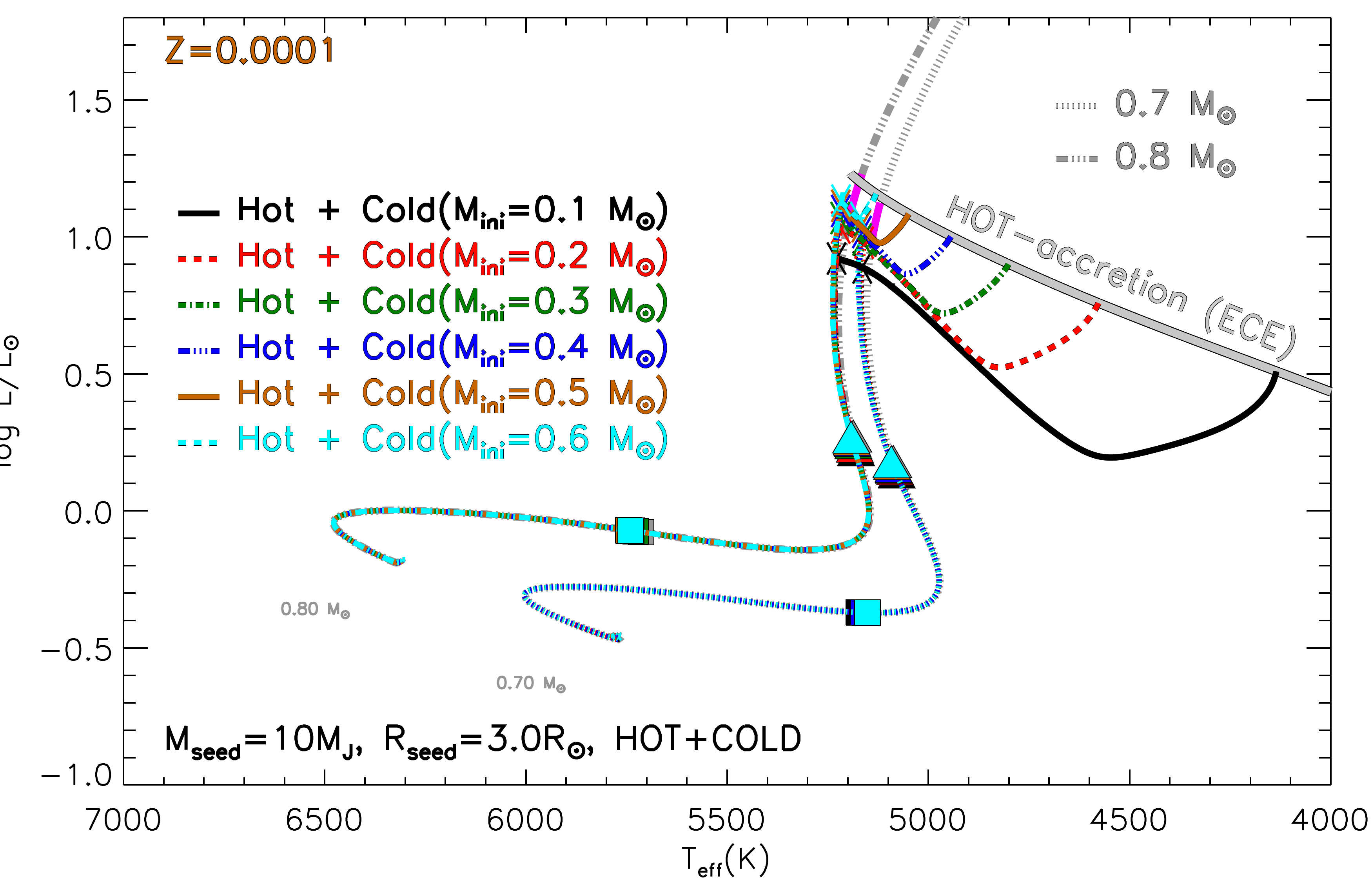}
\includegraphics[width=0.49\linewidth]{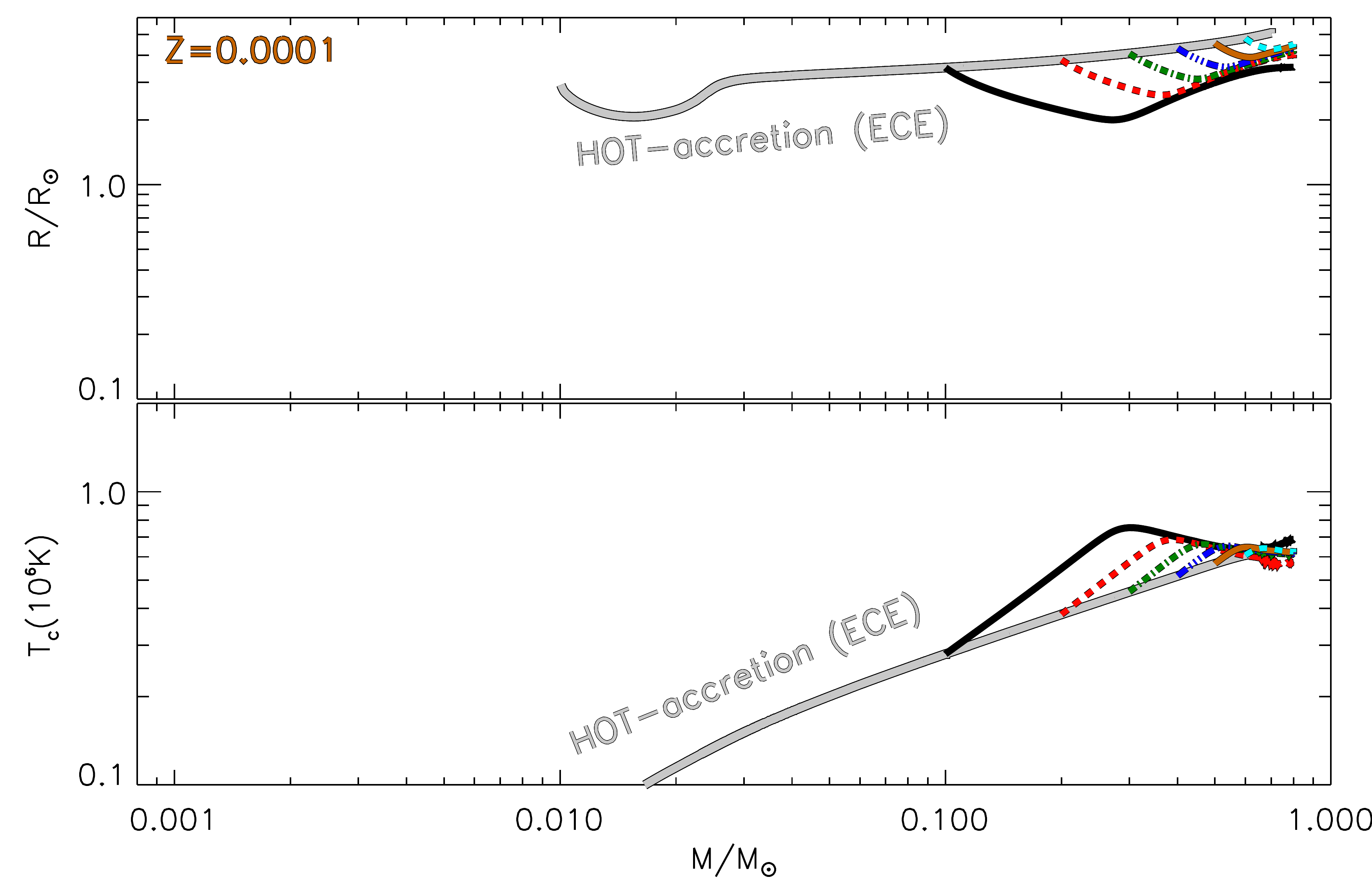}
\includegraphics[width=0.49\linewidth]{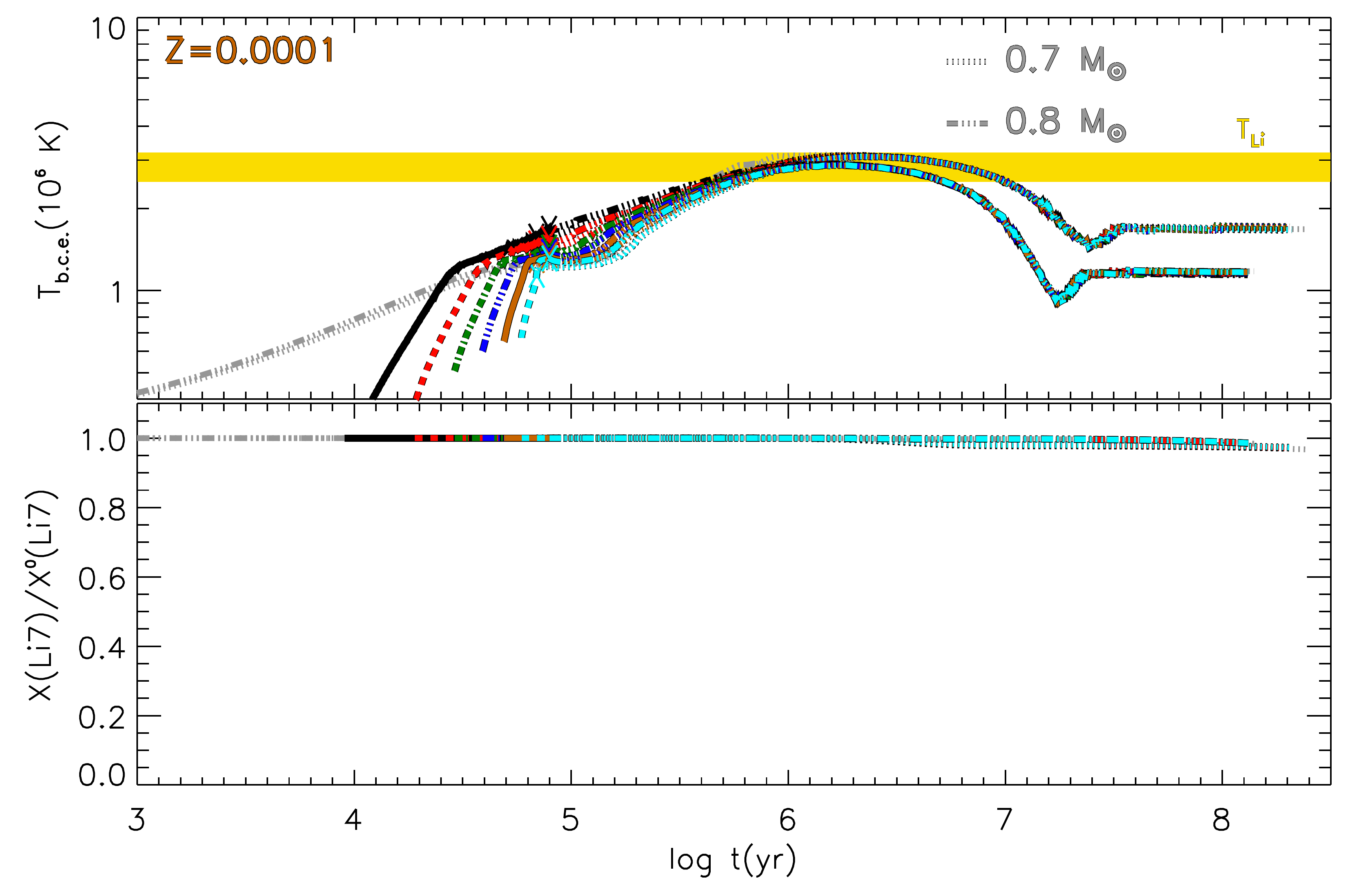}
\includegraphics[width=0.49\linewidth]{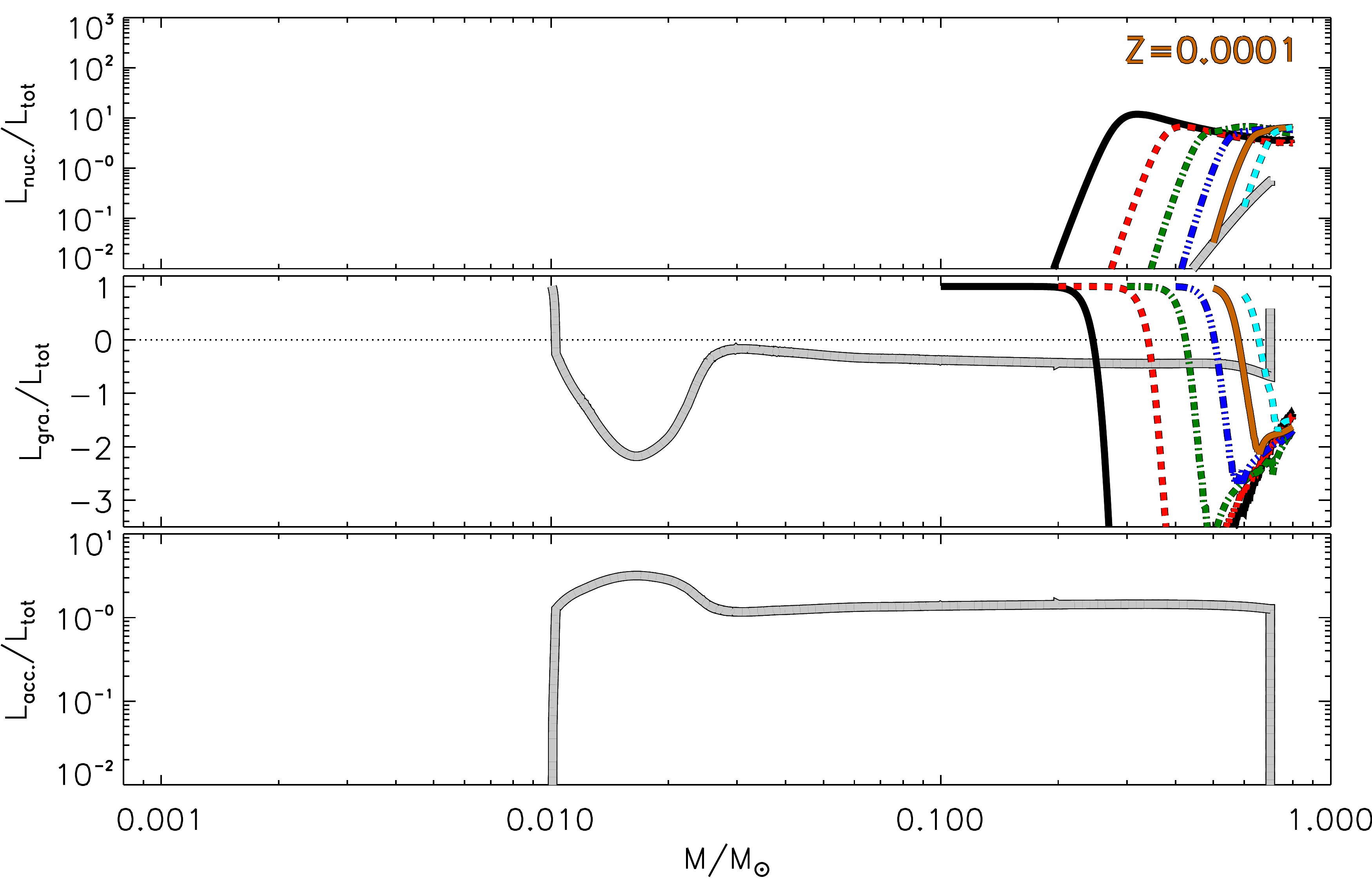}
\caption{Evolution of hot and cold accretion models with $Z=0.0001$, final masses of $M_\mathrm{fin}=0.7$ and 0.8~\msun, \mseed=10~\mj, \rseed=3~\rsun{}, and \mdot=$10^{-5}$\mdotyr. The results are shown for different mass values ($M_\mathrm{ini}$) at the beginning of the cold disc accretion. Standard non-accreting models (grey lines) are also shown for comparison purposes. Top left panel: HR diagram with the position of models at 1~Myr (filled triangles) and 10~Myr (filled squares), deuterium burning region (thick magenta line), and the end of the accretion phase (crosses) are marked. The position of the beginning of the cold disc accretion for the different starting masses is also shown. Top right panel: Evolution of the total radius (in solar radius) and central temperature (in units of $10^6$~K) as a function of the stellar mass during the protostellar phase. Bottom left panel: Temporal evolution of the temperature at the bottom of the convective envelope and surface lithium abundance (normalised to the initial one) for accreting and non-accreting models (grey lines). Bottom right panel: Contribution to the total luminosity of the nuclear ($L_\mathrm{nuc.}$), accretion ($L_\mathrm{acc.}$), and gravitational luminosity ($L_\mathrm{gra.}$) as a function of the actual value for the total mass.}
\label{fig:hot_cold}
\end{figure*}

As an additional test, we analysed a mixed scenario, which consists of a hot and a cold accretion phase. We assumed a simple accretion history. After a first hot accretion phase (\alpacc=1, ECE models described in the previous section), the accretion switches to cold accretion (\alpacc=0) --maintaining the same accretion rate for simplicity-- when a certain value of the mass is reached (hereafter $M_\mathrm{ini}$). Such a configuration mimics the evolution of a star embedded in the cloud that accretes hot material followed by a cold thin disc accretion where most of the accretion energy is radiated before the matter reaches the star. To test the dependence of such a scenario on the value of the mass $M_\mathrm{ini}$ when the transition occurs from hot to cold, we selected the following five values for $M_\mathrm{ini}$ on the ECE sequence, namely $M_\mathrm{ini}$=0.1, 0.2, 0.3, 0.4, 0.5, and 0.6~\msun.

Figure~\ref{fig:hot_cold} shows the results of the computations. Independent of the starting mass, the evolution during the cold accretion phase does not produce large effects on the structure. Concerning the evolution in the HRD (top left panel), all of the models with $M_{ini}  \ge 0.2~$\msun{} converge to a position in the HR diagram that is very close to the D-burning of standard tracks; whereas, for a lower value of the initial mass on the hot sequence, accreting models converge to standard ones still along the Hayashi track, but at a slightly lower luminosity. In all cases, neither the 1~Myr nor the 10~Myr model are appreciably affected by the inclusion of the protostellar accretion phase. 

During the cold accretion phase, models evolve, after a contraction, at an almost constant or slightly increasing radius because of the D-burning. It is important to notice that, before D-burning, the model is still partially radiative, as it was during the hot accretion phase.\ However, as deuterium ignites, models quickly become fully convective. 

Even when a cold phase is introduced, the star remains bright and relatively cold. As such, \tce{} is below $T_\mathrm{Li}$ and consequently no lithium depletion occurs. When the final mass is reached, the structure relaxes to a standard one (on time scales shorter than 1~Myr) and the evolution (also the surface lithium abundance) is essentially the same as for standard tracks (bottom right panel).

\section{Metallicity effects: Models with Z=0.001 and Z=0.005}
\label{metals}
\begin{figure*}
 \centering
 \includegraphics[width=0.49\linewidth]{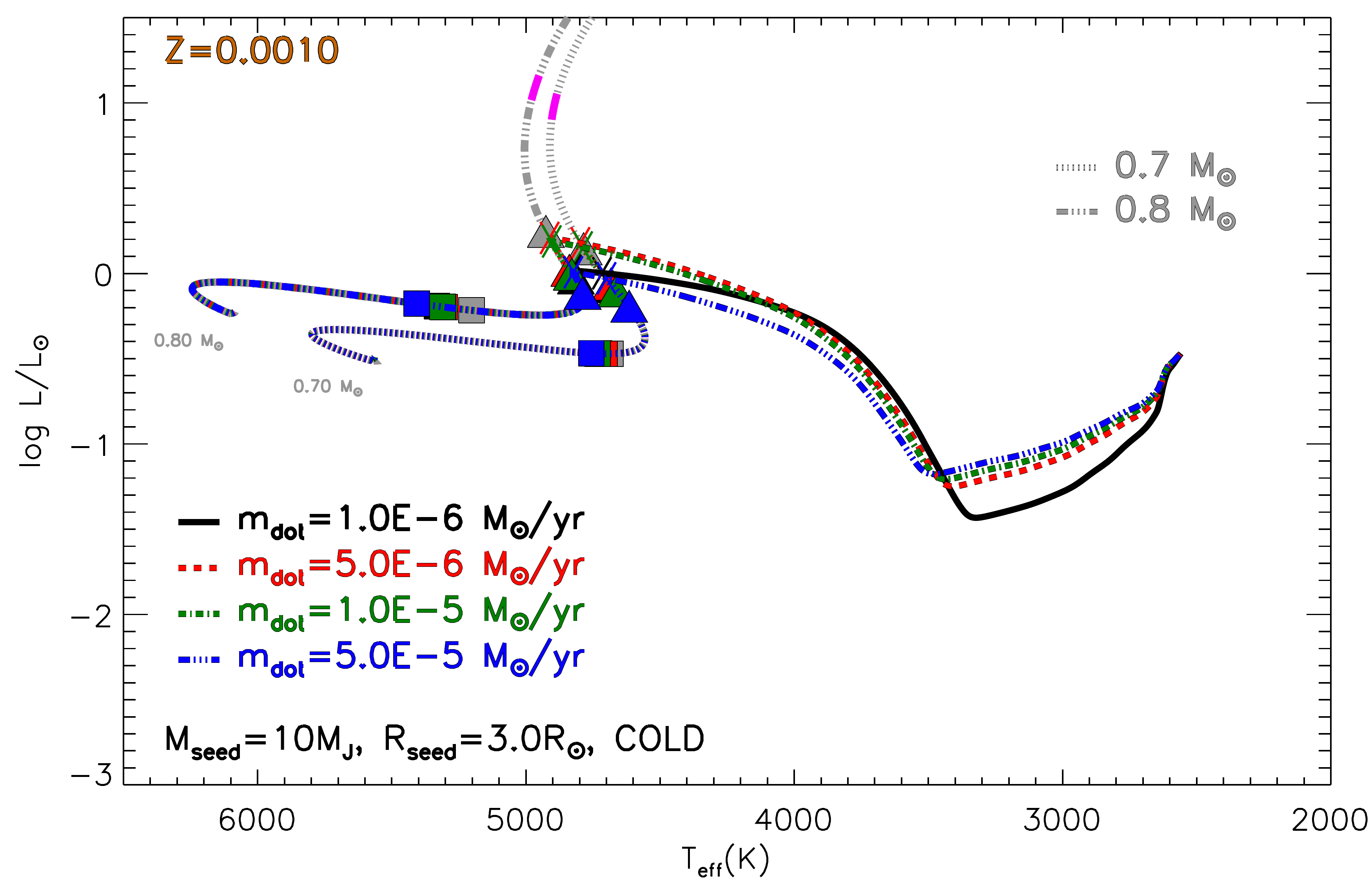}
 \includegraphics[width=0.49\linewidth]{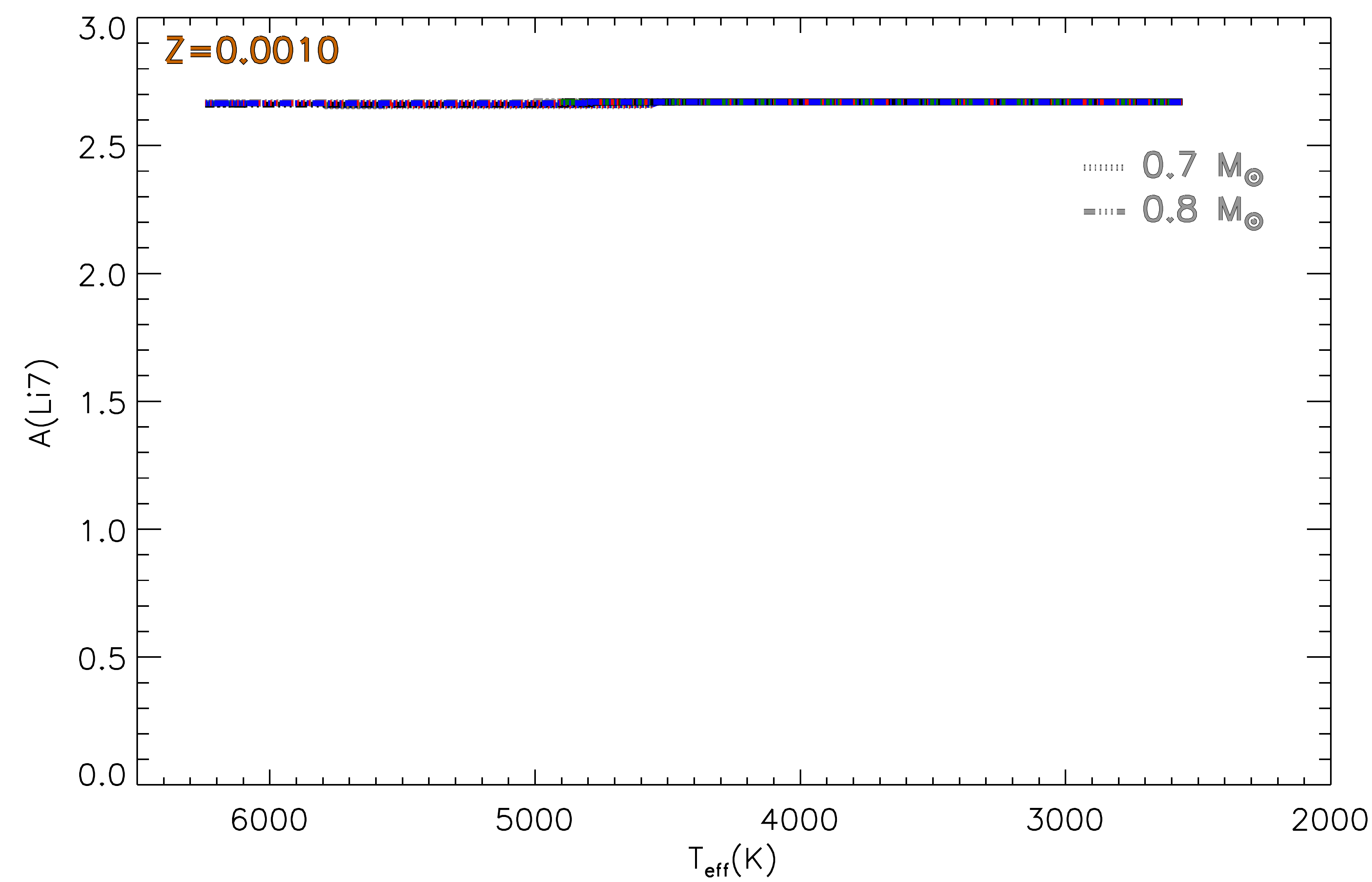}
 \includegraphics[width=0.49\linewidth]{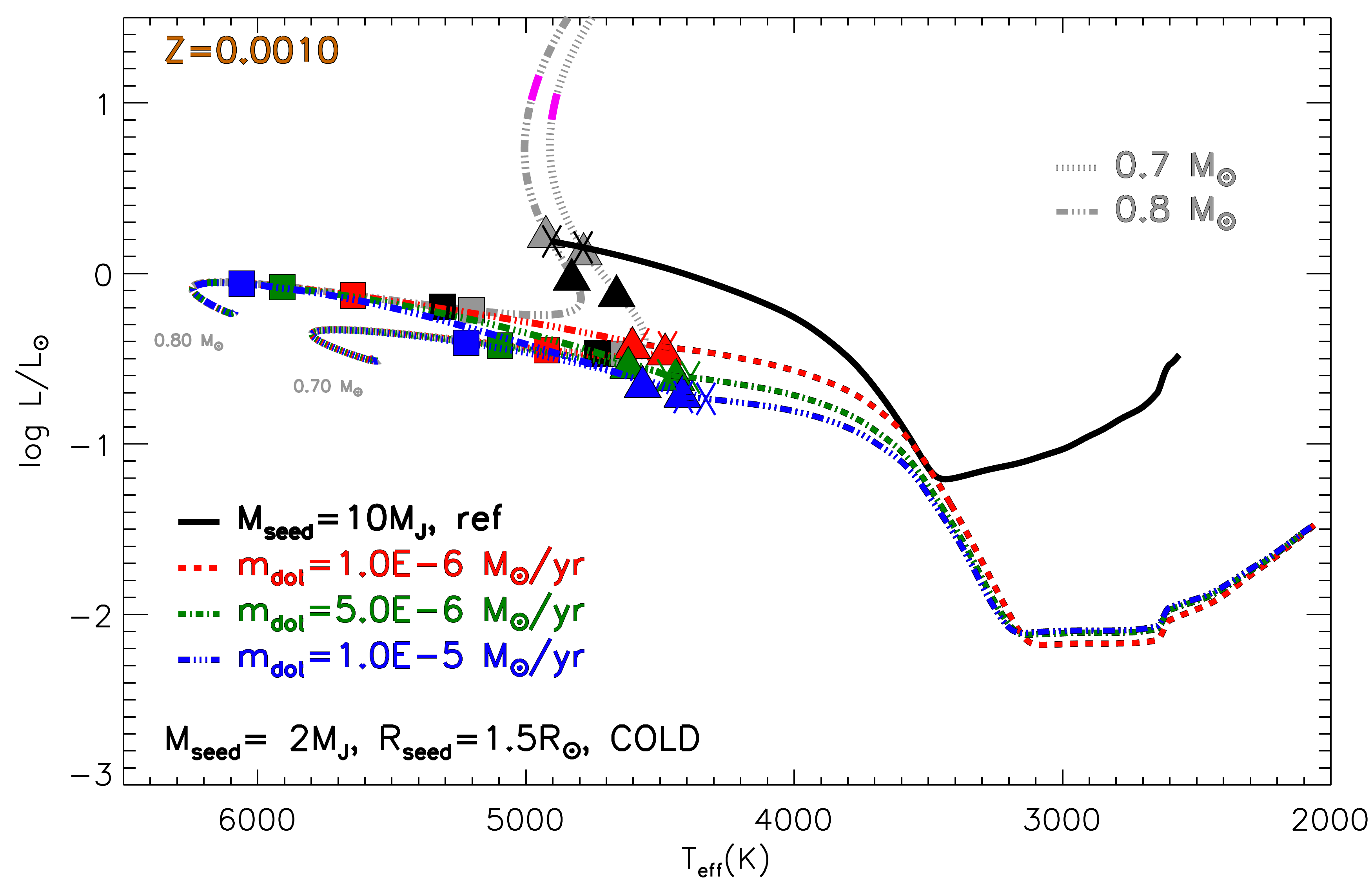}
 \includegraphics[width=0.49\linewidth]{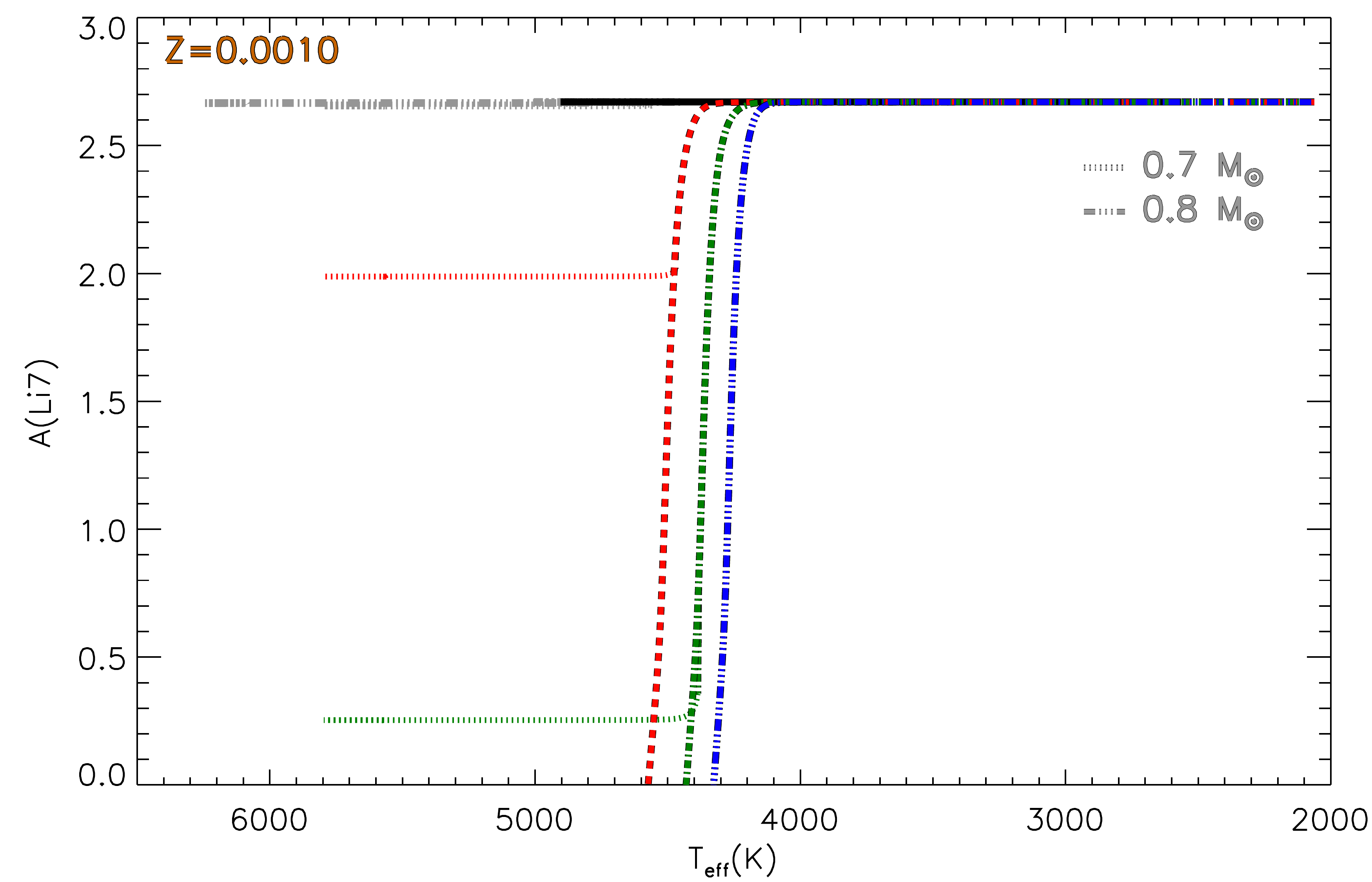}
 \includegraphics[width=0.49\linewidth]{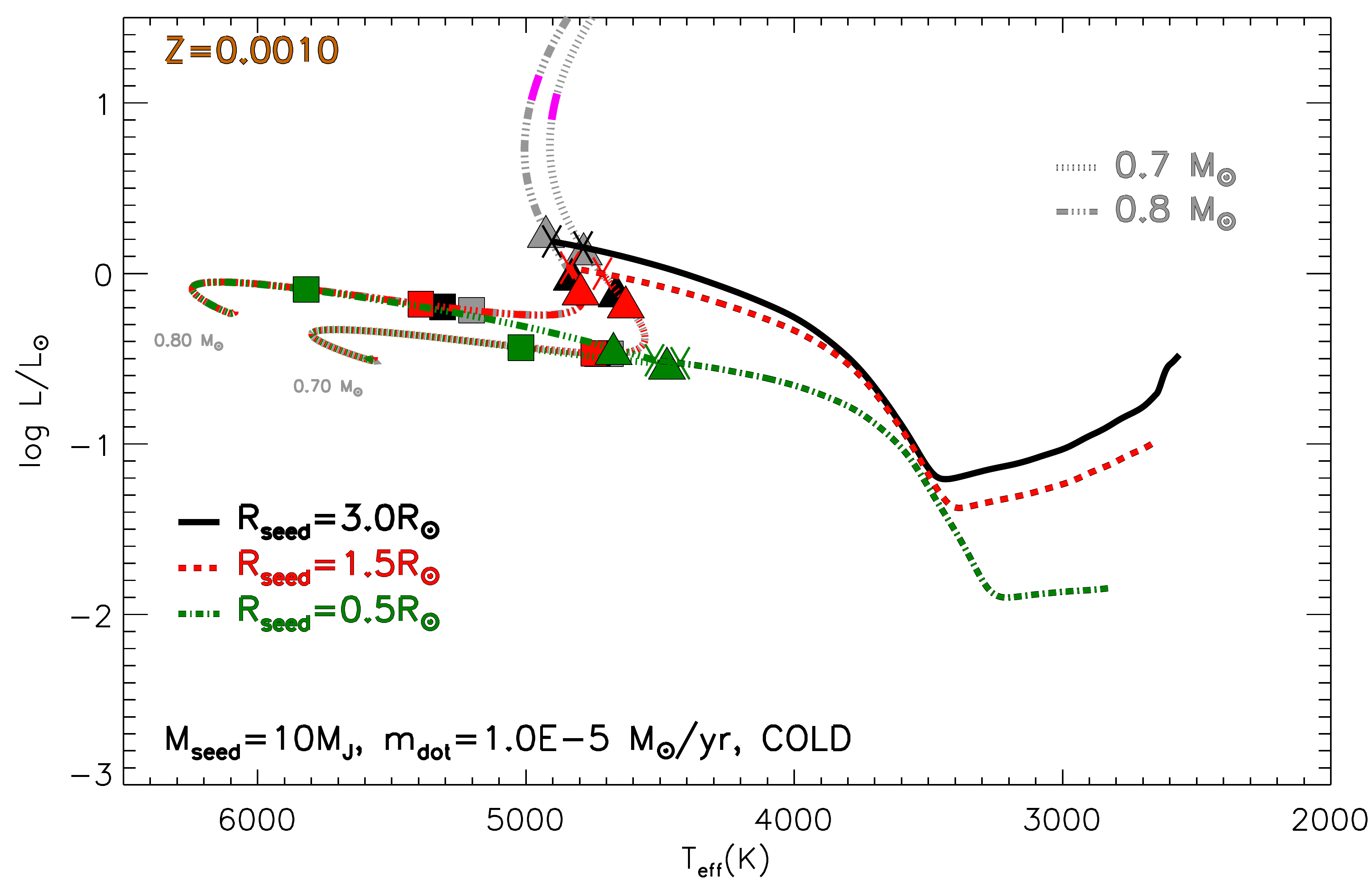}
 \includegraphics[width=0.49\linewidth]{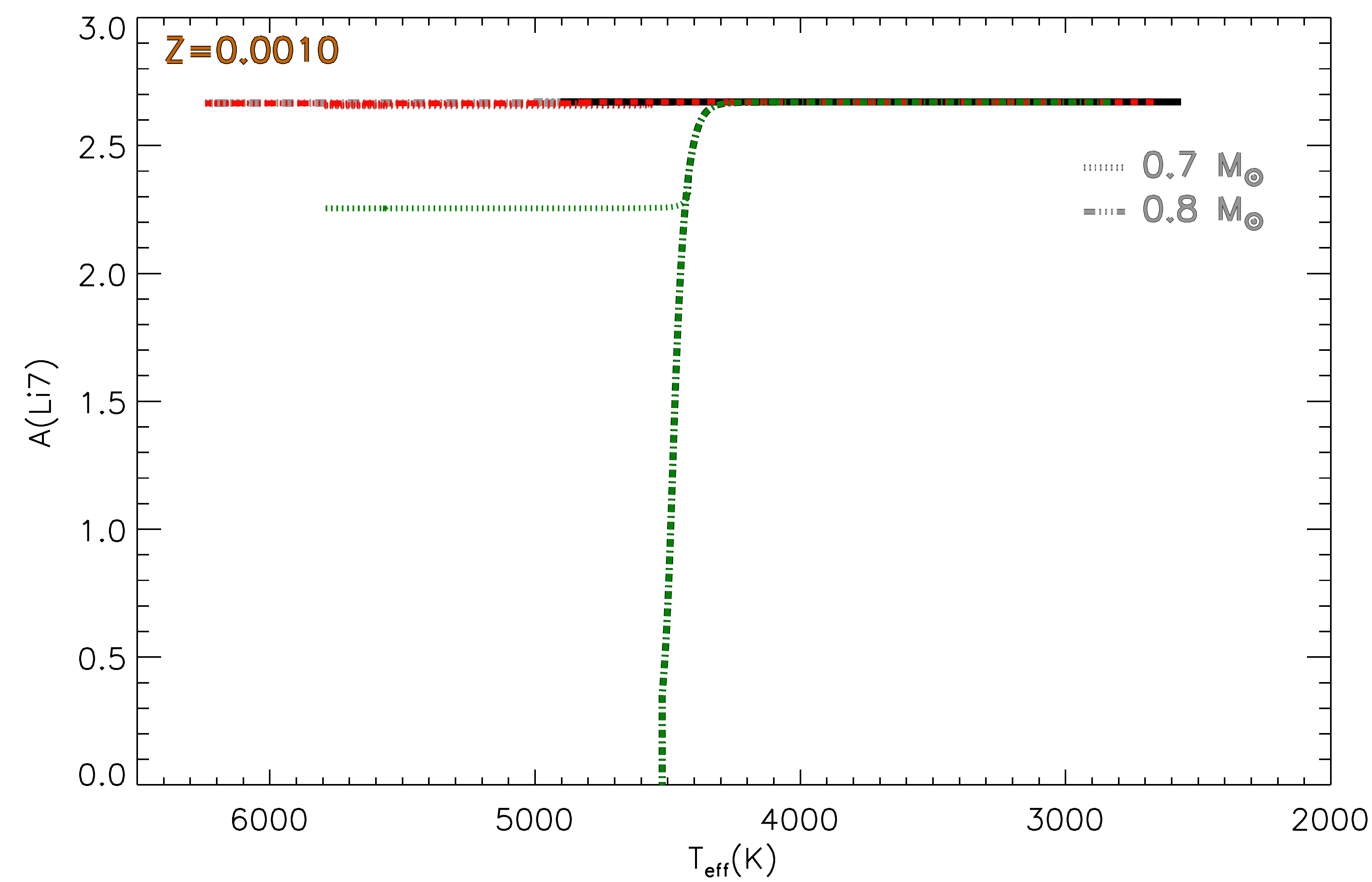}
 \includegraphics[width=0.49\linewidth]{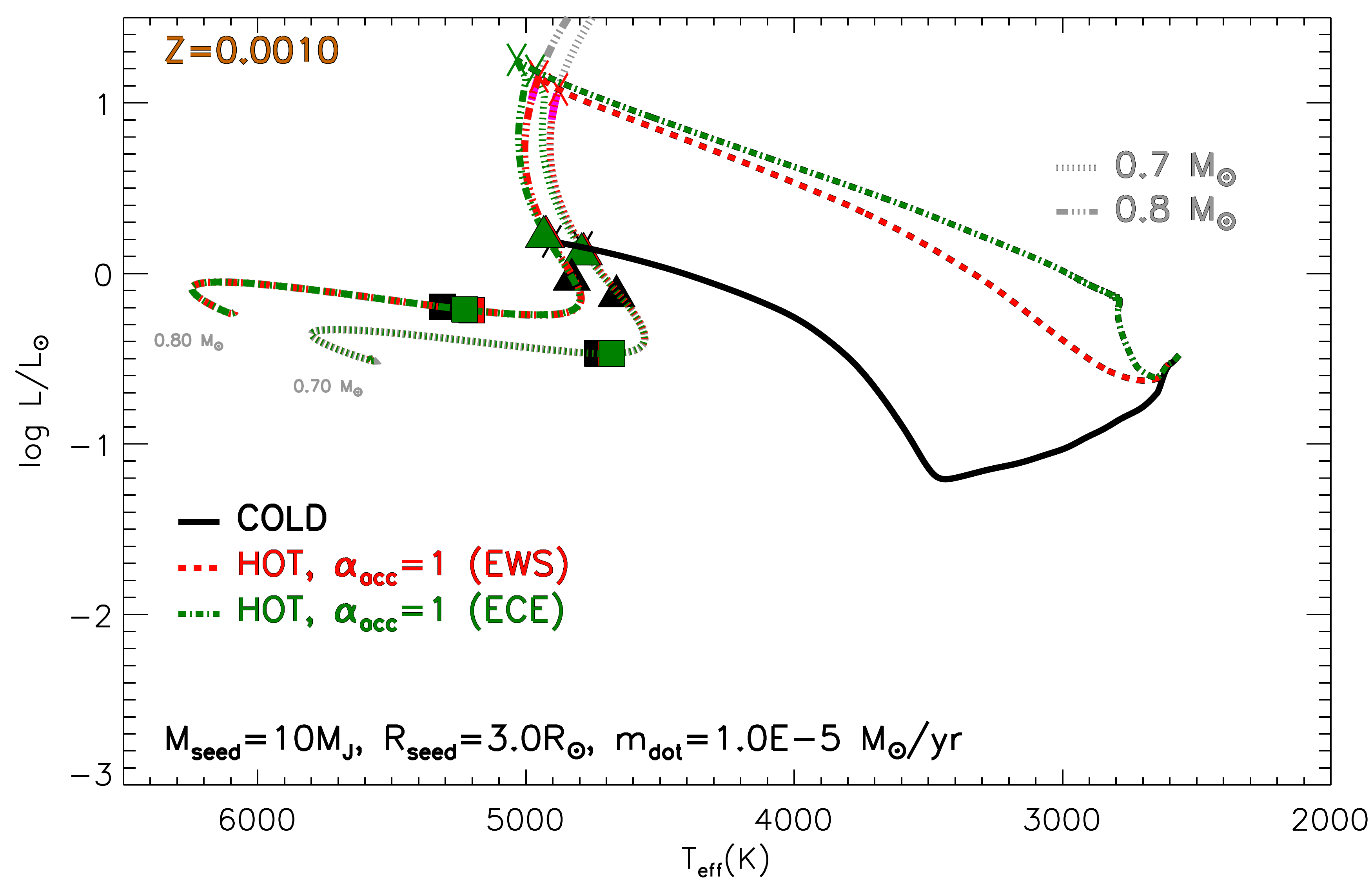}
 \includegraphics[width=0.49\linewidth]{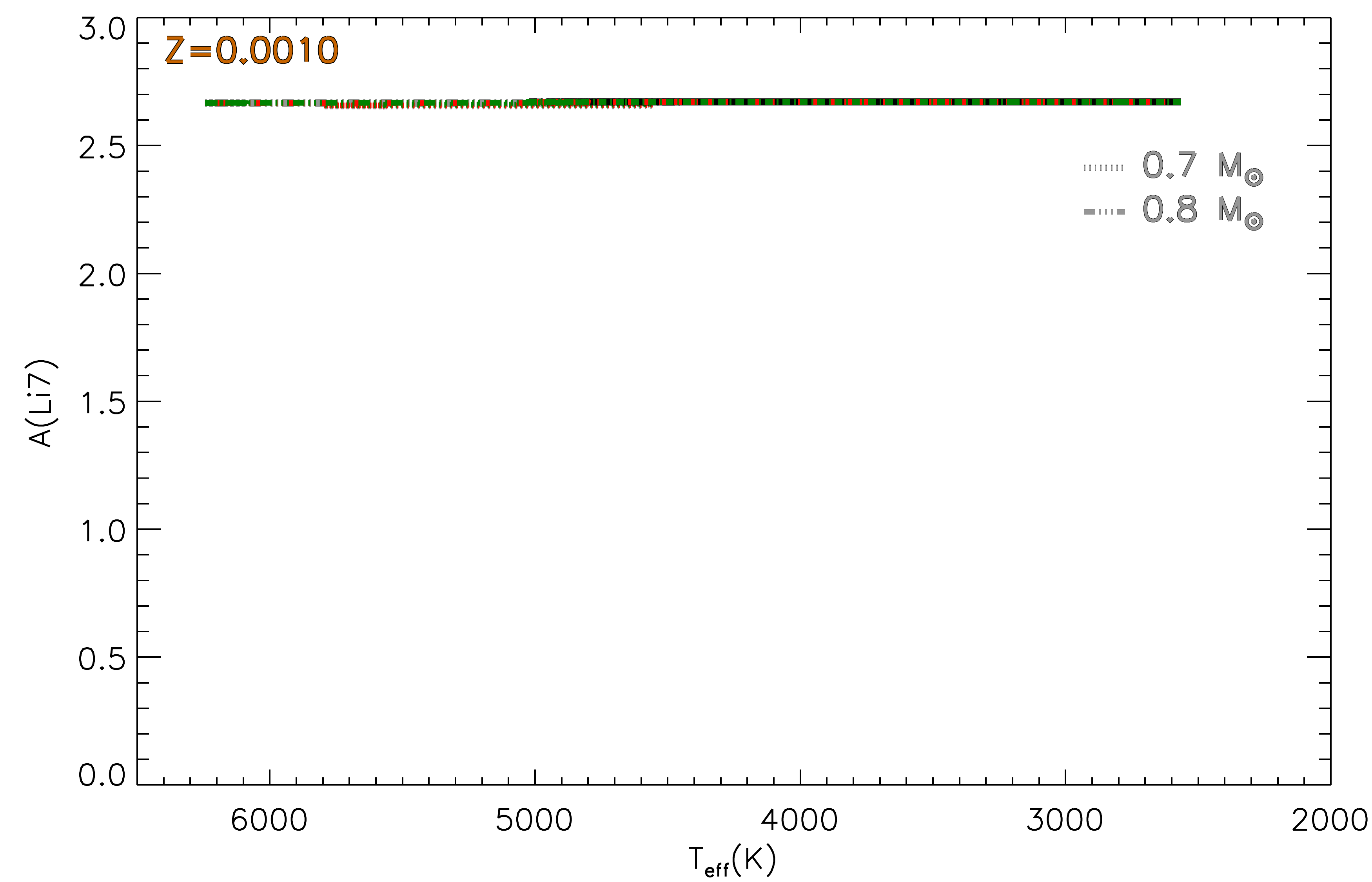}
 \caption{Evolution of accreting models with $Z=0.0010$ in the HR diagram (left panels) and in the $A(Li)$ versus \teff{} diagram (right panels). First row: Dependence on the accretion rate for LLC models (\mseed=10~\mj).  Second row: Dependence on the accretion rate for SLC models (\mseed=1~\mj). Third row: Dependence on \rseed. Bottom row: Hot accretion models.}
 \label{fig:z0p001}
 \end{figure*}
\begin{figure*}
 \centering
 \includegraphics[width=0.49\linewidth]{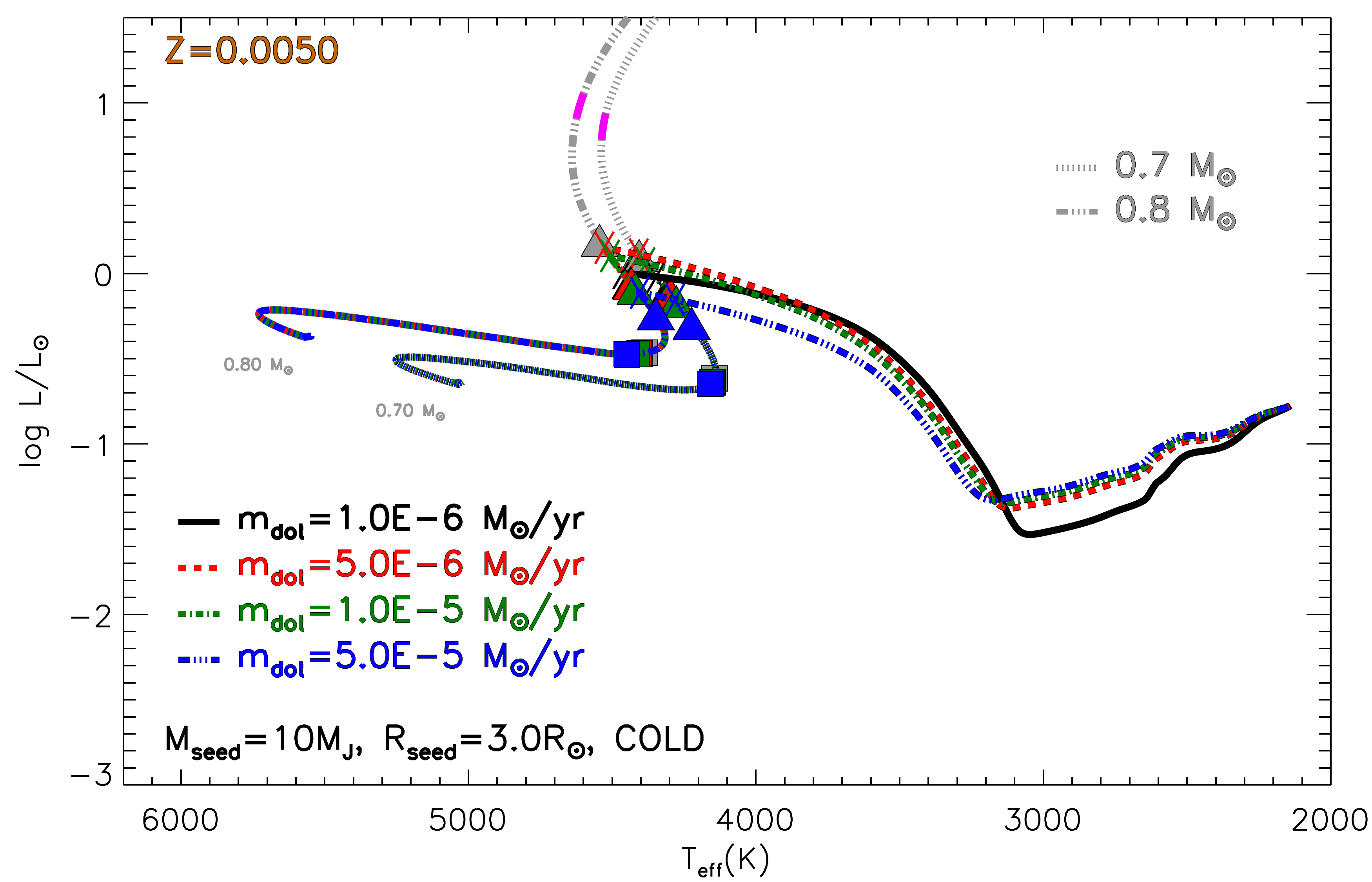}
 \includegraphics[width=0.49\linewidth]{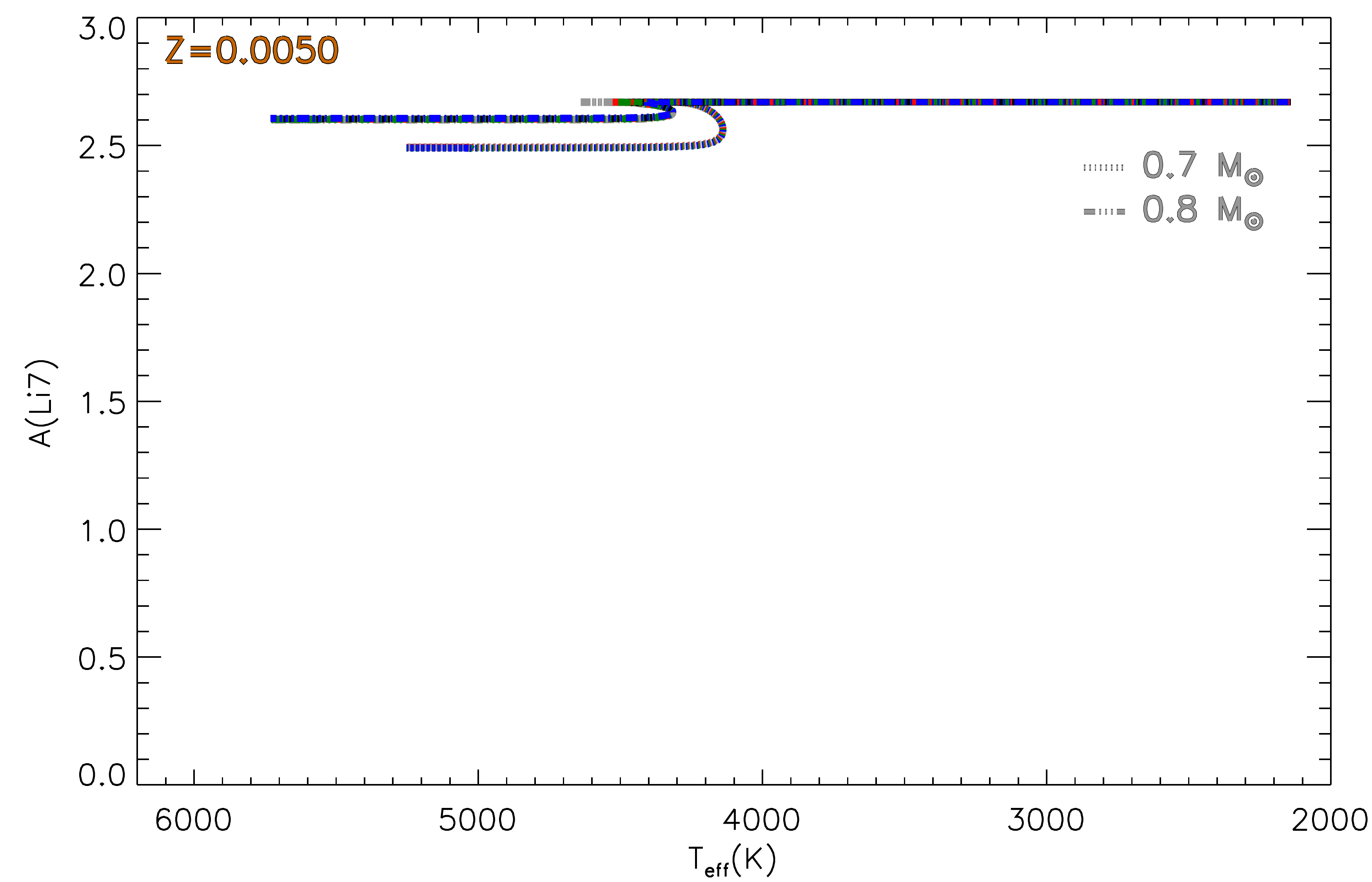}
 \includegraphics[width=0.49\linewidth]{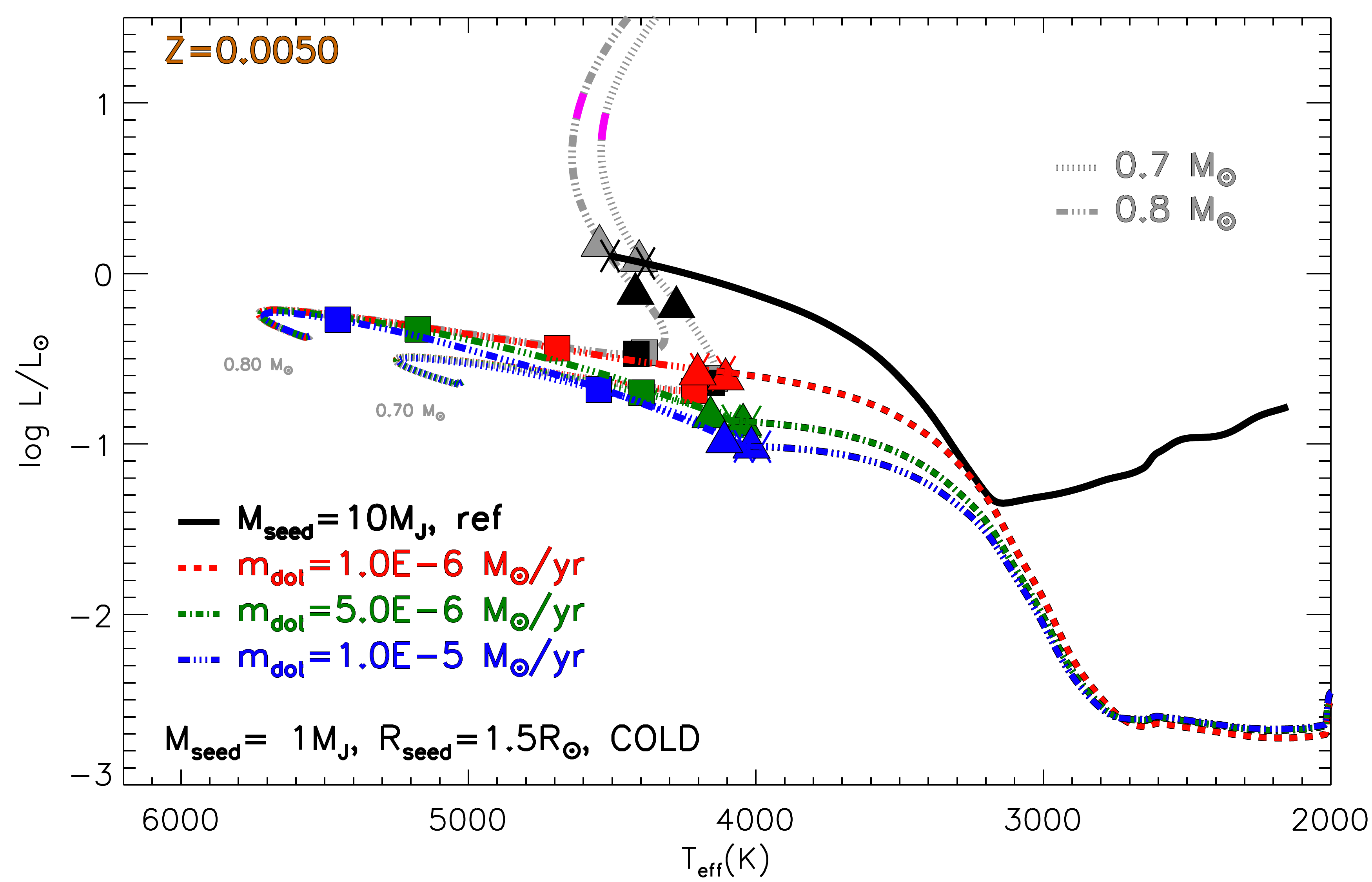}
 \includegraphics[width=0.49\linewidth]{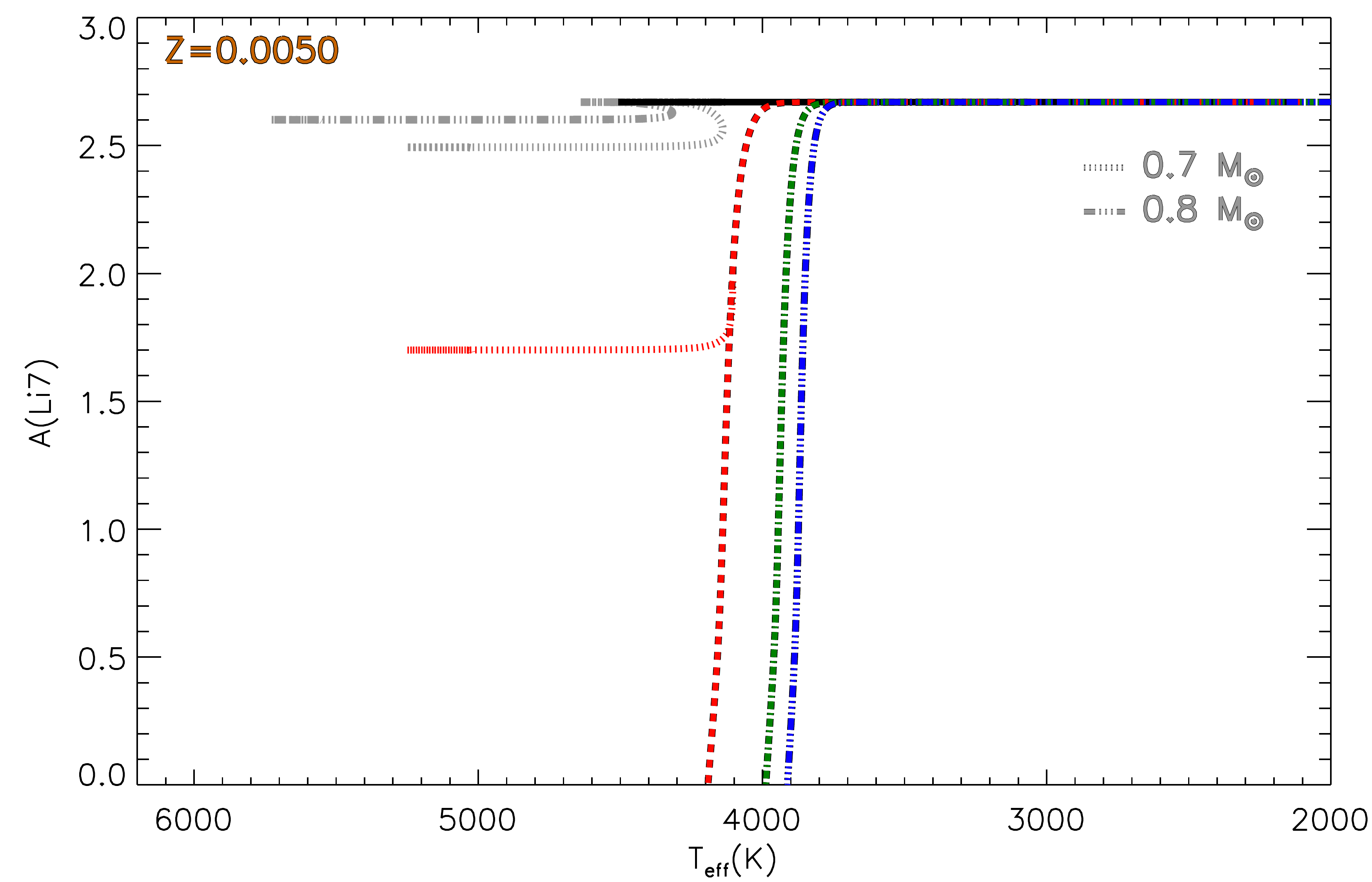}
 \includegraphics[width=0.49\linewidth]{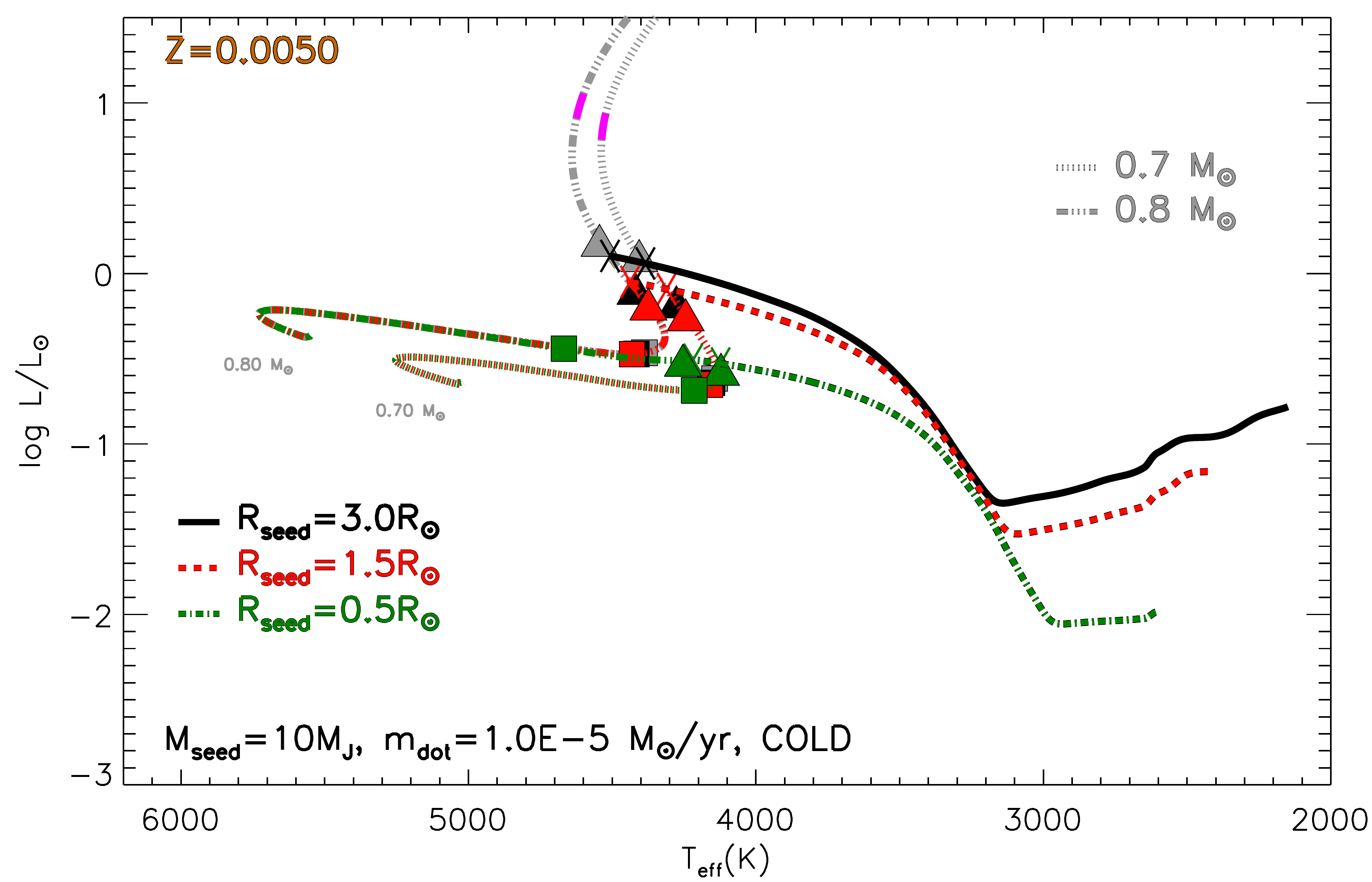}
 \includegraphics[width=0.49\linewidth]{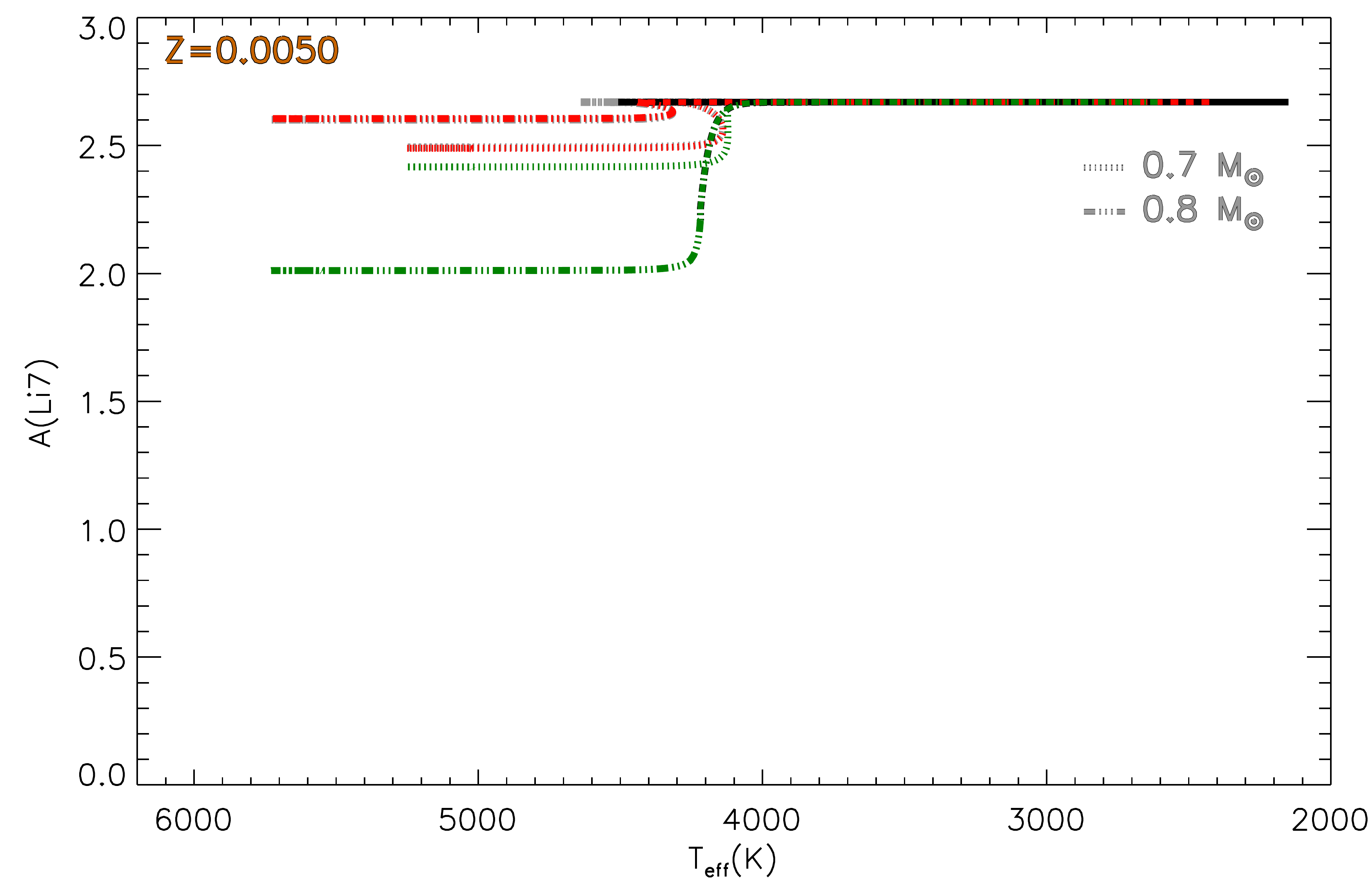}
 \includegraphics[width=0.49\linewidth]{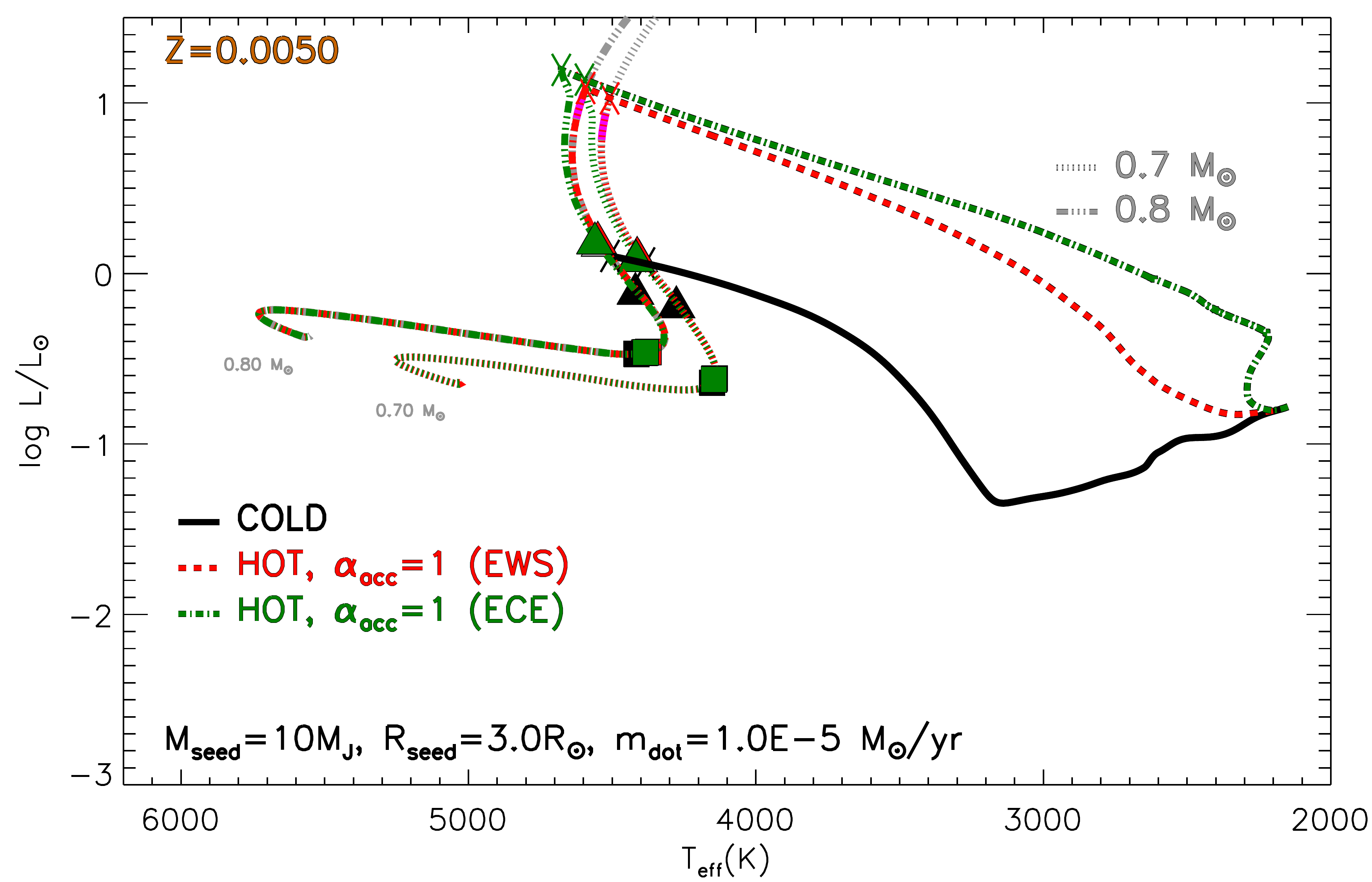}
 \includegraphics[width=0.49\linewidth]{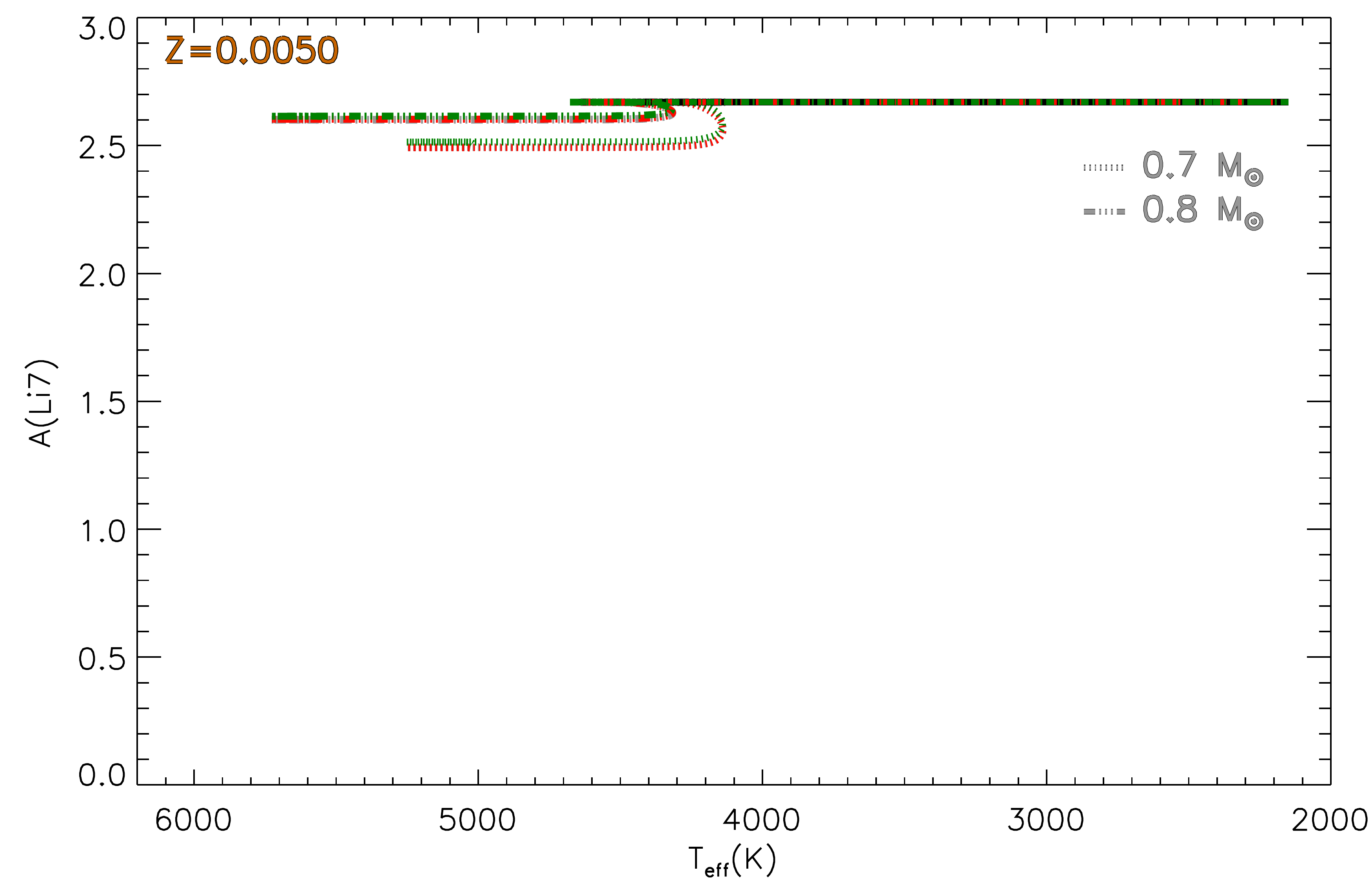}
 \caption{The same as in Fig.~\ref{fig:z0p001}, but for $Z=0.0050$.}
 \label{fig:z0p005}
 \end{figure*}

To quantify the dependence of the model results on the initial metallicity, we repeated the same computations discussed above for both hot and cold accretion cases, assuming larger initial metallicities, that is, $Z=0.0010$ ([Fe/H]$\sim -1.1$) and $Z=0.005$ ([Fe/H]$\sim -0.42$). The first value roughly corresponds to the upper metallicity end of the Spite plateau, while the last one is above the upper limit of the Spite plateau. We also show this latter case to demonstrate that the results found in this work are largely independent of the initial metallicity, over a wide metallicity range, from $Z=1\times10^{-4}$ to $Z=5\times 10^{-3}$ ([Fe/H] from $-2.1$ to $-0.42$).

Figures~\ref{fig:z0p001} and \ref{fig:z0p005} show the HR diagram and the surface lithium abundance $A(Li)$ as a function of \teff{} for the $Z=0.0010$ and $Z=0.005$ models,  respectively. Qualitatively, the effects of varying the accretion parameters are the same as those analysed in the previous sections, thus, we do not repeat the detailed discussion of the results. 

The main point to underline is that in addition to the case with a larger metallicity, we obtained two classes of solutions. For a certain set of parameters, protostellar accretion leads to, at the end of the PMS, depleted surface lithium abundances, and stars skip most of the Hayashi track.\ This class of models generally corresponds to the cold accretion scenario. 

On the other hand, for other parameter choices, it is possible to have an almost standard Hayashi track evolution at the end of accretion.\ Additionally, surface lithium abundances are very close, if not identical, to those predicted by standard non-accreting models. 

\section{Summary and conclusions}
\label{summary}
In taking the entire set of calculations presented in the previous sections into account, a clear picture does emerge. Whenever a model remains  bright and expanded during the accretion process, it eventually resembles, at the end of accretion, a standard non-accreting model of the same final mass as at the beginning of D-burning on the Hayashi track. From this stage onwards, the evolution is the same as for the case without protostellar accretion. For this class of models - which include hot accretion models, or cold accretion ones with a relatively large \mseed{} (about 10~\mj) or \rseed{} (about 3~\rsun) -- the surface lithium abundance is not affected by the phase of protostellar accretion (maximum differences with standard calculations are less than 0.1~dex).

While standard (without accretion) models are fully convective during the protostellar and PMS phase, hot accretion models develop a radiative core. Thus two physical situations are possible: The accretion energy is only deposited into the convective envelope or into the whole structure. In both cases, already at 1 Myr, the stellar characteristics and the surface $^7 Li$ abundance show no appreciable differences with respect to the case of standard models.

In a second class of models, protostellar accretion drastically affects the PMS evolution. During the protostellar phase, these models are more compact and fainter. When accretion ends, they are far from the Hayashi track location of the non-accreting counterparts and they are generally much fainter. These models almost entirely skip the Hayashi track evolution, and they deplete lithium before the end of the accretion phase. The exact amount of depletion depends on the actual combination of the accretion parameters (\mdot, \mseed{},  and \rseed), reaching in some cases the complete exhaustion of lithium in the whole star. Also, the age is significantly affected due to the large thermal timescale of the structures, which makes them very sensitive to the accretion processes that are active on shorter timescales. This class of results is typical of calculations with cold accretion and small \mseed{} (about 1~\mj) and \rseed{} ($< 1~$\rsun). The evolution during the accretion phase is not influenced in a relevant way by the accretion rate \mdot, at least in the explored parameter range. 

The possibility of having burst accretion episodes that are interspersed with quiescent phases during the whole protostellar evolution in the hot accretion scenario has also been analysed. Bursty accretion leaves models on the Hayashi track, but well below the D-burning. They then relax becoming very similar to the standard non-accreting ones. The lithium evolution is thus very similar to that of the standard models.

As a final test, we analysed the case of an initial hot accretion phase until a certain value of the total mass, which could for example mimic the evolution still inside the protostellar cloud, followed by a cold thin disc accretion phase until the prescribed final mass was reached. These models are essentially unaffected by the inclusion of the protostellar accretion phase compared to standard calculations.

Calculations were also repeated for two larger initial metallicities, namely $Z=0.001$ ([Fe/H]$\sim -1.1$, metal-rich stars on the Spite plateau) and $Z=0.005$ ([Fe/H]$\sim -0.42$) above the upper limit of the Spite plateau in order to test the dependence of the results on the various accretion parameters and modes on [Fe/H]. We found that the results are independent of [Fe/H] for the entire explored metallicity range.  

To summarise, different protostellar accretion scenarios have been explored, in the attempt to cover (at least partially) the wide range of accretion parameters suggested by theoretical predictions and observational data of young stars at high metallicity. The results show that a significant reduction of the surface lithium abundance can only be obtained for a restricted range of accretion parameters (cold accretion scenario with small \mseed{}  and \rseed), which however lead to a PMS evolution in the HR diagram that is very different to the one observed for high metallicity PMS stars. The lack of observational data for PMS stars with Spite plateau metallicities makes it impossible to restrict the range of valid accretion parameters and reach firm conclusions. Nonetheless, even from this first exploratory analysis, it seems that fine tuning of the protostellar accretion parameters is required to burn the exact amount of lithium during the protostellar and PMS phase --starting from BBN abundances-- to produce the observed constant abundance for the mass and metallicity ranges typical of the Spite plateau.
 
\begin{acknowledgements}
We thank the anonymous referee for useful and constructive comments. This work has been supported by INFN (iniziativa specifica TAsP) and by PRA (Universit\'a di Pisa 2018-2019, Le stelle come laboratori cosmici di Fisica fondamentale). SC acknowledges support from Premiale INAF MITiC, Progetto Mainstream-INAF, and  grant AYA2013-42781P from the Ministry of Economy and Competitiveness of Spain.
\end{acknowledgements}

\bibliographystyle{aa}
\bibliography{bibliography}
\end{document}